\documentclass[11pt,letterpaper,onecolumn]{article}

\usepackage[letterpaper,margin=1in]{geometry}
\usepackage[T1]{fontenc}
\usepackage[latin9]{inputenc}
\usepackage{amsmath}
\usepackage{amssymb}
\usepackage{mathtools}
\usepackage{amsthm}
\usepackage[linesnumbered,ruled,vlined]{algorithm2e}
\usepackage{graphicx}
\usepackage{pgfplots}
\pgfplotsset{compat=1.18}
\usepackage{subcaption}
\usepackage{appendix}
\usepackage{xcolor}
\usepackage{colortbl}
\usepackage{comment}
\usepackage{thm-restate}
\usepackage[numbers,square,sort]{natbib}
\usepackage{fullpage}
\usetikzlibrary{backgrounds}
\usepackage{array} 
\newcolumntype{C}[1]{>{\centering\arraybackslash}m{#1}}
\usepackage{tikz}
\usepackage{caption}
\usepackage{enumitem}
\usepackage{accents}
\usepackage{bbm}
\usepackage{hhline}

\usepackage{multirow}
\usepackage{bm}
\usetikzlibrary{positioning, fit, calc, backgrounds, shapes.geometric, patterns,arrows.meta,calc}
\usepackage{makecell} 
\usepackage{float}
\usepackage[numbers]{natbib}
\makeatletter
\@ifundefined{date}{}{\date{}}
\makeatother

\usepackage{nicematrix}
\usetikzlibrary{fit}
\usepackage{booktabs}
\usepackage{scalefnt}

\definecolor{BandBlue}{HTML}{0F52C0}
\definecolor{FillBlue}{HTML}{DCEBFF}
\definecolor{Iblue}{HTML}{0E57C9}
\definecolor{Ipurple}{HTML}{6A25B8}
\definecolor{Iteal}{HTML}{078B82}
\definecolor{Imagenda}{HTML}{D81B70}
\definecolor{Iorange}{HTML}{E65A0A}
\definecolor{StatePurple}{HTML}{6B4FA1}
\definecolor{StateFill}{HTML}{F7F5FB}
\definecolor{UBlue}{HTML}{1E5CB3}
\definecolor{VOrange}{HTML}{CC641A}
\definecolor{VerifyGreen}{HTML}{16814B}
\definecolor{VerifyFill}{HTML}{F2F8F4}
\definecolor{AmpRed}{HTML}{BC342D}
\definecolor{AmpFill}{HTML}{FCF2F1}

\definecolor{AnchorGray}{HTML}{8A8A8A}
\definecolor{AnchorFill}{HTML}{F5F5F5}
\usepackage{newtxtext,newtxmath}
\usetikzlibrary{arrows.meta,decorations.pathreplacing,calc}

\usepackage{diagbox}
\usepackage{adjustbox}

\definecolor{headergray}{gray}{0.74}
\definecolor{rowgray}{gray}{0.80}
\definecolor{closuregray}{gray}{0.74}
\definecolor{mainyellow}{RGB}{255,255,188}
\definecolor{linkblue}{RGB}{0,0,185}

\makeatletter
\renewcommand\paragraph{\@startsection{paragraph}{4}{\z@}
  {-2.0ex \@plus -0.5ex \@minus -.2ex}
  {-1em}
  {\normalfont\normalsize\bfseries}}
\makeatother

\newtheorem*{lemma*}{Lemma}

\newcolumntype{L}{>{\raggedright\arraybackslash}p{0.160\textwidth}}

\allowdisplaybreaks
\let\textstyle\relax

\usepackage{tabularx,needspace,xspace}
\usepackage{calc}

\IfFileExists{STOC-version/Macros.tex}{
  
}{}
\newif\ifcleanrender
\cleanrendertrue          

\ifcleanrender
  \usepackage[suppress]{color-edits}
\else
  \usepackage{color-edits}
\fi

\definecolor{uncheckedcolor}{rgb}{0.70,0.35,0.00}

\addauthor{MS}{red}
\newcommand{\msc}[1]{{\MScomment{#1}}}
\newcommand{\mse}[1]{{\MSedit{#1}}}

\addauthor{MF}{blue}

\addauthor{TE}{magenta}

\addauthor{MM}{teal}

\newcommand{\mme}[1]{{\MMedit{#1}}}

\newtheorem{theorem}{Theorem}[section]
\newtheorem{proposition}[theorem]{Proposition}
\newtheorem{lemma}[theorem]{Lemma}

\newtheorem{corollary}[theorem]{Corollary}
\newtheorem{definition}[theorem]{Definition}
\newtheorem{remark}[theorem]{Remark}
\newtheorem{observation}[theorem]{Observation}

\newtheorem*{theorem*}{Theorem}

\newcommand{\reals}{\mathbb{R}}

\newcommand{\eps}{\varepsilon}
\DeclareMathOperator*{\argmax}{arg\,max}
\DeclareMathOperator*{\argmin}{arg\,min}

\newcommand{\agents}{A}
\newcommand{\pequalsnp}{$\textsc{P} = \textsc{NP}$}

\newcommand{\nash}{\mathsf{NE}}
\newcommand{\bne}{\mathsf{NE}^\star}
\newcommand{\noindexcontract}{\boldsymbol{\alpha}}
\newcommand{\icontract}{\alpha_i}

\newcommand{\contract}{\noindexcontract}

\newcommand{\constar}{{\contract^\star}}

\newcommand{\multiInstance}{\langle \agents, \{T_i\}_{i \in \agents}, f, c \rangle}

\newcommand{\maxethree}{\textsc{MAX E3-Set Splitting}\xspace}
\newcommand{\F}{\mathcal{F}}
\newcommand{\I}{\mathcal{I}}

\newcommand{\revision}[1]{{\color{uncheckedcolor}#1}}
\newenvironment{revisionblock}{\par\begingroup\color{uncheckedcolor}}{\par\endgroup}
\newtheorem{introtheorem}{Theorem}

\newcommand{\revisionbibitem}[2][]{
  \color{black}
  \ifboolexpr{test {\ifstrequal{#2}{EisnerSeverance1976}} or
               test {\ifstrequal{#2}{AngluinChen2006}} or
               test {\ifstrequal{#2}{ChandrasekaranXuYu2018}} or
               test {\ifstrequal{#2}{Iwata2009simplecomp}}}{\color{blue}}{}
  \OriginalBibitem[#1]{#2}
}
\AtBeginEnvironment{thebibliography}{
  \let\OriginalBibitem\bibitem
  \let\bibitem\revisionbibitem
}

\ifcleanrender
  
  \renewcommand{\revision}[1]{{#1}}

  \renewcommand{\revisionbibitem}{\OriginalBibitem}
\fi

\usepackage[colorlinks=true,citecolor=blue,linkcolor=blue,urlcolor=black]{hyperref}
\usepackage{cleveref}

\AddToHook{cmd/appendix/before}{
    \crefalias{section}{appendix}
    \crefalias{subsection}{appendix}
    \crefalias{subsubsection}{appendix}
}
\crefname{observation}{observation}{observations}
\Crefname{observation}{Observation}{Observations}

\title{\revision{{Sources of Complexity in Multi-Agent Combinatorial Contracts}}}

\author{Tomer Ezra\thanks{Tel Aviv University, Tel Aviv, Israel. Email: \texttt{tomerezra@tauex.tau.ac.il }} \and Michal Feldman\thanks{Tel Aviv University, Tel Aviv, Israel. Email: \texttt{mfeldman@tauex.tau.ac.il }} \and Mathieu Molina\thanks{Tel Aviv University, Tel Aviv, Israel. Email: \texttt{mathieu.molina.research@gmail.com}} \and Maya Schlesinger \thanks{Tel Aviv University, Tel Aviv, Israel. Email: \texttt{maya.schlesinger@gmail.com}}}

\begin{document}
\maketitle
\thispagestyle{empty}
\msc{TODO: Need to consistentify capitalization of Optimal Contract Problem and Equilibrium Selection Problem}
\msc{TODO: make sure all approximations are $>1$}

\begin{abstract}

We study the computational complexity of multi-agent combinatorial contracts, where a principal incentivizes a team of agents to undertake hidden, costly actions. While optimal contracts can be computed efficiently for several important classes of reward functions in the single-agent setting, the corresponding multi-agent problems are computationally intractable. Existing hardness results, however, rely on a growing number of agents, leaving open whether the difficulty persists for small teams and, more fundamentally, what drives the complexity of multi-agent contract design.

To isolate the sources of complexity, we introduce the critical contracts framework, a multidimensional generalization of the critical points framework for single-agent contracts. This framework reveals three distinct computational challenges: bounding the number of critical contracts, searching over them, and selecting a principal-optimal equilibrium among the potentially many equilibria induced by a given contract.

Our results demonstrate that each of these three challenges can be the sole source of computational hardness, even with only two agents: for each challenge, we identify a prominent class of reward functions, capturing either substitutability or complementarity among actions, for which that challenge is hard while the other two remain tractable. Thus, the source of computational hardness depends fundamentally on the structure of the reward function, revealing that even with just two agents, strategic interactions can make optimal contract design computationally intractable in fundamentally different ways. Complementing these hardness results, we obtain a tight $O(\sqrt{n})$-approximation for subadditive rewards.

\end{abstract}

\clearpage
\pagenumbering{arabic}
\section{Introduction}

Contract theory models the fundamental economic challenge of aligning incentives under asymmetric information.
Its cornerstone is the hidden-action model of contract theory~\cite{ross1973economic,holmstrom1979moral,grossman-hart-1983}, where a principal (she) delegates a project to an agent (he) who chooses an action she cannot observe, and incentivizes him through outcome-contingent payments. A prominent extension of the classical framework is the \emph{single-agent combinatorial-action} model \cite{dutting2022combinatorial}, where the agent selects a \emph{set} $S\subseteq T$ of costly actions, incurring their total cost $c(S)$. A normalized, monotone reward function $f:2^T\to[0,1]$ then maps the chosen action set $S$ to the probability that the project succeeds; the principal receives a reward of $1$ upon success and $0$ otherwise, so $f(S)$ also represents the expected reward. Crucially, the principal observes only the final outcome, not the agent's chosen action set $S$. Thus, to incentivize costly actions from the agent, the principal offers a contract $\alpha\in[0,1]$, committing to pay him $\alpha$ if the project succeeds. The agent then chooses an action set $S$ to maximize his expected utility, defined as the payment from the principal minus his cost, i.e., $\alpha f(S)-c(S)$. The \emph{Optimal Contract Problem} is finding a contract $\alpha$ that maximizes the principal's utility 
$(1-\alpha)f(S)$ where $S$ is the agent's best response to $\alpha$.
We note that the reward function $f$ is the key element in the combinatorial structure of this model. It can capture substitutability and complementarity between the actions' effects, and its structure --- or lack thereof --- is the driving force of the tractability of the Optimal Contract Problem.

More recently, \citet{multimulti} extended this framework to the \emph{multi-agent combinatorial-action} model, in which the principal interacts with $n$ agents who collectively control $m$ costly actions. The project's success probability depends on the union of the actions chosen by all agents. This extension introduces two key differences from the single-agent setting. First, the principal commits to an independent contract with each agent, modeled as a vector $\contract \in [0,1]^n$, where $\alpha_i$ is the payment to agent $i$ upon success. Second, each agent's optimal choice of actions depends not only on the contract but also on the actions chosen by the other agents, giving rise to strategic interactions and equilibria. In this setting, the principal chooses a contract $\contract$ and recommends an equilibrium action profile $S$ for the agents to play. The \emph{Optimal Contract Problem} in this model is to compute a contract $\contract$ and a corresponding equilibrium $S$ that maximize the principal's expected utility.

This multi-agent setting introduces three significant challenges for the Optimal Contract Problem, relative to its single-agent counterpart. The first two arise from the multi-dimensional contract space. First, the number of candidate contracts can grow substantially. Although the contract space is continuous, there is a natural discrete set of candidate contracts to consider ---  we  call these \emph{critical contracts}, defined as the minimal payment vectors required to induce a given action profile (see \Cref{sec:model}). When moving from the single-agent to the multi-agent setting, the number of critical contracts can grow exponentially with the dimension of the contract space. Second, the multi-dimensional contract space lacks the natural (full) ordering of the single-agent setting, where higher payments naturally correlate with higher effort. This loss of structure complicates the identification of critical contracts, and thus the search over them, even when there are only polynomially many such contracts. Finally, strategic interactions among the agents introduce a third challenge, as a fixed contract $\contract\in[0,1]^n$ may induce multiple equilibria. While \emph{some} equilibrium can be computed with a demand query~\cite{multimulti}, it is unclear how to efficiently find one that maximizes the principal's utility.

The computational significance of these challenges is reflected in existing results. For both gross substitutes (GS) and supermodular reward functions, two prominent classes of set functions capturing substitutability and complementarity, respectively, the Optimal Contract Problem admits efficient algorithms in the single-agent setting~\cite{dutting2022combinatorial, contractsBeyondGS, deo2024supermodular}, whereas they become computationally intractable in the multi-agent setting~\cite{deo2024supermodular,equalpay}. However, existing hardness results crucially rely on a growing number of agents, and thus an exponentially exploding number of critical contracts. This focus on arbitrarily large teams not only leaves open the question of tractability in small team settings, a regime of wide practical significance. It also masks our ability to understand the source(s) of complexity in multi-agent contracts; for example, is it still hard when there are few critical contracts? Both our interest in tractability of small team settings and a desire to understand the fundamental complexities of this model lead to our central research question.

\begin{center}
\emph{Where does the computational hardness stem from in multi-agent combinatorial contracts?}
\end{center}

Our results reveal that each of the three challenges can be the sole source of hardness, depending on the setting, and that largely speaking, hardness, and even hardness of approximation, arises already for $n=2$ agents. 

  \medskip

\subsection{Our Framework} 
Our first contribution is the \emph{critical contracts} framework, which generalizes the notion of \emph{critical points} from the single-agent combinatorial-action model. The critical points framework is the only known method for computing optimal contracts in the single-agent setting, and has been widely used in the literature.

Recall that in the single-agent model, a contract is a scalar $\alpha\in [0,1]$, to which the agent best-responds with a subset of actions that maximizes his utility. A \emph{critical point} is a value of $\alpha$ at which the agent's best response changes. Since a larger value of $\alpha$ that elicits the same agent action set is always sub-optimal, the optimal contract must be a critical point. Using this approach, to efficiently compute optimal contracts in the single-agent model we require three ingredients
(i) A polynomial number of critical points, (ii) a way of enumerating the critical points, and (iii) a way of evaluating the principal's utility from a given critical point. In the single-agent model, the agent's best response to a contract can be simply computed by a standard demand query \cite{dutting2022combinatorial}, so the third requirement is satisfied whenever a demand query can be implemented. As to the second requirement, by utilizing the Eisner-Severance algorithm in this setting, enumerating the critical points (in time polynomial in their number) only requires the ability to compute the agent's best response to a contract \cite{contractsBeyondGS, deo2024supermodular}, meaning a demand oracle again suffices. Thus, when a demand query can be implemented, or given as an oracle, the only remaining challenge is the first: bounding the number of critical points by a polynomial.

We extend this approach to multi-agent contracts by defining \emph{critical contracts} --- the natural multi-dimensional parallel of critical points. A critical contract is the coordinate-wise minimal contract vector that incentivizes a specific equilibrium from the agents\footnote{This is well-defined because the set of contracts which incentivize a specific equilibrium is a closed box, see \Cref{sec:model}.}. Now, efficient computation of the optimal contract again requires three components: (i) A polynomial number of critical contracts, (ii) a way of searching over them, and (iii) a subroutine that, given a fixed contract, returns an equilibrium which maximizes the principal's utility. We refer to the third problem as the \emph{Equilibrium Selection Problem}.

All three of these components present significant new challenges relative to their single-agent counterparts.
We investigate these three components and the overall tractability of the Optimal Contract Problem for various classes of reward functions capturing substitutability and complementarity between actions, paying special attention to the regime of small teams.

\subsection{Our Results}
\mse{We present structural, algorithmic, and hardness results for various classes of reward functions with a focus on small teams. Although our results have wider implications on the tractability of the Optimal Contract Problem, we view the isolation of the different sources of hardness as a large part of our contribution, demonstrated in \Cref{table:weakest_link}}.
\paragraph{Supermodular Rewards (Hardness Originating from Many Critical Points).}
When $f$ is supermodular, we observe that for any fixed contract the incentivized equilibria form a lattice structure, allowing us to solve the Equilibrium Selection Problem efficiently (see \Cref{thm:polytime-optimal-equilibrium-supermodular}). However, the Optimal Contract Problem is hard to approximate, even for only $n=2$ agents:
\begin{introtheorem} [Two-agent supermodular hardness of approximation; \Cref{thm:supermodular-hardness}]
For supermodular rewards, no deterministic polynomial-time algorithm guarantees any finite multiplicative approximation to the optimal principal utility, even with two agents, unless \pequalsnp{}.
\msc{can a randomized algorithm do anything? there's a gap here between our positive and negative. Chat says about our proof "The theorem can therefore safely be extended to any finite factor with constant success probability, or any polynomially bounded factor in expectation, unless \(\mathrm{NP}=\mathrm{RP}\)."}
    
\end{introtheorem}
Notably, this result leverages a construction with a super-poly number of critical contracts to derive the underlying hardness. Thus, in contrast to the single-agent setting where the number of critical points under supermodular rewards is bounded by the number of actions, even the addition of only one extra agent blows up the number of critical contracts.
In order to further isolate the source of hardness, we ask whether for instances with poly-many critical contracts the problem becomes tractable. We find that it does; \Cref{thm:polycritical-supermodular} presents a deterministic algorithm which searches for the best critical contract in time polynomial in the input size and the number of critical contracts in $[0,1]^n$.

\paragraph{Subadditive Rewards (Tight Approximation Guarantee).} For subadditive rewards, there is a gap in the literature as to which approximations are possible. \citet{duetting2022multi} rule out a $o(\sqrt{n})$-approximation, while \citet{equalpay} give an $O(n)$-approximation. \mme{We close this gap with a deterministic polynomial-time
$O(\sqrt{n})$-approximation using value and demand queries.
Our algorithm also gives tight $\Theta(\sqrt n)$ bounds on the loss from restricting to equal-pay contracts and on the gap between the principal's optimal utilities under CCE and PNE (\Cref{cor:subadditive-equal-pay,cor:subadditive-cce-pne}, respectively). These improve the previous $O(n)$ upper bounds and match the $\Omega(\sqrt n)$ lower bounds of \citet{equalpay} and \citet{dutting2025black}, respectively.}
\begin{introtheorem}[Tight subadditive approximation; \Cref{thm:subadditive-sqrt-n}]
For subadditive rewards, there is a deterministic polynomial-time $O(\sqrt n)$-approximation algorithm for the Optimal Contract Problem using value and demand queries. 
\end{introtheorem}

\paragraph{WMRF Rewards (Hardness Originating from Equilibrium Selection).} When $f$ is a Weighted Matroid Rank Function (WMRF), we observe that critical contracts can be structurally characterized based on the element weights, which determine the set of possible marginals of $f$. Furthermore, when $n$ is bounded, this characterization gives a polynomial (in the number of actions) bound on the number of critical contracts. Thus, for small teams, any remaining computational hardness must originate from the Equilibrium Selection Problem. We show that indeed, both the Equilibrium Selection Problem and the Optimal Contract Problem admit no PTAS, even for $n=2$, making the known $O(1)$-approximation (for teams of any size) asymptotically tight even for small teams. 
\msc{I think we never say that this is the hardness result known from the large-teams regime (same note for the supermodular hardness}
\begin{introtheorem}
    [No PTAS for WMRF rewards; \Cref{thm:no-ptas}]
For WMRF rewards, neither the Optimal Contract Problem nor the Equilibrium Selection Problem admits a deterministic PTAS, even with two agents, unless \pequalsnp{}. 
\msc{Chat says this too can be done for randomized algorithms}
\end{introtheorem}

In contrast, for {\em unweighted} matroid rank functions, we show that the Equilibrium Selection Problem can be efficiently solved (see \Cref{sec:unweighted}), yielding a polynomial-time algorithm for the Optimal Contract Problem for small teams. This stands in contrast to the setting with an arbitrary number of agents, where the Optimal Contract Problem admits no PTAS \cite{equalpay}.

\paragraph{Gross Substitutes Rewards (Hardness Originating from Search Over Critical Contracts).} 
Recall that in the single-agent setting, enumerating the critical points can be implemented efficiently using the Eisner-Severance approach, and we show that for supermodular rewards in the multi-agent model a search can be efficiently implemented deterministically. It is thus natural to wonder whether a small number of critical contracts and a tractable Equilibrium Selection Problem imply tractability of the Optimal Contract Problem. We answer this question in the negative, for the class of Gross Substitutes (GS) rewards, isolating hardness of searching over critical contracts.
\begin{introtheorem} [No PTAS - hardness of search over critical contracts; \Cref{thm:gs-oracle-lb}]
    For GS rewards, even restricted to two-agent instances with only $O(m)$ critical contracts, there is no randomized PTAS for the Optimal Contract Problem, even given oracle access to the Equilibrium Selection Problem. 
\end{introtheorem}

\begin{table}[!ht]
\centering
\definecolor{rtAlgBg}{HTML}{EAF2F7}
\definecolor{rtHardBg}{HTML}{FAEDE3}
\definecolor{rtAlgFg}{HTML}{24546C}
\definecolor{rtHardFg}{HTML}{8A4829}
\small
\setlength{\tabcolsep}{3pt}
\renewcommand{\arraystretch}{1.15}
\begin{tabular*}{\linewidth}{@{\extracolsep{\fill}}lcccc}
\toprule
Reward class
& \makecell{Number of\\critical\\contracts}
& \makecell{Critical\\contract\\search}
& \makecell{Equilibrium\\Selection\\Problem}
& \makecell{Optimal\\Contract\\ Problem} \\
\midrule
MRF
& \cellcolor{rtAlgBg}Poly & \cellcolor{rtAlgBg} $\mathrm{P}$ & \cellcolor{rtAlgBg}$\mathrm{P}$ & \cellcolor{rtAlgBg}$\mathrm{P}$ \\
\addlinespace[0.6em]
WMRF
& \cellcolor{rtAlgBg} Poly & \cellcolor{rtAlgBg} $\mathrm{P}$ & \cellcolor{rtHardBg} No PTAS
& \cellcolor{rtHardBg}
  \makebox[\widthof{No finite factor}][c]{
    \makecell{$O(1)$~\cite{multimulti}\\No PTAS}
  } \\
\addlinespace[0.6em]
``Easy'' GS
& \cellcolor{rtAlgBg} Poly & \cellcolor{rtHardBg}Query-hard & \cellcolor{rtAlgBg} $\mathrm{P}$
& \cellcolor{rtHardBg}
  \makebox[\widthof{No finite factor}][c]{
    \makecell{$O(1)$~\cite{multimulti}\\No PTAS}
  } \\
\addlinespace[0.6em]
Supermodular
& \cellcolor{rtHardBg}Super-poly & \cellcolor{rtAlgBg} $\mathrm{P}$ & \cellcolor{rtAlgBg}$\mathrm{P}$ & \cellcolor{rtHardBg}No finite factor \\
\bottomrule
\end{tabular*} \caption{An overview of our structural and hardness results for small teams. Hardness results apply even for $n=2$, while positive results apply for any constant $n=O(1)$.
The rows refer to the class of set function to which $f$ belongs. The search column measures running time in the input size and the number of critical contracts, given access to equilibrium selection.
The class ``Easy GS'' refers to restricted instances of Gross Substitutes $f$ where there are polynomially-many critical points and the Optimal Selection Problem is efficiently solvable, so that the critical contract search problem is isolated. The search lower bound holds even given an oracle returning all principal-optimal equilibria. This table demonstrates how each of the three challenges can be isolated as the sole source of complexity.
}\label{table:weakest_link}
\end{table}

\subsection{Our Techniques} 

\paragraph{Supermodular Rewards (Efficient Equilibrium Selection).}
When $f$ is supermodular, we observe that every fixed contract induces a supermodular game among the agents. Thus, this induces a lattice structure over the equilibria, and there is a greatest equilibrium under set inclusion, denoted by $S^+(\contract)$. Since $f$ is monotone, this equilibrium maximizes the principal's utility whenever $\sum_i\alpha_i\leq1$. To compute it, we start from the full action profile and iteratively let each agent choose his greatest best response, which can be computed by submodular minimization. Monotonicity of greatest best responses ensures that the sequence of action profiles only decreases. Whenever the profile changes, at least one action is removed, so the process reaches the greatest equilibrium after at most $m$ strict iterations. This solves the \emph{Equilibrium Selection Problem} efficiently for any number of agents.

\paragraph{Supermodular Rewards (Hardness Originating from Many Critical Contracts).}
Despite the tractability of equilibrium selection, even two agents can generate $2^{\Omega(\sqrt m)}$ critical contracts (\Cref{prop:many-critical-contracts-supermodular}). We use this large family to encode an instance of \textsc{Subset Sum} with $k$ items into the contract space. The construction first defines a quadratic reward on a bipartite graph, where each agent controls the vertices on one side, and an edge contributes its weight to the reward when both endpoints are selected. If either agent's payment is held fixed, increasing the other's can only add actions to the greatest equilibrium, so it changes at most $m$ times, giving only linear complexity along such a slice. We instead move the payments jointly along the anti-diagonal, increasing one agent's payment while decreasing the other's, and create $2^k$ disjoint intervals along this line, each corresponding to a subset of items through the selected actions; see \Cref{fig:hard_construction}. Thus, all subsets are encoded at the same total payment share, so that, from the principal's perspective, the contracts differ only through the rewards they induce.

The main challenge is to construct those $2^k$ disjoint intervals. A naive encoding would let each agent select an item once his payment reaches a fixed threshold, regardless of which other items have been selected. With $O(k)$ thresholds per agent, this creates only $O(k^2)$ regions in payment space, far fewer than the $2^k$ subsets we need to represent. We therefore construct the encoding recursively, in a way reminiscent of the Cantor set construction (\Cref{fig:interval_encoders}). Each interval encoding the choices for the first $i$ items contains two smaller, separated intervals, corresponding to including or excluding item $i+1$. Repeating this refinement yields $2^k$ intervals. The same item must therefore be encoded differently depending on the preceding choices, as earlier choices determine where the next interval splits. This dependence naturally suggests directed edges from earlier items to later ones. However, the two agents share the same reward function, so the graph must be undirected. We overcome this by introducing a three-edge path gadget that simulates a directed edge within the relevant payment range (\Cref{fig:three-edge-path}). The gadget connects source and target through two intermediate actions with carefully chosen costs and increasing edge weights, making the source action necessary for a large gain from selecting the target action, while adding too little incentive to change whether the source itself is selected.

Two verification vertices then test whether the encoded subset has the target weight. If the test succeeds, an amplification gadget activates an edge of very large weight. Overall, a feasible \textsc{Subset Sum} instance admits large principal utility, whereas an infeasible instance has bounded reward at every equilibrium of every feasible contract. Finally, we add a costly action required for any positive reward, preserving supermodularity. Its cost can be covered only in feasible instances, yielding positive versus zero optimal principal utility. Any deterministic polynomial-time approximation within a finite multiplicative factor would decide whether positive utility is achievable, solving \textsc{Subset Sum} in polynomial time, and therefore imply \pequalsnp{}.

\paragraph{Deterministic Exact Optimization for Supermodular Rewards.}

Since equilibrium selection is tractable for supermodular rewards, the remaining difficulty is searching over the $K$ critical contracts. This is an optimization problem over a hidden family, we can evaluate a given contract but we do not know the critical contracts or their coordinate values. The key is that supermodularity, through the lattice structure of the equilibria of supermodular games, gives this search problem a monotone structure. By Topkis's monotonicity theorem, increasing payments coordinatewise can only enlarge the greatest equilibrium. Moreover, lowering a contract to the critical contract of its greatest equilibrium preserves that equilibrium throughout the box between the two contracts. Each hidden critical contract therefore specifies payment requirements that guarantee at least its associated reward.
Writing $\contract^{(1)},\ldots,\contract^{(K)}$ for the critical contracts throughout $[0,1]^n$ and $v_j=f(S^+(\contract^{(j)}))$, we obtain the following representation of the supermodular rewards $f(S^+(\contract))
=\max_{j:\,\contract\geq\contract^{(j)}}v_j$.
Thus, equilibrium computations give access to the maximum reward among the hidden payment requirements satisfied by a proposed contract. We exploit this representation to reconstruct the payment--reward Pareto frontier, proving a discrete optimization result over hidden hypergraphs as our main tool.

We tackle as an intermediate step the following problem which can be of independent interest. Consider a hypergraph whose vertices and their nonnegative weights are known, but whose $K$ hyperedges are hidden. A containment query asks whether a given vertex set contains an hyperedge. Reconstructing the entire hypergraph can require exponentially many queries in $K$, see \cite{AngluinChen2006}. Nevertheless, we show that learning enough to optimize and find the minimum weight hyper-edge can be substantially easier than learning the entire hidden family.

\medskip
\noindent
\textbf{Informal Theorem (Hidden Hypergraph Optimization; \Cref{thm:hidden-edge}).}
Given an hypergraph with $K$ hidden hyperedges and known nonnegative rational vertex weights, a deterministic algorithm finds a minimum-weight hyperedge using a number of containment queries and runtime polynomial in $K$ and the explicit input size.
\medskip

We stress that $K$ is unknown to the algorithm.  Finding one hyperedge is straightforward, starting from a positive query, we delete a vertex whenever the remaining set still contains a hyperedge, continuing until no further deletion is possible. The difficulty is that any query containing a known edge gives a positive answer regardless of whether it also contains a better one. We therefore need query sets that exclude every known edge while retaining a cheaper one, if one exists. Uniformly sampling vertex sets can retain a particular cheaper edge with only exponentially small probability in its size. By derandomizing a sampling argument, we construct a polynomial-size list of vertex sets. Each excludes every known hyperedge, and every hyperedge cheaper than all known ones is contained in at least one set on the list. A positive query lets us extract a new hyperedge. If all queries are negative, a cheapest known hyperedge is optimal.

Using this result, we reconstruct the payment--reward Pareto frontier by repeatedly finding a cheapest contract that improves on the current reward starting from the zero contract. For each search, we build an auxiliary hypergraph whose edges represent hidden payment requirements rounded down to the shares discovered so far. A minimum-weight edge therefore gives a lower bound on the payment needed to improve the reward. We either find a contract that achieves the improvement at that payment or discover a new critical share by binary search and refine the hypergraph. After at most $K$ reward improvements, we discard dominated pairs and select a feasible frontier contract maximizing the principal's utility. We return this contract and its greatest equilibrium.

\paragraph{A Tight Approximation for Subadditive Rewards.}
Our algorithm compares single-agent contracts with contracts offering a uniform payment to a small group of agents. Fix an optimal contract $\alpha$ and equilibrium $S$, and first suppose that every payment share is at most $1/2$. Equilibrium and subadditivity give $c(S_i)\leq\alpha_i f(S_i)$. If some agent has standalone reward at least a suitable constant times $f(S)/\sqrt n$, this bound ensures that a single-agent contract can recover a constant fraction of that reward. The known single-agent approximation~\cite{multimulti} then gives the desired guarantee.

Otherwise, every agent has small standalone reward. Since the payment shares sum to at most one,
\[
c(S)\leq\sum_i\alpha_i f(S_i)\leq\max_i f(S_i).
\]
Thus, small individual rewards force the total equilibrium cost to be small compared with $f(S)/\sqrt n$. We exploit this by partitioning the agents arbitrarily into blocks of at most $\lceil\sqrt n\rceil$ agents. For each block, we offer every member the same share $1/(4\sqrt n)$, chosen so that the principal retains a constant fraction of the reward, and pay everyone else zero.

A demand query finds a profile within the block maximizing the potential $f(X)-4\sqrt{n} \cdot c(X)$. This profile is an equilibrium, since any profitable unilateral deviation would increase the potential. To see why one block performs well, consider the optimal profile restricted to each block. Subadditivity ensures that these block rewards sum to at least $f(S)$, while the total-cost bound ensures that their combined cost penalty is only a small fraction of $f(S)$. Their potentials therefore sum to a constant fraction of $f(S)$. Since there are only $O(\sqrt n)$ blocks, one has potential $\Omega(f(S)/\sqrt n)$. Its demand query returns an equilibrium with at least this potential, and hence at least this reward, giving the desired principal utility.

Finally, at most one agent can receive a share above $1/2$. Either a single-agent contract for him recovers a constant fraction of the optimal principal utility, or the remaining agents' actions generate reward at least a constant fraction of that utility. In the latter case, the preceding argument applies to the remaining agents with the same blocks, evaluating each block's potential at the profile in which the high-share agent plays nothing.

\paragraph{WMRF Rewards (Hardness of Equilibrium Selection).}
The construction proves this hardness even for the restricted class of weighted graphic matroid-rank functions. The reduction is from \textsc{Max E3-Set Splitting}: given a ground set and a family of triplets, the goal is to 2-color the ground set so as to maximize the number of non-monochromatic triplets. This problem is known to admit no PTAS \cite{Guruswami2004InapproximabilityRF}.

Given such an instance, we construct a two-agent contract instance whose reward is the weighted graphic matroid-rank function of a multigraph; see \Cref{fig:graphic_matroid_instance}. The agents control disjoint sets of edges, colored red and blue by ownership, and the value of a chosen edge set is the maximum weight of a forest it contains. The graph contains two isolated high-weight, high-cost edges, one for each agent, which pin every near-optimal contract to the neighborhood of $(\contract_1,\contract_2)=(1/4,1/4)$. Below this threshold an agent does not take its isolated edge, while above it the principal gives away too large a share of the reward. This reduces the hardness of the \emph{contract-design problem} to the hardness of finding the best equilibrium at this specific contract.  

For weighted graphic matroids, finding \emph{some} equilibrium under a fixed contract is not the main difficulty, instead the hardness comes from the fact that a fixed contract may admit many different spanning-tree equilibria. For each element $i$ of the ground set, the graph contains a pair of parallel red and blue assignment edges from a root to an element vertex $v_i$. If the red assignment edge for element $i$ is selected, we color $i$ red; if the blue assignment edge is selected, we color $i$ blue. Since the two edges are parallel, no equilibrium can contain both, and the construction ensures that every equilibrium at a contract in this neighborhood contains one of them. Thus every such equilibrium induces a well-defined $2$-coloring of the ground set.

A triple gadget then turns the encoded coloring into reward. Each triplet is represented by two set vertices, one for each color. The construction ensures that a split triplet can attach both vertices through high-weight edges, whereas a monochromatic triplet forces one of them to use a lower-weight fallback edge. The spanning-tree equilibrium constraints enforce this correspondence: action profiles that would obtain the high reward without respecting the encoded coloring are not equilibria. Thus the reward of an equilibrium is larger exactly when the corresponding coloring splits more triplets, and finding the principal's favorite equilibrium would recover a good solution to the original \textsc{Max E3-Set Splitting} instance.

\paragraph{Gross Substitutes Rewards (Hardness of Searching over Critical Contracts).}
Finally, we show that a small number of critical contracts and oracle access to the \emph{Equilibrium Selection Problem} do not suffice for efficient contract search. We construct an exponentially large family of assignment valuations, and hence GS rewards, indexed by hidden sets $H$. Action costs and ownership are identical across the family.
With a single agent or supermodular rewards, equilibria that exist only at indifference points do not create isolated gains for the principal, since increasing payment weakly raises the reward obtained at the optimal equilibrium. For general GS rewards, however, paying one agent more can make the other drop actions, leaving no equilibrium with the same high reward. Our construction exploits this possibility by giving each instance a distinguished equilibrium that is sustained only by exact indifference and yields a constant-factor advantage over every other contract. Moreover, because only the presence of the hidden set $H$ and the cardinality of the action set matters, this yields only $O(m)$ distinct equilibrium regions, and hence
$O(m)$ critical contracts.

The main challenge is to prevent oracle queries from revealing the profitable contract. We compare the family with a common reference instance. Any fixed value query differs from its reference answer on only polynomially many hidden instances, while a query to the Equilibrium Selection Problem differs on at most one. This holds even when the oracle returns all equilibria maximizing the principal's utility, including ties. Along a deterministic algorithm's reference transcript, each query therefore rules out only polynomially many candidates, while its output can achieve the required approximation on at most one remaining candidate. A counting argument followed by Yao's principle gives the exponential query lower bound for randomized algorithms.

\section{Model and Preliminaries}\label{sec:model}

We begin by describing the multi-agent combinatorial-action contract design setting of \citet{multimulti}.

There's a single principal and a team of $n$ agents, denoted by $A=[n]$.
The principal runs a project with a binary outcome---success or failure---that gives her a \mse{normalized} reward \mse{of} $1$ on success and $0$ on failure. 
The agents influence the project's success through their choice of actions; every agent $i$ has an action set $T_i$, and may select any subset of this set.
\mse{The agents' action sets} are pairwise disjoint: $T_i \cap T_{i'} = \emptyset$ for $i \neq i'$, and we write $T = \biguplus_{i \in A} T_i$ for the set of all actions\mse{, and $m=|T|$ for its size}. 
For an action profile $S \subseteq T$ we write $S_i = S \cap T_i$ for agent $i$'s chosen actions and $S_{-i} = \biguplus_{i' \neq i} S_{i'}$ for the actions chosen by the other agents.
Taking actions is costly to the agents; 
each action $j \in T_i$ carries a nonnegative cost $c_j \ge 0$ to agent $i$. Costs are additive, so that the cost of selecting $S_i\subseteq T_i$ is $c(S_i) = \sum_{j \in S_i} c_j$, with $c(\emptyset)=0$.

The project's success probability is determined by the agents' actions, through a monotone set function $f : 2^T \to [0,1]$, which maps any action profile $S = \biguplus_{i \in A} S_i$ to a probability of success.
Because the principal's reward for success is normalized to $1$, $f(S)$ also equals her expected reward; we therefore call $f$ the \emph{reward function}.
We assume $f$ is normalized, i.e., $f(\emptyset) = 0$.

As standard in contract theory, the agents' actions are assumed to be hidden from the principal. Thus, in order to incentivize the agents to take costly actions, the principal designs a  \emph{contract}, which is a payment made to each agent, conditioned on the project's success. More formally, a contract is a vector $\contract = (\alpha_1, \dots,\alpha_n) \in [0,1]^n$, where $\icontract$ is the fraction of the reward given to agent $i$.

This type of contract is known in the literature as a \emph{linear contract}.

An instance of the model is a tuple $\multiInstance$, where $A$ is the agent set, $\{T_i\}_{i \in A}$ are their action sets, $f\mse{:2^T\rightarrow [0,1]}$ is the reward function, and $c = \{c_j\}_{j \in T}$ specifies action costs.

\paragraph{Utilities and Equilibria.}
Under a contract $\noindexcontract$ and an action profile $S \subseteq T$, the principal's utility is her expected reward minus the expected payments to the agents, i.e.,
\[
u_P(\noindexcontract, S)
=
\textstyle \left(1 - \sum_{i \in A} \icontract\right) f(S).
\]
Each agent $i$'s expected utility is his expected payment minus the cost of his actions, given by,
\[
u_i(\noindexcontract, S)
=
\icontract \cdot f(S_i \uplus S_{-i}) - c(S_i).
\]
Observe that agent $i$'s utility depends both on his choice of actions, i.e., $S_i$, and on the other agents' choices, i.e., $S_{-i}$.

Every contract $\noindexcontract$ induces a game among the agents. We say that the contract $\noindexcontract$ \emph{incentivizes} the action profile $S$ if $S$ is a pure Nash equilibrium of the induced game: for every agent $i$ and every deviation $S'_i \subseteq T_i$, it holds that $u_i(\noindexcontract, S_i \cup S_{-i}) \ge u_i(\noindexcontract, S_i'\cup S_{-i})$.
By \cite{multimulti}, for every contract the induced game admits at least one pure Nash equilibrium.
A contract may induce several equilibria; we write $\nash(\noindexcontract)$ for the set of all equilibria induced by $\alpha$.

\paragraph{Optimization Problems.} Our goal is to maximize the principal's utility. 
\begin{definition} [Optimal Contract Problem]
    Given a problem instance $\multiInstance$, we wish to compute a contract $\contract$ and an equilibrium $S\in \nash(\contract)$ which maximize the principal's utility, i.e.,
    \[
    u_P(\contract, S) \ge u_P(\contract', S'),
    \]
    for any contract $\alpha'$ and equilibrium $S'\in \nash(\contract')$.
\end{definition}

A closely related optimization problem that emerges is the \emph{equilibrium selection problem} which seeks to find the optimal equilibrium for a given contract.
\begin{definition} [Equilibrium Selection Problem]
    Given a problem instance $\multiInstance$ and a contract $\contract$, we wish to find an equilibrium $S \in \nash(\contract)$ which maximizes the principal's utility, i.e.,
    \[
    u_P(\contract, S) \ge u_P(\contract, S'),
    \]
    for any equilibrium $S'\in \nash(\contract)$.
\end{definition}

In some settings, computing an exact optimal solution to these problems is hard, and we turn to approximately optimal solutions. For any $\gamma>1$, we say that a feasible solution is a
$\gamma$-approximation if its principal's utility is at least
a $(1/\gamma)$-fraction of the optimal value of the corresponding
problem.

\paragraph{\mse{Critical Contracts. }}
We also introduce formally the notion of critical contracts, which provides a useful lens for identifying where computational hardness arises in different settings. \mse{Critical contracts are a natural multi-dimensional generalization of the notion of \emph{critical points}, a prevalent approach in the computation of optimal contracts in the single-agent combinatorial model\cite{dutting2022combinatorial}.}

\begin{definition}[Critical Contracts]\label{def:critical-contracts}
For an action profile $S$, we say that a contract $\contract$ incentivizes $S$ if $S\in \nash(\contract)$. A contract $\contract$ is critical for $S$ if it
incentivizes $S$ and is coordinate-wise minimal\footnote{This is well defined since, for any equilibrium $S$ and agent $i\in [n]$, $\alpha_i$ is only required to satisfy $\forall S_i'\subseteq T_i.~\alpha_i \cdot f(S_i \uplus S_{-i} ) - c(S_i) \ge \alpha_i \cdot f(S_i' \uplus S_{-i}) - c(S_i')$, and thus some minimal value of $\alpha_i$ exists, independently of the other coordinates of $\contract$.    } among contracts that incentivize $S$; that is, no $\contract' \le \contract$, $\contract'\ne \contract$, also incentivizes $S$. We denote the critical contract for a profile $S$ by $\underline{\contract}(S)$ \msc{searching through overleaf this notation is rarely used. Do we want to keep it?}.  We call $\contract$ a critical contract of the instance if it is critical for some 
\mse{action}
profile $S$.
\end{definition}

\mse{The critical points approach from the single-agent model thus has a multi-dimensional parallel: For a given instance of the optimal contract problem, compute the set of critical contracts and evaluate the equilibrium selection problem at each, taking the best resulting contract and equilibrium. This yields a polynomial-time algorithm under three conditions: (i) the equilibrium selection problem is poly-time solvable, (ii) there is a polynomial bound on the number of critical contracts, and (iii) the set of critical contracts can be identified in polynomial time. \msc{Note, condition (ii) is implied by (iii) but I thought it might be nice to make it explicit, let's consider.} }

\paragraph{Reward Functions Classes and Oracle Access.}
We restrict attention to monotone reward functions $f$ throughout.
For sets $S, S'$, write $f(S' \mid S) = f(S' \cup S) - f(S)$ for the marginal gain of adding $S'$ to $S$. For presentation simplicity, when $S'=\{j\}$ is a singleton we omit the curly brackets and write $f(j\mid S)$.
\mse{We consider reward functions which belong to the following classes:
\begin{itemize}
    \item \emph{Weighted matroid rank functions (WMRF)}: A set function $f:2^T\rightarrow [0,1]$ is a \emph{weighted matroid rank function} if there exist weights $\mse{\{w_i\}_{i\in T}}$ and a matroid $\mathcal{M} = (T, \I)$ with ground set $T$ and independent sets $\I\subseteq 2^T$, such that
    $f(S) = \max_{S' \in \I: S'\subseteq S} \sum_{j\in S'} w_j$.
    \item \emph{Gross substitutes}: A set function $f:2^T\rightarrow [0,1]$ is \emph{gross substitutes} if for any two price vectors
    $p,q \in \mathbb{R}^n$ such that $q \geq p$,  and any set $S\in \arg\max_{X \in 2^T}
        \left\{ f(X) - \sum_{i \in X} p_i \right\}$,
    there is a $U \in 2^T$ such that
    $U
        \in \arg\max_{X \in 2^T}
        \left\{ f(X) - \sum_{i \in X} q_i \right\}$
    as well as
    $U \supseteq \{i \in S \mid q_i = p_i\}$.
    In words, when prices weakly increase, the demand for items whose prices are unchanged weakly increases.
    \item \emph{Subadditive: } A set function $f:2^{T}\rightarrow [0,1]$ is \emph{subadditive} if it holds that $f(S\cup S')\le f(S)+f(S')$ for any two sets $S,S'\subseteq T$.
    \item \emph{Supermodular}: A set function $f:2^T\rightarrow [0,1]$ is \emph{supermodular}, if marginals are (weakly) increasing, i.e., it holds that 
    $f(j\mid S) \le f(j\mid S')$ for any two sets $S\subseteq S'\subseteq T$ and $j\in T\setminus S'$.
\end{itemize}

It is well-known that 
\[
\text{weighted matroid rank} \subsetneq \text{gross substitutes} \subsetneq \text{subadditive},
\]
and these are incomparable to supermodular.
}

Because $f$ may require exponential space in $|T|$, algorithms access it through a value oracle: given $S \subseteq T$, a value oracle returns $f(S)$.
For subadditive rewards, we will assume access to a \emph{demand oracle} \mse{  as well}, which is stronger than value-oracle access. Given nonnegative prices $q\in \reals_{\ge 0}^{T}$, a demand query returns a set
$S\in \argmax_{X\subseteq T}\left\{ f(X)-\sum_{j\in X} q_j \right\}$.

Demand queries are useful because they can compute \emph{some} pure equilibrium for a fixed contract: given $\contract$, set $q_j=c_j/\alpha_i$ for every action $j\in T_i$ with $\alpha_i>0$; then a demand query returns a global maximizer of $f(X)-\sum_{i:\alpha_i>0}\frac{c(X_i)}{\alpha_i}$
and such a maximizer is a pure Nash equilibrium under $\contract$ \cite{multimulti}. If $\alpha_i=0$, agent $i$'s positive-cost actions can be assigned a prohibitively large price, while zero-cost actions may be priced at zero. 
We note that, for weighted matroid rank\mse{, gross substitutes, }and supermodular functions, demand queries can be implemented using polynomially many value queries --- via 
\mse{a}
greedy optimization \mse{ for WMRF and gross substitutes, and by submodular function} minimization for supermodular\cite{leme2017gross,schrijver2000combinatorial}. Hence, demand oracle access does not strengthen the model for these classes. However, for subadditive functions, demand queries cannot even be approximated efficiently \cite{feige2014demand}.

\section{Complements: Supermodular Rewards}\label{sec:complements}
We first study supermodular rewards. In \Cref{sec:optimal-equilibrium-supermodular}, we show that the equilibrium selection problem can be solved in polynomial time. Nevertheless, \Cref{sec:supermodular-main-hardness} shows that the optimal contract problem admits no finite multiplicative approximation, even for two agents, and gives the full construction and proof. Finally, \Cref{sec:supermodular-search} gives a deterministic exact algorithm whose running time is polynomial in the number of critical contracts. Together, these results isolate the size of the critical family as a source of complexity for supermodular rewards.

\subsection{The Equilibrium Selection Problem}
\label{sec:optimal-equilibrium-supermodular}

For a fixed contract and a supermodular reward function, the equilibrium selection problem can be solved in polynomial time, for any number of agents.

\begin{theorem}
\label{thm:polytime-optimal-equilibrium-supermodular}
Given an instance $\multiInstance$, a
contract $\contract$ such that $\sum_{i \in [n]} \alpha_i \leq 1$, and value-oracle access to a supermodular reward
function $f$, the equilibrium selection problem can be solved in polynomial
time.
\end{theorem}

We first recall the relevant
supermodular definitions.

\begin{definition}\label{def:supermodular_game}
    A finite game with agents $A$ is a \emph{supermodular game} if each
    agent $i\in A$ has a finite lattice of actions $X_i$, and, writing
    $x=(x_i,x_{-i})$ for an action profile, each utility function
    $v_i:X_i\times X_{-i}\rightarrow \mathbb{R}$ satisfies:
    \begin{enumerate}
        \item for every $x_{-i}\in X_{-i}$, the function $x_i \longmapsto v_i(x_i,x_{-i})$ is supermodular on $X_i$;
        \item $v_i$ has increasing differences in $(x_i,x_{-i})$, i.e.,
        for every $x_i\le y_i$ and $x_{-i}\le y_{-i}$,
        \begin{equation*}
            v_i(y_i,y_{-i})-v_i(x_i,y_{-i})
            \ge
            v_i(y_i,x_{-i})-v_i(x_i,x_{-i}).
        \end{equation*}
    \end{enumerate}
\end{definition}

The key observation is that, for supermodular reward functions, every
contract induces a supermodular game among the agents.

\begin{lemma}
\label{lem:lattice-equilibria-supermodular}
Let $f$ be supermodular. For every contract $\contract$, the game induced
among the agents is a supermodular game. Consequently,
$\nash(\contract)$ is nonempty and forms a complete lattice under set
inclusion. In particular, it has a least and a greatest equilibrium.
\end{lemma}

\begin{proof}
Fix an agent $i$ and the actions $S_{-i}$ of the other agents. The set 
function $ S_i\longmapsto
    u_i(\noindexcontract,S_i\uplus S_{-i}) =
    \icontract f(S_i\uplus S_{-i})-c(S_i)$
is supermodular: the first term is supermodular because
$\icontract\ge 0$, and the second term is modular.

It remains to verify increasing differences. Consider
$S_i\subseteq S_i'$ and $S_{-i}\subseteq S_{-i}'$. Since agents'
action sets are pairwise disjoint, supermodularity of $f$ gives
\begin{equation*}
    f(S_i'\uplus S_{-i})
     +f(S_i\uplus S_{-i}') \le
    f(S_i'\uplus S_{-i}')
     +f(S_i\uplus S_{-i}).
\end{equation*}
Rearranging, multiplying by $\icontract\ge 0$, and observing that the
cost difference $c(S_i')-c(S_i)$ is independent of $S_{-i}$ yields
\begin{equation*}
u_i(\noindexcontract,S_i'\uplus S_{-i}')
    -u_i(\noindexcontract,S_i\uplus S_{-i}') \qquad\ge
    u_i(\noindexcontract,S_i'\uplus S_{-i})
    -u_i(\noindexcontract,S_i\uplus S_{-i}).
\end{equation*}
Thus, the induced game is supermodular. Since the agents' action spaces
are finite lattices, the complete-lattice characterization of equilibria
in supermodular games implies the result by
\cite{ZHOU1994295}.
\end{proof}

When $\sum_i\alpha_i\leq1$, this greatest equilibrium maximizes the principal's utility by monotonicity of the reward function. For any contract $\contract$, let $S^+(\contract)$ denote greatest equilibrium of $\contract$, according to the subset order guaranteed by the lattice characterization from \Cref{lem:lattice-equilibria-supermodular}. The following observation is immediate from monotonicity of the reward function.

\begin{observation}
\label{obs:optimal-extremal-equilibrium}
If $\sum_{i \in [n]} \alpha_i \leq 1$ then $S^+(\contract)$ maximizes the
    principal's utility over $\nash(\contract)$ .
\end{observation}

We are now ready to prove \Cref{thm:polytime-optimal-equilibrium-supermodular}.
\begin{proof} [Proof of \Cref{thm:polytime-optimal-equilibrium-supermodular}]
Fix an agent $i$ and an action profile
$S_{-i}$ of the other agents. We have 
\begin{equation*}
    \arg\max_{S_i'\subseteq T_i}
    u_i(\noindexcontract,S_i'\uplus S_{-i}) =
    \arg\min_{S_i'\subseteq T_i}
    \left\{
    c(S_i')-\icontract f(S_i'\uplus S_{-i}) 
    \right\}.
\end{equation*}
Because $c$ is modular and $f$ is supermodular, the objective on the
right-hand side is submodular. It can therefore be minimized in
polynomial time in the value-oracle model, and the greatest minimizer in the inclusion sense, equivalently, the greatest best-response, can also be recovered in polynomial time, see
\citet{Iwata2009simplecomp}.

We now compute the greatest equilibrium. Set $S^0=T$,
and, for every $r\ge 0$ and every agent $i$, let $S_i^{r+1}$ be the
greatest best response of agent $i$ to $S_{-i}^r$. By \Cref{lem:lattice-equilibria-supermodular}, the induced game is supermodular. Hence, by Topkis' monotonicity theorem \cite{Topkis1979equilibriumsub}, the greatest best-response maps are nondecreasing. Since
$S^1\subseteq T= S^0$, it follows inductively that $S^{r+1}\subseteq S^r$ for every $r \geq0$.
Whenever the profile changes, at least one action is removed. The
sequence therefore stabilizes after at most $|T|$ strict iterations.
Let $S^\star$ denote the profile at which it stabilizes. Every
$S_i^\star$ is a best response to $S_{-i}^\star$, and hence
$S^\star\in\nash(\contract)$.

To see that $S^\star$ must be the greatest equilibrium, fix any
$S\in\nash(\contract)$. We prove inductively that $S\subseteq S^r$.
The claim is immediate for $r=0$. Suppose that $S\subseteq S^r$.
Because $S_i$ is a best response to $S_{-i}$, it is contained in the
greatest best response to $S_{-i}$. Moreover,
$S_{-i}\subseteq S_{-i}^r$, so monotonicity of greatest best responses
implies that this best response is contained in $S_i^{r+1}$. Thus,
$S_i\subseteq S_i^{r+1}$ for every $i$, and therefore
$S\subseteq S^{r+1}$. Passing to the stabilized profile gives
$S\subseteq S^\star$. Hence $S^\star=S^+(\contract)$.

Each iteration requires one extremal best-response computation for each
agent, and there are at most $|T|$ iterations. The overall running time
is consequently polynomial. Outputting $S^+(\contract)$ yields the best equilibrium according to \Cref{obs:optimal-extremal-equilibrium}.
\end{proof}

\par
\begingroup
\definecolor{STOCHardAgentOne}{HTML}{315E91}
\definecolor{STOCHardAgentTwo}{HTML}{B46527}
\definecolor{STOCHardEncode}{HTML}{6B4FA1}
\definecolor{STOCHardVerify}{HTML}{26755A}
\definecolor{STOCHardAmplify}{HTML}{A43E48}
\definecolor{STOCHardInk}{HTML}{35414B}
\captionsetup{font=small,labelfont=bf,skip=6pt}

\subsection{Inapproximability for Two Agents}
\label{sec:supermodular-main-hardness}

We next show that the Optimal Contract Problem remains hard even with two
agents, despite the tractability of equilibrium selection.

\begin{theorem}\label{thm:supermodular-hardness}
For normalized, monotone, supermodular rewards, no deterministic polynomial-time
algorithm approximates the optimal principal utility within any finite
multiplicative factor, even for two agents, unless $\mathrm{P}=\mathrm{NP}$.
\end{theorem}

We reduce from \textnormal{\textsc{Subset Sum}}. Given positive integer weights $w_1,\ldots,w_k$ and a target $T$, the goal is to determine whether some subset $H\subseteq[k]$ has total weight $w(H)\coloneqq \sum_{i\in H}w_i=T$. The construction first encodes every subset
of items at a different interval of contracts. Two additional actions test
whether the encoded subset has weight $T$, and successful verification makes
an edge of very large weight available at equilibrium. This gives a gap
between large principal utility in feasible \textnormal{\textsc{Subset Sum}}-instances and bounded reward in
unfeasible instances. A final action turns this gap into positive versus zero optimal utility. We describe the key steps in the construction and its main guarantees below. The full construction is quite complex, as every parameter needs to be balanced together, so we defer the full detailed construction and its proof to \Cref{app:superhard}. 

\paragraph{Quadratic rewards on a bipartite graph.}
Until the final step, the reward is defined by a bipartite graph with
nonnegative edge weights. Each agent controls the vertices on one side,
each vertex is a costly action, and an edge contributes its weight when both
endpoints are selected. This reward is monotone and supermodular. Given the other agent's actions,
an agent's utility is additive over his own vertices. In particular, selecting
a vertex of cost $c$ is worthwhile precisely when the agent's payment share
times the total weight of its edges to selected neighbors is at least $c$.
This simple condition will implement both the encoding and the verification.

\paragraph{Encoding all subsets.}
For each item $i\in[k]$, we introduce an inclusion action $q_i$ owned by agent~1 and an exclusion action $p_i$ owned by agent~2. Although both actions could in principle be selected, we choose the edge weights and action costs so that every equilibrium selects at most one of $p_i,q_i$ for each item. An action profile $S$ then encodes a subset $H\subseteq[k]$ if, among these item actions, it contains exactly $q_i$ for $i\in H$ and $p_i$ for $i\notin H$. In the final construction, each of the $2^k$ subsets is represented by disjoint contracts intervals along the line $\alpha_1+\alpha_2=1-\varepsilon$, for some small $\varepsilon>0$. At every contract in the interval associated with $H$, the greatest equilibrium encodes $H$.

Suppose for now that directed edges are allowed. We explain below how to simulate their effect at equilibrium using undirected edges. To encode the subsets, we make the payment required to include or exclude item $i$ depend on the choices made for the preceding items. Specifically, for every $j<i$, we add an edge $q_j\to p_i$ of weight $2\cdot3^{-j}$ and an edge $p_j\to q_i$ of the same weight. These directed dependencies allow us to determine sequentially whether an item is included or not, since later choices do not affect the incentives for earlier items. Suppose that the choices for the first $i-1$ items agree with a subset $H\subseteq[k]$, selecting $q_j$ and excluding $p_j$ when $j\in H$, and doing the reverse otherwise. The incoming contributions from these earlier items to $p_i$ and $q_i$ are then, respectively,
\begin{equation*}
Q_i(H)\coloneqq \sum_{j<i:j\in H}2\cdot3^{-j},\quad \mbox{and} \quad
\sum_{j<i:j\notin H}2\cdot3^{-j}=\sum_{j<i}2\cdot3^{-j}-Q_i(H)
=1-3^{-(i-1)}-Q_i(H).
\end{equation*}

Now suppose that $\alpha_1+\alpha_2\approx 1$, with both payment shares close to $1/2$. Give each item action a fixed baseline contribution $M_0$, where $M_0$ is sufficiently large, and parameterize the payment segment by $\tau\in[0,1]$ so that $\alpha_1\approx \frac12+\frac{\tau}{2M_0}$ and $\alpha_2\approx \frac12-\frac{\tau}{2M_0}$. 
Thus, increasing $\tau$ makes inclusion actions more attractive and exclusion actions less attractive. Choose the costs $c(p_i)=\frac{M_0-3^{-i}}{2}$ and $c(q_i)=\frac{M_0+1-3^{-i}}{2}$.
Using the incoming contributions computed above, the incentive constraints for selecting $p_i$ and $q_i$ become
\[
\alpha_2\bigl(M_0+Q_i(H)\bigr)\ge c(p_i),
\qquad
\alpha_1\bigl(M_0+1-3^{-(i-1)}-Q_i(H)\bigr)\ge c(q_i).
\]
Ignoring the small error terms arising from the large baseline, these conditions simplify to
\[
p_i \in S \Leftrightarrow  \tau\le Q_i(H)+3^{-i},
\qquad
q_i \in S \Leftrightarrow  \tau\ge Q_i(H)+2\cdot3^{-i}.
\]
This has subdivided the interval $[Q_i(H),\,Q_i(H)+3^{-(i-1)}]$ into equal thirds, the left part corresponding to including $p_i$, the right part to including $q_i$, and the middle part where no level $i$ actions are incentivized. Inside each surviving interval, the next item makes the same left-versus-right choice, and so on. This is the recursive refinement that fully encodes $H$, and is illustrated in \Cref{fig:main-intervals} for $k=3$.

\begin{figure}[!ht]
\centering
\begin{tikzpicture}[x=1cm,y=1cm,font=\footnotesize,
  line cap=round,line join=round,>={Stealth[length=1.5mm]}]
 \def\L{1.4}\def\W{13.5}
 \node[anchor=east,text=STOCHardInk] at (1.12,2.52) {item 1};
 \node[anchor=east,text=STOCHardInk] at (1.12,1.5) {item 2};
 \node[anchor=east,text=STOCHardInk] at (1.12,.48) {item 3};
 \foreach \q in {0,2}{
   \filldraw[draw=STOCHardInk!55,fill=STOCHardInk!7] ({\L+\W*\q/3},2.43)
       rectangle ({\L+\W*(\q+1)/3},2.61);
 }
 \filldraw[draw=STOCHardEncode,fill=STOCHardEncode!18,line width=.9pt]
    (\L,2.43) rectangle ({\L+\W/3},2.61);
 \node[text=STOCHardEncode] at ({\L+\W/6},2.93) {exclude item 1};
 \node[text=STOCHardInk] at ({\L+5*\W/6},2.93) {include item 1};
 \foreach \q in {0,2,6,8}{
   \filldraw[draw=STOCHardInk!55,fill=STOCHardInk!7] ({\L+\W*\q/9},1.41)
       rectangle ({\L+\W*(\q+1)/9},1.59);
 }
 \filldraw[draw=STOCHardEncode,fill=STOCHardEncode!18,line width=.9pt]
    ({\L+2*\W/9},1.41) rectangle ({\L+3*\W/9},1.59);
 \foreach \q in {0,2}{
   \draw[STOCHardInk!28] ({\L+\W*(\q+.5)/3},2.40)
       --({\L+\W*(3*\q+.5)/9},1.92);
   \draw[STOCHardInk!28] ({\L+\W*(\q+.5)/3},2.40)
       --({\L+\W*(3*\q+2.5)/9},1.92);
 }
 \node[text=STOCHardInk,font=\scriptsize] at ({\L+\W/18},1.85) {exclude 2};
 \node[text=STOCHardEncode,font=\scriptsize] at ({\L+5*\W/18},1.85) {include 2};
 \node[text=STOCHardInk,font=\scriptsize] at ({\L+13*\W/18},1.85) {exclude 2};
 \node[text=STOCHardInk,font=\scriptsize] at ({\L+17*\W/18},1.85) {include 2};
 \foreach \q in {0,2,6,8}{
   \draw[STOCHardInk!28] ({\L+\W*(\q+.5)/9},1.38)
       --({\L+\W*(3*\q+.5)/27},.69);
   \draw[STOCHardInk!28] ({\L+\W*(\q+.5)/9},1.38)
       --({\L+\W*(3*\q+2.5)/27},.69);
 }
 \foreach \q/\lab in {0/000,2/001,6/010,8/011,18/100,20/101,24/110,26/111}{
   \filldraw[draw=STOCHardInk!55,fill=STOCHardInk!7] ({\L+\W*\q/27},.39)
       rectangle ({\L+\W*(\q+1)/27},.57);
   \node[font=\scriptsize,text=STOCHardInk] at ({\L+\W*(\q+.5)/27},.14) {$\lab$};
 }
 \filldraw[draw=STOCHardEncode,fill=STOCHardEncode!35,line width=.9pt]
    ({\L+8*\W/27},.39) rectangle ({\L+9*\W/27},.57);
 \draw[STOCHardEncode,line width=1pt] ({\L+\W/6},2.40)
       --({\L+5*\W/18},1.92);
 \draw[STOCHardEncode,line width=1pt] ({\L+5*\W/18},1.38)
       --({\L+8.5*\W/27},.69);
 \draw[->,STOCHardInk!75] (\L,-.25)--({\L+\W},-.25);
 \node[fill=white,inner sep=2pt,text=STOCHardInk,font=\scriptsize]
       at ({\L+\W/2},-.25) {$\alpha_1$ increases and $\alpha_2$ decreases};
\end{tikzpicture}
\caption{Recursive encoding along a constant-total-payment line. A bit is
$1$ when its item is included. The highlighted branch encodes $H=\{2,3\}$.
Each next choice occurs inside the interval selected by the earlier choices.}
\label{fig:main-intervals}
\end{figure}
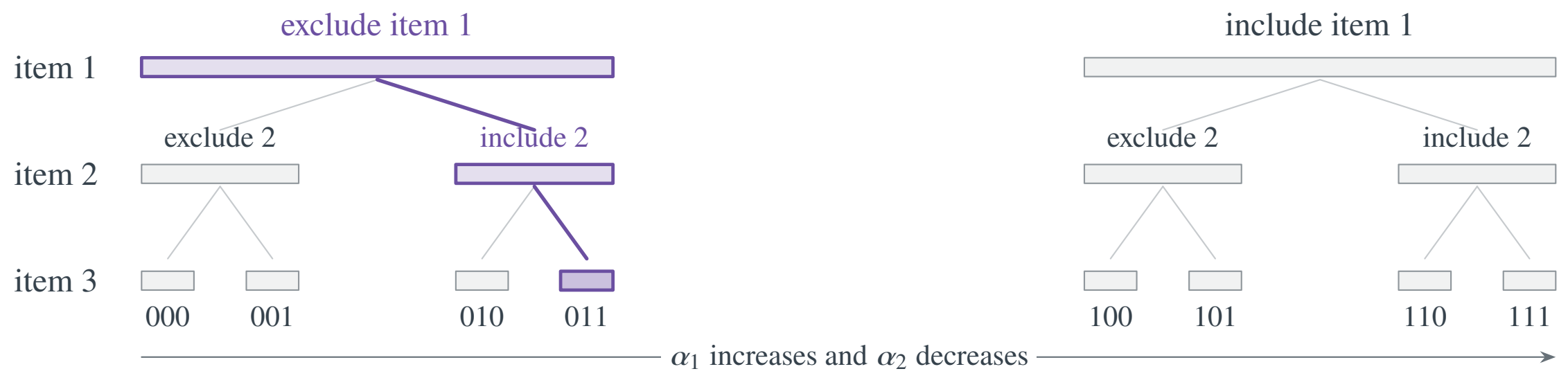

\paragraph{Simulating directed edges.}
To implement this refinement, an item action must change the incentives for
later items without those later items changing the choice already encoded.
The directed edges described above provide this dependence. The shared
reward function, however, requires an undirected graph.

We simulate the required direction using the three-edge path gadget in
\Cref{fig:main-three-edge-path}. Its endpoints are a source action $s$ and a
target action $t$ on opposite sides of the graph. Two intermediate actions
connect them, with edge weights $\eta^2M$, $\eta M$, and $M$, for a sufficiently
small $\eta>0$. The large last edge makes the path useful to the target,
whereas the first edge has only a small effect on the source's incentives.

\begin{figure}[!ht]
\centering
\begin{tikzpicture}[
 x=1cm,y=1cm,font=\small,line cap=round,line join=round,
 >={Stealth[length=1.7mm,width=1.2mm]},
 one/.style={circle,draw=STOCHardAgentOne,fill=STOCHardAgentOne!8,line width=.75pt,minimum size=7mm,inner sep=0pt},
 two/.style={circle,draw=STOCHardAgentTwo,fill=STOCHardAgentTwo!8,line width=.75pt,minimum size=7mm,inner sep=0pt},
 lab/.style={font=\footnotesize,inner sep=1pt},
 gadgetedge/.style={draw=black!78,line width=.7pt},
 dashedpath/.style={draw=black!78,dash pattern=on 2.6pt off 2pt,
   -{Stealth},line width=.7pt,shorten <=2.6pt,shorten >=2.6pt}]
 \node[one] (sabs) at (.65,0) {$s$};
 \node[two] (tabs) at (3.45,0) {$t$};
 \draw[dashedpath] (sabs)--node[lab,above=3pt] {$M$} (tabs);

 \node[one] (s) at (5.9,.6) {$s$};
 \node[one] (y) at (5.9,-.6) {$y$};
 \node[two] (t) at (10.3,.6) {$t$};
 \node[two] (z) at (10.3,-.6) {$z$};

 \begin{scope}[on background layer]
  \draw[gadgetedge]
    (s)--node[lab,sloped,above=3pt,pos=.22] {$\eta^2M$} (z);

  \draw[gadgetedge,preaction={draw=white,line width=3pt}]
    (y)--node[lab,sloped,above=3pt,pos=.78] {$M$} (t);

  \draw[gadgetedge]
    (y)--node[lab,below=3pt] {$\eta M$} (z);
\end{scope}
\end{tikzpicture}
\caption{A directed edge of value $M$ (left) and its undirected
three-edge implementation (right). Colors indicate ownership.}
\label{fig:main-three-edge-path}
\end{figure}
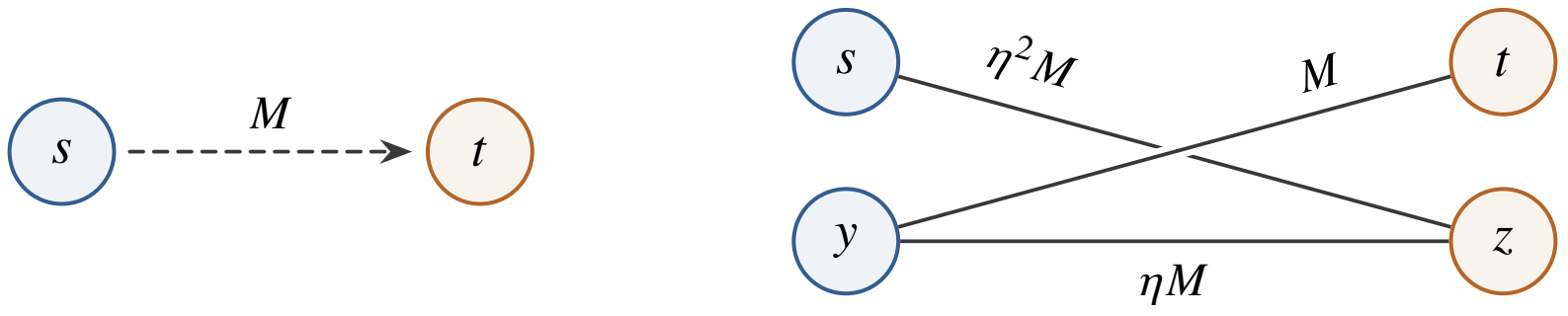

The idea is to make a tiny contribution from the source necessary for
sustaining a large contribution to the target. We achieve this by placing
the intermediate actions close to indifference. For this illustration, let
agent~1 control $s,y$ and agent~2 control $z,t$. Give the intermediate actions costs $c(z)=h(\eta^2+\eta)M$ and $c(y)=h(1+\eta)M$. Thus, each costs $h$ times the total weight of its incident edges. Choose
$h$ slightly below $1/2$, so that $2h<1<h(2+\eta)$. We examine contracts where both payment shares are at least $h$, while their sum is at most one.

Suppose first that both source and target belong to the greatest equilibrium. If we also select $z,y$, their marginal rewards are $(\eta^2+\eta)M$ and $(1+\eta)M$, respectively. Both actions cover their costs because each agent's payment share is at least $h$. Adding them only increases the rewards of the actions already selected. Thus, the intermediate actions can be sustained together with the endpoints, and they belong to the greatest equilibrium. The target receives the contribution $M$ from the last edge, and the source almost no contribution for $\eta$ small enough. The source's incentives are determined elsewhere in the construction, and the first edge is too light to change the encoded choice.

Now suppose that the source is absent. For the target to receive $M$, action $y$ must still be selected. There are two possibilities. $1)$ If $z$ is absent, action $y$ receives only $M$, but still costs
$h(1+\eta)M$. It therefore requires $\alpha_1\ge h(1+\eta)$.
Together with $\alpha_2\ge h$, this gives total payment at least
$\alpha_1+\alpha_2 \geq h(2+\eta)>1$, which makes the principal utility negative. $2)$ If $z$ is selected, it receives only $\eta M$, because its edge to
the source contributes nothing. Its cost requires $\alpha_2\eta M\ge h(\eta+\eta^2)M$, so $\alpha_2\ge h(1+\eta)$. Together with $\alpha_1\ge h$, we again need total payment greater than one. Thus, without the source, neither possibility can sustain the large edge within the payment budget. This explains why even the tiny first edge matters. Removing it makes one of the intermediate actions slightly too expensive. Increasing that agent's payment enough to compensate leaves too little payment for the other agent.

With both the encoding and directed edge construction, we obtain the following intermediate result. 
\begin{lemma*}[Subset Encoding; Informal]
Each subset $H\subseteq[k]$ is encoded by the greatest equilibrium throughout a nonempty interval of contracts. These intervals are disjoint and share the same total payment below one. At any feasible contract, the greatest equilibrium selects at most one of $p_i,q_i$ per item.
\end{lemma*}

The same construction also gives the claimed lower bound on the number of
critical contracts. There are only $\Theta(k^2)$ actions, since each pair of items uses a constant number of intermediate actions, yet the $2^k$ subsets give distinct greatest equilibria. Lowering a contract to the critical contract of its greatest equilibrium preserves that equilibrium, so these equilibria have distinct critical contracts.

\begin{proposition}\label{prop:many-critical-contracts-supermodular}
For infinitely many $m$, there is a two-agent instance with $m$ actions and a normalized, monotone, quadratic supermodular reward having
$2^{\Omega(\sqrt m)}$ distinct critical contracts.
\end{proposition}

\paragraph{Testing whether the encoded sum is exact.}
We now use the encoded subsets to test the \textnormal{\textsc{Subset Sum}} instance, and extract some value from this test. Fix a designated contract whose greatest equilibrium encodes $H$. It selects $q_i$ for every $i\in H$ and $p_i$ for every $i\notin H$.
Write $w(H)=\sum_{i\in H}w_i$, $W=w([k])$, and $d=w(H)-T$.

Introduce an inclusion verifier action owned by agent~2 and an exclusion verifier action  owned by agent~1. For a baseline $b>W$, connect the inclusion verifier to each $q_i$ with weight $w_i$, and give it a baseline\footnote{Each baseline is implemented by an edge to a zero-cost action on the opposite side, which is always selected in the greatest equilibrium.} contribution $b-T$. Similarly, connect the exclusion verifier to each $p_i$ with weight $w_i$, and give it a baseline contribution $b-(W-T)$.  The marginal rewards of the two
verifiers are therefore
\begin{equation*}
\text{inc. verif. action:}\quad (b-T)+w(H)=b+d, \quad 
 \text{excl. verif. action:}\quad (b-(W-T))+(W-w(H))=b-d.
\end{equation*}
Encoding a subset $H$ with larger total weight $w(H)$ gives the inclusion verifier a larger marginal reward and the exclusion verifier a smaller one, since the excluded items have total weight $W-w(H)$.
Give both verifiers cost $c$. For them both to be selected, their additional payments must cover their costs, so $ \alpha_2(b+d)\ge c$ and $\alpha_1(b-d)\ge c$. Consequently, the total payment share must satisfy
\[
 \alpha_1+\alpha_2
 \ge \frac{c}{b+d}+\frac{c}{b-d}
 =\frac{2cb}{b^2-d^2}.
\]
Choose $c$ such that $(b^2-1)/(2b)<c<b/2$. If the encoded subset $H$ has the target weight $T$, then $d=0$ and the verifiers require total payment only $2c/b<1$, leaving a positive share for the principal. Otherwise, integrality gives $|d|\ge1$, so both verifiers would require total payment at least $\alpha_1+\alpha_2\geq 2cb/(b^2-1)>1$. Thus, no feasible contract can sustain both verifiers for an incorrect sum.

\paragraph{Making a successful test valuable.}
We now turn a successful test into a large reward. Add two actions, one per agent, joined by an edge of very large weight. Connect each verifier to the new action owned by the opposite agent
via a simulated directed edge. We choose the costs so that the large edge alone is not enough to sustain both new actions. However, at a designated contract encoding a subset of the target weight, the contributions from the two verifiers allow both actions to cover their costs. They therefore belong to the greatest equilibrium, and the large edge contributes its weight to the reward.

If either verifier is absent, the corresponding new action loses this additional contribution. Covering its cost would then require a payment share so large that the other agent could no longer afford his new action. Since neither new action can cover its cost without the other, both are absent. For the full construction, see \Cref{fig:main-construction}.

\definecolor{AgentOne}{HTML}{315E91}
\definecolor{AgentTwo}{HTML}{B46527}
\definecolor{Encode}{HTML}{6B4FA1}
\definecolor{Verify}{HTML}{26755A}
\definecolor{Amplify}{HTML}{A43E48}
\definecolor{Ink}{HTML}{35414B}

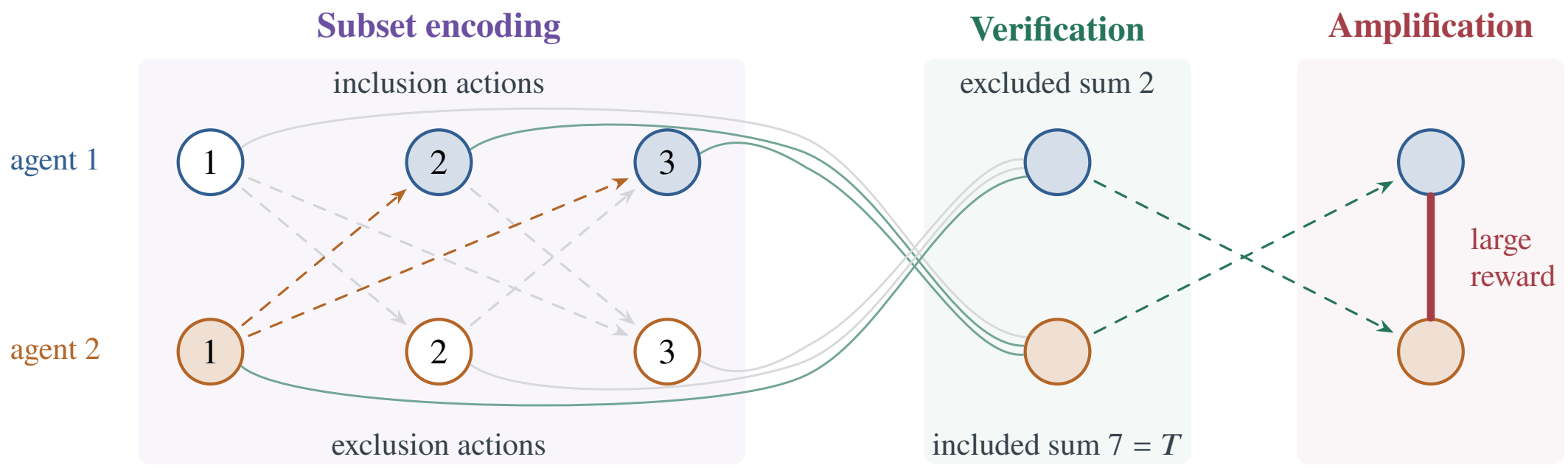
\begin{figure}[!ht]
\centering
\begin{tikzpicture}[x=1cm,y=1cm,font=\footnotesize,
  line cap=round,line join=round,>={Stealth[length=1.5mm]},
  one/.style={circle,draw=AgentOne,fill=white,line width=.8pt,minimum size=5.8mm,inner sep=0pt},
  two/.style={circle,draw=AgentTwo,fill=white,line width=.8pt,minimum size=5.8mm,inner sep=0pt},
  influence/.style={->,dashed,line width=.6pt,shorten <=2pt,shorten >=2pt},
  ordinary/.style={draw=Verify!65,line width=.55pt},
  title/.style={font=\footnotesize\bfseries},
  little/.style={font=\scriptsize,text=Ink}]
  \begin{scope}[on background layer]
    \fill[Encode!5,rounded corners=3pt] (1.6,-.16) rectangle (7.05,3.48);
    \fill[Verify!5,rounded corners=3pt] (8.65,-.16) rectangle (11.05,3.48);
    \fill[Amplify!5,rounded corners=3pt] (12.0,-.16) rectangle (14.4,3.48);
  \end{scope}
  \node[title,text=Encode] at (4.3,3.75) {Subset encoding};
  \node[title,text=Verify] at (9.85,3.75) {Verification};
  \node[title,text=Amplify] at (13.2,3.75) {Amplification};
  \node[anchor=east,little,text=AgentOne] at (1.4,2.55) {agent 1};
  \node[anchor=east,little,text=AgentTwo] at (1.4,.85) {agent 2};
  \node[little] at (4.3,3.27) {inclusion actions};
  \node[little] at (4.3,.02) {exclusion actions};
  \node[one] (q1) at (2.25,2.55) {$1$};
  \node[one,fill=AgentOne!20] (q2) at (4.3,2.55) {$2$};
  \node[one,fill=AgentOne!20] (q3) at (6.35,2.55) {$3$};
  \node[two,fill=AgentTwo!20] (p1) at (2.25,.85) {$1$};
  \node[two] (p2) at (4.3,.85) {$2$};
  \node[two] (p3) at (6.35,.85) {$3$};
  \draw[influence,Ink!22] (q1)--(p2);
  \draw[influence,Ink!22] (q2)--(p3);
  \draw[influence,Ink!22] (q1)--(p3);
  \draw[influence,AgentTwo] (p1)--(q2);
  \draw[influence,Ink!22] (p2)--(q3);
  \draw[influence,AgentTwo] (p1)--(q3);
  \node[one,fill=AgentOne!20] (r0) at (9.85,2.55) {};
  \node[two,fill=AgentTwo!20] (r1) at (9.85,.85) {};
  \node[little] at (9.85,3.27) {excluded sum $2$};
  \node[little] at (9.85,.02) {included sum $7=T$};
  \begin{scope}[on background layer]
    \draw[ordinary,draw=Ink!20] (q1.25) .. controls (3.0,3.10) and (6.6,3.15) .. (7.5,2.8)
      .. controls (8.3,2.52) and (8.8,1.05) .. (r1.155);
    \draw[ordinary] (q2.22) .. controls (5.0,2.98) and (6.7,2.95) .. (7.55,2.65)
      .. controls (8.35,2.35) and (8.9,.9) .. (r1.170);
    \draw[ordinary] (q3.15) .. controls (6.9,2.79) and (7.2,2.75) .. (7.6,2.5)
      .. controls (8.4,2.1) and (9.0,.78) .. (r1.185);
    \draw[ordinary] (p1.-25) .. controls (3.0,.30) and (6.6,.25) .. (7.5,.6)
      .. controls (8.3,.88) and (8.8,2.35) .. (r0.205);
    \draw[ordinary,draw=Ink!20] (p2.-22) .. controls (5.0,.42) and (6.7,.45) .. (7.55,.75)
      .. controls (8.35,1.05) and (8.9,2.5) .. (r0.190);
    \draw[ordinary,draw=Ink!20] (p3.-15) .. controls (6.9,.61) and (7.2,.65) .. (7.6,.9)
      .. controls (8.4,1.3) and (9.0,2.62) .. (r0.175);
  \end{scope}
  \node[one,fill=AgentOne!20] (a) at (13.2,2.55) {};
  \node[two,fill=AgentTwo!20] (b) at (13.2,.85) {};
  \draw[influence,Verify] (r0)--(b);
  \draw[influence,Verify] (r1)--(a);
  \draw[Amplify,line width=2pt] (a)--(b);
  \node[little,text=Amplify,anchor=west,align=left] at (13.42,1.7) {large\\reward};

\end{tikzpicture}
\caption{The quadratic construction for weights $(2,3,4)$ and target $T=7$.
Filled nodes correspond to selected actions. The item actions encode
$H=\{2,3\}$, and both verifiers and the heavy edge are selected. Colors
indicate ownership, faint connections contribute zero. Arrows represent
directed edge gadgets. Two zero-cost baseline actions and their edges are omitted.}
\label{fig:main-construction}
\end{figure}

\paragraph{A gap in utility and reward.}
Let $B$ be the total weight of all edges introduced before amplification.
The amplification edge is chosen sufficiently large that a successful test
leaves the principal utility at least $8B$, even after accounting for the
payment shares. Recall that all designated contracts have the same total
payment, strictly below one, so the principal retains a fixed positive
fraction of this large reward.

If the \textnormal{\textsc{Subset Sum}} instance has no solution, at least one
verifier is absent at every feasible contract. No amplification edge then
contributes, and the greatest equilibrium has reward at most $B$. Every
other equilibrium is contained in the greatest one and has no greater
reward. We obtain the following guarantee for the quadratic construction.

\begin{lemma}[Utility Gap; Informal]
Given a \textnormal{\textsc{Subset Sum}} instance, the above construction is such that
\begin{enumerate}
\item if the \textnormal{\textsc{Subset Sum}} instance has a solution, the optimal principal utility is at least $8B$;
\item otherwise, the reward is at most $B$ at every equilibrium of every feasible contract.
\end{enumerate}
\end{lemma}

The distinction between utility and reward is useful here. In a yes-instance, the principal has enough utility to give up part of her share and still benefit. In a no-instance, even giving an agent the entire reward cannot cover an action whose cost exceeds $B$. We use these two properties to complete the reduction.

\paragraph{From the gap to inapproximability.}
Add an action $g$ of cost $2B$, owned by agent~1, and require it for any positive reward. Thus, the new reward is $f(X\setminus\{g\})$ when $g\in X$, and zero otherwise. This preserves monotonicity and supermodularity.

In a no-instance, any equilibrium selecting $g$ would induce an equilibrium of the original instance with reward at most $B$, insufficient to cover the cost of $g$. Hence the optimal principal utility is zero. In a yes-instance, take a contract $\contract$ and its greatest equilibrium $S$ yielding principal utility at least $8B$, and write $r=1-\alpha_1-\alpha_2$. Increase agent~1's share by $r/2$. By Topkis's monotonicity theorem, the greatest equilibrium can only increase. Agent~1 can still choose his old actions, so his utility at the new equilibrium is at least $(r/2)f(S)\ge4B$, enough to cover the additional cost of $g$. Adding $g$ therefore gives an equilibrium in which the principal retains utility at least $4B$.

Thus, the optimal principal utility is positive exactly when the \textnormal{\textsc{Subset Sum}} instance has a solution. Any deterministic polynomial-time approximation within a finite multiplicative factor would decide this, proving \Cref{thm:supermodular-hardness}.

\par\endgroup

\paragraph{Additive approximation for a constant number of agents.}

The preceding result rules out every finite multiplicative approximation, even for two agents. For a fixed number of agents, perhaps surprisingly, an additive approximation can be obtained by discretizing the contract space. See the proof \Cref{app:supermodular-additive-fptas}

\begin{proposition}\label{prop:supermodular-additive-fptas}
For a constant number of agents, the optimal contract problem with monotone supermodular rewards $0\leq f\leq1$ admits an additive FPTAS in the value-oracle model. 
\end{proposition}

\subsection{Optimization with Few Critical Contracts}
\label{sec:supermodular-search}
The preceding subsection shows that two-agent instances can have super-polynomially many critical contracts and that unrestricted supermodular instances are hard to approximate. We now show that a bound on the number of critical contracts, even if it is not known to the principal, suffices for exact optimization. 

\begin{theorem}\label{thm:polycritical-supermodular}
Consider an $n$-agent instance with a normalized, monotone, supermodular reward given by a value oracle and at most $K$ critical contracts in $[0,1]^n$.
If costs and oracle answers are rationals encoded in at most $L$ bits, then a deterministic algorithm, given the bound $L$, finds an optimal contract and equilibrium in time polynomial in $n,m,K,L$.
\end{theorem}

The algorithm does not need to know $K$ in advance. The assumption concerns the full cube, not just contracts with $\sum_i\alpha_i\leq1$, the algorithm does evaluate auxiliary contracts outside this simplex. Equilibrium computation at these points returns the greatest equilibrium, whether or not it is principal-optimal there. Only contracts in the simplex are considered as output candidates.

We first use the lattice structure and Topkis's theorem to obtain a monotone representation of the reward of the greatest equilibrium through at most $K$ hidden payment requirements. We then prove a hypergraph optimization result of independent interest, showing that containment queries suffice to find a minimum-weight hidden hyperedge in deterministic polynomial time. Combining this result with the hidden payment representation allows us to reconstruct the payment--reward Pareto frontier. Enumerating over each Pareto point and applying our polynomial-time equilibrium selection algorithm yields an optimal contract and equilibrium.

\subsubsection{From Monotonicity to a Hidden Payment Representation} 

We will prove a key property for supermodular rewards, namely that increasing the agents' payment shares can only enlarge the greatest equilibrium. We then use this monotonicity to express the reward at any contract in terms of the critical contracts coordinatewise below it.

We first establish monotonicity of best responses. For an action profile $S\subseteq T$ and a payment share $t\in[0,1]$, let $\operatorname{BR}_i^+(t,S_{-i})$ denote agent~$i$'s greatest best response to $S_{-i}$ when his share is $t$. We prove this result in \Cref{app:greatest-monotone} via Topkis' theorem. 
\begin{lemma}
\label{lem:greatest-monotone}
For every agent $i$, the greatest best response $\operatorname{BR}_i^+(t,S_{-i})$ is nondecreasing both in $t$ and in $S_{-i}$. Consequently, if $\contract,\beta\in[0,1]^n$ and $\contract\leq\beta$ coordinate-wise, then $S^+(\contract)\subseteq S^+(\beta)$.
\end{lemma}
Recall the following observation for critical contracts.
\begin{observation}
\label{obs:incentive-interval}
For every action profile $S$ and agent $i$, the set of shares at which $S_i$ is a best response to $S_{-i}$ is a closed interval, possibly empty. Consequently, the set of contracts that incentivize $S$ is a box. Whenever this box is nonempty, its unique coordinate-wise smallest element is the critical contract $\underline{\contract}(S)$.
\end{observation}

\begin{proof}
A share $t$ belongs to $\left\{t\in[0,1]:S_i\in\operatorname{BR}_i(t,S_{-i})\right\}$ if and only if, for every $R\subseteq T_i$, $t f(S)-c(S_i)\geq t f(R\sqcup S_{-i})-c(R)$.
Each inequality defines a closed interval in $[0,1]$, so their intersection $\left\{t\in[0,1]:S_i\in\operatorname{BR}_i(t,S_{-i})\right\}$ is also a closed interval, possibly empty. Because agent $i$'s best-response condition depends only on his own share, the contracts incentivizing $S$ are exactly $\prod_{i\in[n]} \left\{t\in[0,1]:S_i\in\operatorname{BR}_i(t,S_{-i})\right\}$.
Whenever this product is nonempty, let $\underline{\alpha}_i(S)$ be the left endpoint of $\left\{t\in[0,1]:S_i\in\operatorname{BR}_i(t,S_{-i})\right\}$. Closedness implies that $\underline{\contract}(S)=\bigl(\underline{\alpha}_1(S),\ldots,\underline{\alpha}_n(S)\bigr)$
incentivizes $S$ and is the unique coordinate-wise smallest contract that does so.
\end{proof}

We next prove the structural property of the contract region induced by supermodular rewards that allows the algorithm to discard an entire region of the contract space after processing a single query.

Each contract $\contract\in[0,1]^n$ determines an associated critical contract $\underline{\contract}(S^+(\contract))$, namely the critical contract of the greatest equilibrium induced by $\contract$. The resulting mapping $
\contract\longmapsto\underline{\contract}(S^+(\contract))$
induces a partition of $[0,1]^n$ through its pre-image: two contracts belong to the same part if and only if they have the same associated critical contract. We call the parts of this partition \emph{critical regions}. The next lemma highlights some structure on these critical regions. 

\begin{lemma}
\label{lem:box-certificate}
Let $q\in[0,1]^n$. Then, for every $\contract\in[0,1]^n$ satisfying $\underline{\contract}(S^+(q))\leq\contract\leq q$, we have $S^+(\contract)=S^+(q)$. 
\end{lemma}

\begin{proof}
Both $q$ and $\underline{\contract}(S^+(q))$ incentivize $S^+(q)$, by definition of the critical mapping. By \Cref{obs:incentive-interval}, the contracts incentivizing $S^+(q)$ form a box. Therefore, every $\contract$ satisfying $\underline{\contract}(S^+(q))\leq\contract\leq q$
also incentivizes $S^+(q)$. Since $S^+(\contract)$ is the greatest equilibrium at $\contract$, it follows that $S^+(q)\subseteq S^+(\contract)$.
Conversely, $\contract\leq q$ and \Cref{lem:greatest-monotone} imply $S^+(\contract)\subseteq S^+(q)$.
The two inclusions prove the result.
\end{proof}
Thus, if $q$ belongs to the critical region associated with $\underline{\contract}(S^+(q))$, then the entire coordinate-wise order interval (equivalently, the box) $\left[\underline{\contract}(S^+(q)),q\right]$ belongs to the same critical region. Hence, although a critical region need not be convex, it becomes downward closed when its critical contract is shifted to the origin.

We now use these properties to represent the reward of the greatest
equilibrium through hidden payment requirements. Let
$\contract^1,\ldots,\contract^K$ be the distinct critical contracts in
$[0,1]^n$, and set $v_j=f(S^+(\contract^j))$. 

\begin{lemma}\label{eq:reward-frontier}
For every
$\contract\in[0,1]^n$, $\displaystyle f(S^+(\contract))
=
\max_{j:\,\contract^j\leq\contract} v_j$.
\end{lemma}

\begin{proof}
 Indeed, if $\contract^j\leq\contract$, monotonicity gives
$S^+(\contract^j)\subseteq S^+(\contract)$, so
$v_j\leq f(S^+(\contract))$. Conversely, the critical contract of
$S^+(\contract)$ belongs to this family and lies below $\contract$.
By \Cref{lem:box-certificate}, its greatest equilibrium is still
$S^+(\contract)$, so the maximum attains $f(S^+(\contract))$.
\end{proof}
Each pair $(\contract^j,v_j)$ therefore specifies payment requirements
that guarantee a reward of at least $v_j$ at the greatest equilibrium.
These $K$ pairs are hidden, but we can evaluate the reward at given any
contract using the polynomial-time equilibrium algorithm. Moreover,
every positive critical share is a ratio of a cost difference to a
reward difference, so the coordinates of the critical contracts have
encoding length polynomial in $m$ and $L$.

This hidden rewards representation leads us to study the following independent problem.

\subsubsection{Optimization over a Hidden Hypergraph}
\label{sec:hidden-optimization}

Consider a known vertex set $[d]$ with nonnegative rational weights
$w_a$, and an unknown family $\mathcal E\subseteq 2^{[d]}$ of at most
$K$ hyperedges. A containment query asks whether a set $X\subseteq[d]$
contains at least one hyperedge,
\[
h(X)=\mathbf{1}\!\left\{\exists E\in\mathcal E:\ E\subseteq X\right\}.
\]
We write $w(X)=\sum_{a\in X}w_a$. Since weights are nonnegative, discarding
hyperedges that contain another hyperedge changes neither the oracle
nor the minimum weight. We may therefore restrict attention to
inclusion-minimal hyperedges.

Reconstructing the entire hypergraph can require $2^{\Omega(K)}$
containment queries~\cite[Theorem~22]{AngluinChen2006}. Nevertheless, we show
that finding a minimum-weight hyperedge requires only polynomially
many queries.

\begin{theorem}\label{thm:hidden-edge}
Given a nonempty hidden hypergraph with $d$ vertices, at most $K$ hyperedges, and known nonnegative rational vertex weights, a minimum-weight hyperedge can be found deterministically using polynomially many containment queries and time polynomial in $K$ and the explicit input size.
\end{theorem}
We stress that the algorithm does not need $K$ as input. 

Finding one hyperedge is straightforward. Starting from a positive
query, we delete vertices whenever the answer remains positive.
After at most $d$ queries, the remaining set is an inclusion-minimal
hyperedge. The difficulty is finding further hyperedges, since a
query containing a known hyperedge gives a positive answer regardless
of whether it also contains a cheaper one. To construct a query containing no known edge, we must delete at least one vertex from each known edge. Thus, the deleted vertices form a vertex cover of the known edges. The more vertices we delete, the easier it is to exclude all the known edges, but the greater is the risk of losing a minimum-weight edge. A single deletion from that \emph{unknown} edge is enough to exclude it from the query. The challenge is to make enough deletions to rule out every known edge while preserving the  minimum-weight edge in at least one query.\medskip 

The following example helps to build intuition on why the weights help: let us compare a known edge $E$ to a cheaper vertex set $X^*$. The vertices they share contribute the same weight to both, so cancelling that common weight gives $w(E\setminus X^*)>w(X^*\setminus E)$. Now imagine choosing a vertex that belongs to exactly one of the two sets, with probability proportional to its weight. The probability of choosing a vertex in $E\setminus X^*$ is $w(E\setminus X^*)/(w(E\setminus X^*)+w(X^*\setminus E))>\frac12$. By repeating the draw independently and keeping each deletion as a separate candidate, we would obtain with high probability at least one candidate that rules out $E$ and leaves $X^*$ intact. 

There are unfortunately three issues with this naive approach. First, $X^*$ is unknown, so we cannot identify the vertices that belong to exactly one of $E$ and $X^*$ and sample from them. Second, we must rule out all known edges in the same query while leaving $X^*$ intact. Even with just two candidate deletions for each of $r$ known edges, trying all combinations would take $2^r$ steps. Finally, this remains a randomized sampling procedure, which can still fail with some probability, while we want a deterministic guarantee. 

The idea is to use one known edge as a reference. Suppose $E$ is the known edge with the least weight outside $X^*$. By the comparison above, even this minimum exceeds the weight of $X^*\setminus E$. Thus the weight we need to protect outside $E$ is smaller than the weight available for safe deletions in every known edge. Although $X^*$ is hidden, we can try each of the $r$ known edges in the role of $E$. For each choice, we use a derandomized sampling argument to construct a short list of sets of candidate vertices. For a suitable set on the list, we can safely delete all its vertices outside $E$ at once, and we can find a way to rule out the remaining known edges while preserving $X^*$ among only polynomially many choices of deletions inside $E$.

The key technical result is therefore a covering lemma that, given a set of known hyperedges, produces query sets containing no known hyperedge while preserving every cheaper set in at least one query. See the proof in \Cref{app:cover-hidden}.

\begin{lemma}\label{lem:cover-hidden}
Given nonempty $E_1,\ldots,E_r\subseteq[d]$ and nonnegative rational vertex
weights, a deterministic algorithm outputs polynomially many sets,
each containing no $E_j$. Each $X^*\subseteq[d]$ with
$w(X^*)<\min_j w(E_j)$ is contained in some output. The running
time is polynomial in $d,r$, and the encoding length of the weights.
\end{lemma}

Armed with this covering guarantee, we can now finish proving the optimization result. 

\begin{proof}[Proof of \Cref{thm:hidden-edge}]
Since the hidden family is nonempty, $[d]$ is a positive query. From any positive query, we can extract a hyperedge by testing each vertex once and deleting it whenever the remaining set is positive. A deletion that fails cannot become possible as the set shrinks, so the final set is inclusion-minimal among positive sets and hence is a hyperedge. Apply this procedure to $[d]$ to obtain the first edge. If this edge is empty, return it, since it has weight zero and all vertex weights are nonnegative.

Given the edges found so far, construct the list from \Cref{lem:cover-hidden}. If a listed query is positive, extract a hyperedge from it and repeat. This edge is new because the query contains no known edge. If all listed queries are negative, return a minimum-weight known edge. Any strictly cheaper hidden edge would be contained in some listed query, making it positive. There are at most $K$ rounds, each using one polynomial-time covering construction, polynomially many containment queries, and at most $d$ additional queries for extraction. This gives the claimed bounds without requiring the algorithm to know $K$.
\end{proof}

We next extend this search to fractional payment requirements and
use it to reconstruct the payment--reward Pareto frontier.

\subsubsection{Pareto Frontier Reconstruction}
\label{sec:hidden-frontier}

We return to the hidden pairs $(\contract^j,v_j)$ from
\Cref{eq:reward-frontier}. The payment--reward Pareto frontier consists
of the nondominated pairs
\[
\big(\sum_i\contract_i,\ f(S^+(\contract))\big),
\qquad \contract\in[0,1]^n,
\]
where smaller payments and larger rewards are preferred. By
\Cref{lem:box-certificate}, every contract can be lowered to a critical
contract without changing its greatest equilibrium. The frontier
therefore contains at most $K$ points. We show how to reconstruct it
without learning all critical contracts. 

To recover the payment--reward frontier, we repeatedly find the cheapest contract that improves on the current best recorded reward. For each search, we build an auxiliary hypergraph from the critical payment shares identified so far. Its hidden edges represent critical contracts that improve on the current reward. By applying the algorithm from \Cref{thm:hidden-edge} on this auxiliary graph, we identify a lower bound on the payment needed for such an improvement. We then either find a contract attaining this bound or discover a new critical share and use it to refine the hypergraph. See the proof in \Cref{app:hidden-frontier}.
\begin{lemma}\label{lem:hidden-frontier}
A deterministic algorithm computes the payment--reward Pareto frontier and a witness contract for each point in time polynomial in $n,m,K$, and $L$.
\end{lemma}

The reason why what matters is the number of critical contract in $[0,1]^n$, is because in order to use the hypergraph result, we need to always start with a first positive query, which here is $\mathbf 1$ that dominates all other contracts in terms of rewards. 
We can now conclude with the supermodular contract optimization result. 

\begin{proof}[Proof of \Cref{thm:polycritical-supermodular}]
By \Cref{eq:reward-frontier}, the reward of the greatest equilibrium
has a representation through at most $K$ hidden payment requirements,
whose encoding length is polynomial in $m$ and the input encoding size. Moreover,
\Cref{thm:polytime-optimal-equilibrium-supermodular} allows us to
evaluate this reward in polynomial time throughout $[0,1]^n$.

Apply \Cref{lem:hidden-frontier}, retain the reported contracts with
$\sum_i\contract_i\leq1$, and choose one maximizing $(1-\sum_i\contract_i)f(S^+(\contract))$.
Lower payments and higher rewards can only improve this objective
on the feasible simplex. A frontier point dominating a feasible
contract also lies in this simplex, and contracts with total
payment greater than one give nonpositive utility while the zero
contract gives nonnegative utility. Thus the frontier contains
a feasible optimum. By \Cref{obs:optimal-extremal-equilibrium},
the greatest equilibrium of the selected contract is principal-optimal.
Returning this contract and equilibrium proves the claimed guarantee.
The running time is polynomial in $n,m,K,L$ by the frontier and
equilibrium algorithms.
\end{proof}

\newcommand{\PrintComplementsAppendix}{

\section{Supermodular Hardness}\label{app:superhard}

This section provides the fully detailed construction and proof for \Cref{thm:supermodular-hardness} of the $n=2$ agents supermodular hardness result. 

\subsection{Inapproximability for Supermodular Rewards}
\label{sec:supermodular-hardness}

 We first construct a quadratic reward given by induced edge weight in a bipartite graph, and then add a gating action to obtain the final supermodular instance.

Agent~1 has ground set $U$, agent~2 has ground set $V$, with
$U\cap V=\emptyset$, and the reward function of this intermediate instance has the
form
\begin{equation}\label{eq:quadratic-reward}
        f(A,B)=\sum_{u\in A,\,v\in B} f_{uv},
        \qquad f_{uv}\ge0.
\end{equation}
This corresponds to the total weights of edges fully covered on both ends by vertices in a bipartite graph, where non-existent edges are interpreted to have weight $0$.  

For $B\subseteq V$ and $u\in U$, let
\begin{equation*}
d_B(u)=\sum_{v\in B}f_{uv}.
\end{equation*}
For $A\subseteq U$ and $v\in V$, define $d_A(v)$ symmetrically

We shall also
use payment vectors $\lambda=(\lambda_1,\lambda_2)\in\mathbb{R}_+^2$ that are
not necessarily feasible contracts, but which will be rescaled at the end.

A profile $(A,B)$ is called $\lambda$-stable if 

\begin{equation*}
\lambda_1 d_B(u)\ge c(u)
        \quad\text{for every }u\in A,
        \qquad
        \lambda_2 d_A(v)\ge c(v)
        \quad\text{for every }v\in B.
\end{equation*}
Let $S(\lambda)$ be the union of all $\lambda$-stable profiles. If a vertex
belongs to the profile under discussion, we call it active. The support of a
vertex is the total weight of its edges to active vertices on the opposite side.

\begin{lemma}\label{lem:selected-quadratic}
For every instance of the form \eqref{eq:quadratic-reward} and every $\lambda\in\mathbb{R}_+^2$, the profile $S(\lambda)$ is a pure Nash equilibrium at payments $\lambda$ and  $\max_{S \in \nash(\lambda)} f(S)=f(S(\lambda))$.

\end{lemma}

\begin{proof} 

Stable profiles are closed under coordinate-wise union. Indeed, if $(A,B)$ and
$(A',B')$ are stable, then every vertex in $A\cup A'$ has at least as much
support against $B\cup B'$ as it had in one of the original profiles, and
non-negativity of the edge weights therefore preserves its stability. The same
argument applies on side $V$. Hence $S(\lambda)$ is itself stable and is the
unique inclusion-wise maximal stable profile.

Write $S(\lambda)=(A^*,B^*)$. An omitted vertex cannot have strictly positive
marginal utility: if, for example, $u\notin A^*$ satisfied
$\lambda_1d_{B^*}(u)>c(u)$, then adding $u$ would preserve the stability of all
old vertices and make $u$ stable, contradicting maximality. Thus active
vertices have nonnegative marginal utility and inactive vertices have
nonpositive marginal utility. For a fixed opponent set, each agent's utility is
additive over her own vertices. Consequently, deleting active vertices or
adding inactive vertices cannot increase utility, even as part of an arbitrary
set deviation. Therefore $S(\lambda)$ is a pure Nash equilibrium.

Conversely, every pure Nash equilibrium is stable, since an active vertex with
negative marginal utility could be profitably deleted. Every pure Nash
equilibrium is therefore contained in $S(\lambda)$. The reward is monotone, so
$S(\lambda)$ maximizes reward among pure Nash equilibria.
\end{proof}

\subsubsection{Construction of the Bipartite Graph Instance}

Fix a \textnormal{\textsc{Subset Sum}} instance with positive integer weights
$w_1,\ldots,w_k$, target $T$, and total weight $W=\sum_{i=1}^k w_i$. We assume without loss of generality that $k\ge1$ and $1\le T\le W-1$. We define the following parameter for the reward instance:
\begin{equation*}
L=3^{-k},
        \qquad
        \Delta=\frac{3^{-3k}}{10^{20}kW^4},
        \qquad
        \varepsilon=\frac{\Delta L}{8}.
\end{equation*}
We note in particular that $\Delta=3^{-3k}/(10^{20}kW^4) \leq 10^{-20}$. We now start from an empty graph, and add vertices, edge weights, and costs step by step so as to construct a valid polynomial-size weighted bipartite graph reward function. We will make it so that if $\lambda_1+\lambda_2 >1+\varepsilon$, after rescaling, $\alpha_1+\alpha_2>1$, which would yield negative utility for the principal. We give a high-level overview of the full bipartite graph construction in \Cref{fig:hard_construction} for $k=3$. The bold lines correspond to standard edges, whereas the dashed arrows represent a three-edge path gadget described in \Cref{fig:three-edge-path}.  

\begin{figure}
    \centering
    \begin{tikzpicture}[scale=0.8, transform shape,
  x=1cm,y=1cm,
  font=\small,
  line cap=round,
  line join=round,
  >={Stealth[length=1.7mm,width=1.2mm]},
  anchornode/.style={circle,draw=AnchorGray,fill=white,line width=.75pt,minimum size=9.5mm,inner sep=0pt},
  qnode/.style={circle,draw=UBlue,fill=white,line width=.85pt,minimum size=9.5mm,inner sep=0pt},
  pnode/.style={circle,draw=VOrange,fill=white,line width=.85pt,minimum size=9.5mm,inner sep=0pt},
  rnode/.style={circle,draw=VerifyGreen,fill=white,line width=.85pt,minimum size=9.5mm,inner sep=0pt},
  anode/.style={circle,draw=AmpRed,fill=white,line width=.85pt,minimum size=9.5mm,inner sep=0pt},
  boxgray/.style={rounded corners=3pt,draw=AnchorGray!75,line width=.5pt,fill=AnchorFill},
  boxpurple/.style={rounded corners=3pt,draw=StatePurple,line width=.6pt,fill=StateFill},
  boxgreen/.style={rounded corners=3pt,draw=VerifyGreen,line width=.6pt,fill=VerifyFill},
  boxred/.style={rounded corners=3pt,draw=AmpRed,line width=.6pt,fill=AmpFill},
  ge/.style={draw=AnchorGray!78,line width=.48pt},
  ve/.style={draw=VerifyGreen,line width=.62pt},
  qp/.style={draw=UBlue,dash pattern=on 3pt off 2pt,-{Stealth},line width=.78pt,shorten <=3pt,shorten >=3pt},
  pq/.style={draw=VOrange,dash pattern=on 3pt off 2pt,-{Stealth},line width=.78pt,shorten <=3pt,shorten >=3pt},
  ap/.style={draw=AmpRed,dash pattern=on 3.1pt off 2.2pt,-{Stealth},line width=.85pt,shorten <=3pt,shorten >=3pt},
  ae/.style={draw=AmpRed,line width=1.15pt},
  title/.style={font=\footnotesize\bfseries},
  lab/.style={font=\scriptsize,inner sep=.5pt},
  legendtxt/.style={font=\scriptsize,align=left},
  gadgetedge/.style={draw=black!78,line width=.65pt},
  gadgetpath/.style={draw=black!78,dash pattern=on 2.6pt off 2.0pt,-{Stealth},line width=.7pt,shorten <=2.6pt,shorten >=2.6pt}
]

\def\yU{4.25}
\def\yV{0.60}

\begin{pgfonlayer}{background}
  \path[boxgray]   (0.20,-1.15) rectangle (2.05,6.05);
  \path[boxpurple] (2.60,-1.15) rectangle (8.30,6.05);
  \path[boxgreen]  (10.45,-1.15) rectangle (12.70,6.05);
  \path[boxred]    (13.55,-1.15) rectangle (15.20,6.05);

\end{pgfonlayer}

\node[title,text=AnchorGray] at (1.13,6.48) {Anchors};
\node[title,text=StatePurple] at (5.45,6.48) {State encoding};
\node[title,text=VerifyGreen] at (11.58,6.48) {Verification};
\node[title,text=AmpRed] at (14.38,6.48) {Amplification};
\node[font=\footnotesize\bfseries,text=UBlue,anchor=east] at (-0.04,\yU) {side $U$};
\node[font=\footnotesize\bfseries,text=VOrange,anchor=east] at (-0.04,\yV) {side $V$};
\node[font=\footnotesize,text=UBlue] at (5.45,5.35) {$q_i$: offset $1-3^{-i}$};
\node[font=\footnotesize,text=VOrange] at (5.45,-0.45) {$p_i$: offset $-3^{-i}$};

\node[anchornode] (u0) at (1.13,\yU) {$u_0$};
\node[anchornode] (v0) at (1.13,\yV) {$v_0$};

\node[qnode] (q1) at (3.50,\yU) {$q_1$};
\node[qnode] (q2) at (5.35,\yU) {$q_2$};
\node[qnode] (q3) at (7.20,\yU) {$q_3$};
\node[pnode] (p1) at (3.75,\yV) {$p_1$};
\node[pnode] (p2) at (5.60,\yV) {$p_2$};
\node[pnode] (p3) at (7.45,\yV) {$p_3$};

\node[rnode] (r0) at (11.58,\yU) {$r_0$};
\node[rnode] (r1) at (11.58,\yV) {$r_1$};

\node[anode] (a) at (14.38,\yU) {$a$};
\node[anode] (b) at (14.38,\yV) {$b$};

\begin{pgfonlayer}{background}

\draw[ge] (u0.-30) to[bend right=5]
   (p1.168);
\draw[ge] (u0.-20) to[bend right=10] (p2.170);
\draw[ge] (u0.-10) to[bend right=15] node[lab,text=AnchorGray,pos=.1,sloped,above] {$1/(2\Delta)$} (p3.172);

\draw[ge] (u0.30) .. controls (2.28,5.70) and (7.30,6.03) .. (9.82,5.88)
               .. controls (10.42,5.85) and (10.42,1.50) .. (r1.148);

\draw[ge] (v0.30) to[bend left=5]
   (q1.192);
\draw[ge] (v0.20) to[bend left=10] (q2.190);
\draw[ge] (v0.10) to[bend left=15] node[lab,text=AnchorGray,pos=.1,sloped,below] {$1/(2\Delta)$} (q3.188);

\draw[ge] (v0.-30) .. controls (2.28,-0.70) and (7.30,-1.03) .. (9.82,-0.88)
               .. controls (10.42,-0.85) and (10.42,3.50) .. (r0.212);

\draw[ve] (q1.22) .. controls (4.20,5.18) and (7.40,5.34) .. (9.16,5.12)
                  .. controls (9.84,5.04) and (9.94,1.54) .. (r1.154);
\draw[ve] (q2.18) .. controls (5.80,4.96) and (7.54,4.96) .. (9.16,4.78)
                  .. controls (9.84,4.68) and (9.96,1.34) .. (r1.166);
\draw[ve] (q3.14) .. controls (7.12,4.70) and (7.80,4.58) .. (9.16,4.42)
                  .. controls (9.84,4.30) and (9.98,1.14) .. (r1.178);

\draw[ve] (p1.-22) .. controls (4.20,-0.18) and (7.40,-0.34) .. (9.16,-0.12)
                   .. controls (9.84,-0.04) and (9.94,3.46) .. (r0.206);
\draw[ve] (p2.-18) .. controls (5.80,0.04) and (7.54,0.04) .. (9.16,0.22)
                   .. controls (9.84,0.32) and (9.96,3.66) .. (r0.194);
\draw[ve] (p3.-14) .. controls (7.12,0.30) and (7.80,0.42) .. (9.16,0.58)
                   .. controls (9.84,0.70) and (9.98,3.86) .. (r0.182);
\end{pgfonlayer}

\draw[qp] (q1) --  (p2);
\draw[qp] (q1) -- node[lab,text=UBlue,sloped,above=1pt,pos=0.15] {$2\cdot 3^{-i}$} (p3);
\draw[qp] (q2) --  (p3);

\draw[pq] (p1) --  node[lab,text=VOrange,sloped,above=1pt,pos=0.15] {$2\cdot 3^{-1}$} (q2);
\draw[pq] (p1) --   (q3);
\draw[pq] (p2) --  (q3);

\draw[ae] (a) -- node[lab,text=AmpRed,right=2pt,rotate=0,pos=.52] {$D$} (b);
\draw[ap] (r1) -- node[lab,text=AmpRed,sloped,above=2pt,pos=.18] {$12\Delta D$} (a);
\draw[ap] (r0) -- node[lab,text=AmpRed,sloped,below=2pt,pos=.18] {$12\Delta D$} (b);

\node[font=\footnotesize\bfseries,anchor=west] at (0,-1.5) {Edges};
\draw[ge] (0.20,-1.9) -- (0.9,-1.9);
\node[legendtxt,anchor=west] at (1.1,-1.9) {anchor edges};
\draw[ve] (0.20,-2.3) -- (0.9,-2.3);
\node[legendtxt,anchor=west] at (1.1,-2.3) {state-verification edges};
\draw[ae] (0.20,-2.7) -- (0.9,-2.7);
\node[legendtxt,anchor=west] at (1.1,-2.7) {amplification edge};

\node[font=\footnotesize\bfseries,anchor=west] at (6,-1.5) {Gadget Paths};
\draw[qp] (6,-1.9) -- (7.3,-1.9);
\node[legendtxt,anchor=west] at (7.5,-1.9) {paths $q_j\to p_i$, for $j<i$};
\draw[pq] (6,-2.3) -- (7.3,-2.3);
\node[legendtxt,anchor=west] at (7.5,-2.3) {paths $p_j\to q_i$, for $j<i$};
\draw[ap] (6,-2.7) -- (7.3,-2.7);
\node[legendtxt,anchor=west] at (7.5,-2.7) {verification-amplification paths};

\end{tikzpicture}
    \caption{High level representation of the reduction from \textnormal{\textsc{Subset Sum}} to weighted bipartite graph for $k=3$.}
    \label{fig:hard_construction}
\end{figure}

\medskip
\noindent\textbf{Anchors and comparator vertices.}
Add two zero-cost vertices $u_0\in U$ and $v_0\in V$. For $\theta\in[-1,1]$ and $s \geq 0$, define the following threshold function
\begin{equation*}
T_\theta(s)=\frac{\frac12+\Delta\theta}{1+2\Delta s}.
\end{equation*}
A comparator of offset $\theta$ on side $V$ is defined as a vertex
$v\in V$ with cost
\begin{equation*}
c(v)=\left(\frac12+\Delta\theta\right)\frac{1}{2\Delta} \quad  \mbox{and} \quad f_{u_0v}=\frac{1}{2\Delta}.
\end{equation*}
If its non-anchor support is $s$, then stability means
\begin{equation*}
\lambda_2\left(\frac{1}{2\Delta}+s\right)
        \ge
        \left(\frac12+\Delta\theta\right)\frac{1}{2\Delta},
\end{equation*}
which is equivalent to $\lambda_2\ge T_\theta(s)$. Comparators on side $U$ are
defined identically with anchor weight coming from $v_0$: $c(u)=\left(\frac12+\Delta\theta\right)\frac{1}{2\Delta}$ and $f_{v_0 u} =1/(2\Delta)$, with stability being equivalent to $\lambda_1 \geq T_\theta(s)$ with a non-anchor support of $s$. 

\medskip
\noindent\textbf{Three-edge paths.} One limitation of the two agents having the same reward, is that we cannot add directed edges to the graph, which are needed for the encoding of the chosen subset in the subset-sum problem. We circumvent this issue by constructing a three-edge gadget that simulates a directed edge when the construction parameters are tuned correctly. 
For vertices $s$ and $t$ on opposite sides and for a value $M>0$, add two
internal vertices: $z$ on the side of $t$ and $y$ on the side of $s$. Add three
edges with weights $f_{sz}=1024 \Delta^2M$, $f_{zy}=32\Delta M$, and $f_{yt}=M$. 
Set
\begin{equation*}
c(z)=\left(\frac12-4\Delta\right)
        (1024\Delta^2M+32\Delta M), \quad \mbox{and} \quad c(y)=\left(\frac12-4\Delta\right)(32\Delta M+M).
\end{equation*}
We shall say only for orientation that the path goes from $s$ to $t$; the graph
remains undirected. The edge incident to $s$ has weight $1024\Delta^2M$, which will be very small and negligible even when activated, and
the edge incident to $t$ has the desired weight $M$. See \Cref{fig:three-edge-path} for a representation of this gadget. 

\begin{figure}[t]
\centering
\begin{tikzpicture}[
  x=1cm,y=1cm,
  font=\small,
  line cap=round,
  line join=round,
  >={Stealth[length=1.7mm,width=1.2mm]},
  nodev/.style={
    circle,draw=AnchorGray,fill=white,line width=.75pt,
    minimum size=7mm,inner sep=0pt
  },
  lab/.style={font=\scriptsize,inner sep=.5pt},
  gadgetedge/.style={draw=black!78,line width=.65pt},
  dashedpath/.style={
    draw=black!78,dash pattern=on 2.6pt off 2.0pt,
    -{Stealth},line width=.7pt,
    shorten <=2.6pt,shorten >=2.6pt
  }
]

\node[nodev] (sabs) at (.65,0) {$s$};
\node[nodev] (tabs) at (3.45,0) {$t$};
\draw[dashedpath] (sabs) -- node[lab,above=2pt] {$M$} (tabs);

\node[nodev] (s) at (5.65,.48) {$s$};
\node[nodev] (y) at (5.65,-.48) {$y$};
\node[nodev] (t) at (9.35,.48) {$t$};
\node[nodev] (z) at (9.35,-.48) {$z$};

\draw[gadgetedge] (s) -- node[lab,sloped,above=2pt,pos=.2] {$1024\Delta^2M$} (z);
\draw[gadgetedge] (y) -- node[lab,sloped,above=2pt,pos=.75] {$M$} (t);
\draw[gadgetedge] (y) -- node[lab,below=2pt] {$32\Delta M$} (z);

\end{tikzpicture}

\caption{The three-edge path of value $M$ from $s$ to $t$. The graph is
undirected; the orientation records only which endpoint is intended to provide
support to the other.}
\label{fig:three-edge-path}
\end{figure}

We prove the following Lemma, which essentially states that the target of the constructed gadget receives the support $M$ if and only if the source is activated as well. Together with the negligible reward that the source receives, this proves that the gadget functions properly as a directed edge. See the proof in \Cref{app:three-edge-path}.

\begin{lemma}\label{lem:three-edge-path}
Consider one three-edge path $s-z-y-t$ of value $M$ from $s$ to $t$. Let $\lambda_s$ and
$\lambda_t$ denote the payment coordinates on the sides of $s$ and $t$,
respectively, and suppose that $\lambda_s+\lambda_t \leq 1+\varepsilon$. 
\begin{enumerate}
    \item If $s,t\in S(\lambda)$ and $\lambda_s,\lambda_t\geq 1/2-4\Delta$, then
$y,z\in S(\lambda)$, and the edge of weight $M$ incident to $t$ is
active.
    \item  If $s\notin S(\lambda)$ and $\lambda_t\geq 1/2-4\Delta$, then
$y\notin S(\lambda)$, and the edge of weight $M$ incident to $t$ is
inactive.
\end{enumerate}
\end{lemma}

\medskip
\noindent\textbf{State vertices.}
We would like to map a subset of $[k]$ for the subset sum problem to some segments of $[0,1]$. Intuitively, we can represent such a subset via a binary encoding, $1$ at coordinate $i$ meaning that item $i$ is added to the subset, and $0$ meaning that it is not added to the subset. We want to further map these bits to vertices on the bipartite graph. For each item $i\in\{1,\ldots,k\}$, add two comparator vertices:
\begin{equation*}
p_i\in V \quad\text{with offset }-3^{-i},
        \qquad
        q_i\in U \quad\text{with offset }1-3^{-i}.
\end{equation*}
For every pair $j<i$, add one three-edge path of value $2\cdot3^{-j}$ from
$q_j$ to $p_i$, and one three-edge path of value $2\cdot3^{-j}$ from $p_j$ to
$q_i$. For $p_i$ and $q_i$ to be active, the corresponding payment coordinates
$\lambda_2$ and $\lambda_1$ must be large enough, respectively. See the proof in \Cref{app:active-state-large-payment}.

\begin{lemma}\label{lem:active-state-large-payment}
For all $i \in [k]$, if $q_i \in S(\lambda)$ then $\lambda_1 \geq 1/2-4\Delta$, and if $p_i \in S(\lambda)$ then $\lambda_2 \geq 1/2-4\Delta$. 
\end{lemma}

Intuitively, $p_i$ active represents that the bit at coordinate $i$ is $0$ and therefore the item is not included in the subset, while $q_i$ active represents that the bit at coordinate $i$ is $1$ and therefore the item is included in the subset. For $\sigma\in\{0,1\}^k$, define the ternary mass of the binary vector and the encoder interval as
\begin{equation*}
Q(\sigma)=\sum_{i:\,\sigma_i=1}2\cdot3^{-i}, \quad \mbox{and} \quad I_\sigma=
        \left[
        \frac12+\Delta Q(\sigma)+\frac{\Delta L}{4},
        \frac12+\Delta Q(\sigma)+\frac{3\Delta L}{4}
        \right].
\end{equation*}

\begin{figure}
    \centering
    \resizebox{\textwidth}{!}{\begin{tikzpicture}[
  x=1cm,y=1cm,
  font=\small,
  line cap=round,
  line join=round,
  >={Stealth[length=2.1mm,width=1.45mm]},
  axis/.style={black,line width=.65pt,-{Stealth[length=2.1mm,width=1.45mm]}},
  tick/.style={black,line width=.65pt},
  endpoint/.style={black,fill=black},
  bdash/.style={BandBlue,dashed,line width=.75pt,dash pattern=on 3.2pt off 3.2pt},
  brace/.style={BandBlue,line width=.75pt,decorate,decoration={brace,amplitude=5pt,mirror}},
  label/.style={font=\small,inner sep=1pt},
  ilabel/.style={font=\large,inner sep=1pt},
  qlabel/.style={font=\scriptsize,inner sep=1pt},
  interval/.style={line width=.8pt},
]

\def\xzero{0.8}
\def\xone{16.2}
\def\xhalf{8.5}
\def\xleft{6.35}
\def\xright{11.70}

\def\ytop{5.8}
\draw[axis] (0.05,\ytop) -- (17.15,\ytop);
\fill[endpoint] (\xzero,\ytop) circle (1.5pt);
\fill[endpoint] (\xone,\ytop) circle (1.5pt);
\draw[tick] (\xhalf,\ytop-.12) -- (\xhalf,\ytop+.12);
\node[label,anchor=north] at (\xzero,\ytop-.15) {$0$};
\node[label,anchor=north] at (\xhalf,\ytop-.24) {$1/2$};
\node[label,anchor=north] at (\xone,\ytop-.15) {$1$};

\fill[FillBlue] (\xleft,\ytop+.03) rectangle (\xright,\ytop+.27);
\draw[bdash] (\xleft,\ytop-.4) -- (\xleft,\ytop+.4);
\draw[bdash] (\xright,\ytop-.4) -- (\xright,\ytop+.4);

\node[label,anchor=south] (topL) at (\xleft,\ytop+0.5)
  {$\displaystyle \frac12+\frac{\Delta L}{4}$};

\node[label,anchor=south] (topR) at (\xright,\ytop+0.5)
  {$\displaystyle \frac12+\Delta\!\left(1-\frac{L}{4}\right)$};

\draw[brace] (\xleft,\ytop-.55) -- (\xright,\ytop-.55);
\node[label,text=BandBlue,anchor=north] at ($(\xleft,\ytop-.86)!.5!(\xright,\ytop-.86)$)
  {all $2^3$ encoder intervals $I_\sigma$ lie in this band};

\def\ybot{2.80}
\draw[axis] (0.05,\ybot) -- (17.15,\ybot);
\draw[tick] (\xhalf,\ybot-.12) -- (\xhalf,\ybot+.12);
\node[label,anchor=north] at (\xhalf,\ybot-.18) {$\displaystyle \frac12$};

\draw[bdash] (\xzero-.18,\ybot-.95) -- (\xzero-.18,\ybot+1.15);
\draw[bdash] (\xone+.12,\ybot-.95) -- (\xone+.12,\ybot+1.15);

\node[label,anchor=north] at (\xzero-.18,\ybot-1.20)
  {$\displaystyle \frac12+\frac{\Delta L}{4}$};
\node[label,anchor=north] at (\xone+.12,\ybot-1.20)
  {$\displaystyle \frac12+\Delta\!\left(1-\frac{L}{4}\right)$};

\def\iw{1.32}
\def\ih{0.18}
\def\yrect{2.80}
\def\ydashTop{3.5}
\def\ylabel{3.7}
\def\yq{2.32}

\foreach \x/\name/\q/\col/\fillcol in {
  1.45/000/{0}/Iblue/Iblue!10,
  3.25/001/{\frac{2}{27}}/Ipurple/Ipurple!10,
  5.15/010/{\frac{2}{9}}/Iteal/Iteal!12,
  7.05/011/{\frac{8}{27}}/Imagenda/Imagenda!10,
  9.35/100/{\frac{2}{3}}/Iorange/Iorange!10,
  11.15/101/{\frac{20}{27}}/Iblue/Iblue!10,
  12.95/110/{\frac{8}{9}}/Ipurple/Ipurple!10,
  14.75/111/{\frac{26}{27}}/Iteal/Iteal!12
} {
  \draw[interval,draw=\col,fill=\fillcol]
    ($ (\x,\yrect) + (-.5*\iw,0) $) rectangle
    ($ (\x,\yrect) + (.5*\iw,\ih) $);
  \draw[draw=\col,dashed,line width=.7pt,dash pattern=on 3pt off 3pt]
    (\x,\yrect+\ih) -- (\x,\ydashTop);
  \node[ilabel,text=\col,anchor=south] at (\x,\ylabel) {$I_{\name}$};
  \node[qlabel,anchor=north] at (\x,\yq) {$Q(\name)=\q$};
}

\draw[<->,line width=.6pt] (9.35-.66,1.72) -- (9.35+.66,1.72);
\node[label,anchor=north] at (9.35,1.58) {$\displaystyle \frac{\Delta L}{2}$};

\end{tikzpicture}}
    \caption{Ternary Encoding Intervals $I_\sigma$, $\sigma \in \{0,1\}^k$, for $k=3$.}
    \label{fig:interval_encoders}
\end{figure}

See \Cref{fig:interval_encoders} for a representation of the $I_\sigma$ for $k=3$. In effect, $\Delta$ is extremely small, so the encoder intervals only occupy a very small band of $[0,1]$. 
We now make the above intuition formal via the following lemma, proved in \Cref{app:exact-encoding}.

\begin{lemma}\label{lem:exact-encoding}
Let $\sigma\in\{0,1\}^k$, $x\in I_\sigma$, and set
$\lambda=(x,1-x)$. Among the state vertices, $S(\lambda)$ contains exactly $\{p_i:\sigma_i=0\}\cup \{q_i:\sigma_i=1\}$.
\end{lemma}
This result proves that the intervals $I_\sigma$ are able to fully encode the $2^k$ sets. The next lemma proves the complementary result on the whole feasible payment region, even outisde of the $I_\sigma$, that at most one $p_i$ and $q_i$ can be active simultaneously. See the proof in \Cref{app:no-double-state}.

\begin{lemma}\label{lem:no-double-state}
For every $\lambda\in\mathbb{R}_+^2$ satisfying
$\lambda_1+\lambda_2\leq 1+\varepsilon$, and for every item $i\in[k]$,
at most one of $p_i$ and $q_i$ belongs to $S(\lambda)$.
\end{lemma}

\paragraph{Verification Vertices.}
Add $r_0\in U$ and $r_1\in V$, such that
\begin{align*}
&c(r_0)=\frac{L}{4}, \quad f_{r_0v_0}=\frac L2-\frac{L(W-T)}{100W^2},  \quad f_{r_0p_i}=\frac{Lw_i}{100W^2},\\
\mbox{and} \quad &c(r_1)=\frac{L}{4}-\Delta L, \quad f_{r_1 u_0}=\frac L2-\frac{LT}{100W^2},  \quad f_{r_1q_i}=\frac{Lw_i}{100W^2}
\end{align*}
These two vertices implement the subset-sum target test: they can both be active only when the active $q_i$-vertices encode a subset whose total weight is exactly $T$.

\paragraph{Amplification.}
Let $B$ be the total edge weight introduced so far, before the final two
vertices are added. Define
\begin{equation*}
D=\frac{8\cdot3^kB}{\varepsilon}.
\end{equation*}
Add $a\in U$ and $b\in V$, add $f_{ab}=D$, and set $c(a)=c(b)=\left(\frac12+4\Delta\right)D$.
Finally, add one three-edge path of value $12\Delta D$ from $r_1$ to $a$, and
one three-edge path of value $12\Delta D$ from $r_0$ to $b$. 
The amplification gadget turns successful verification into a large reward. The edge $ab$ has very large weight $D$, but the costs of $a$ and $b$ are set so that this edge alone is insufficient to make them stable. The additional support transmitted from $r_1$ to $a$ and from $r_0$ to $b$ is exactly what allows both endpoints of $ab$ to become active. We make this intuition formal in the next lemma, with the proof in \Cref{app:amplification}. 

\begin{lemma}[Amplification]\label{lem:amplification}
For $x\in[\frac12+\Delta L/4,\frac12+\Delta]$, if $r_0,r_1 \in S(x,1-x)$, then $a, b \in S(x,1-x)$, and in particular $f(S(x,1-x)) \geq D$. For $\lambda_1+\lambda_2 \leq 1+\varepsilon$, if at least one of $(r_0,r_1)$ is absent from $S(\lambda)$, then neither $a$ nor $b$ belongs to $S(\lambda)$, and we have $f(S(\lambda))\leq B$. 

\end{lemma}

\subsubsection{Completeness}

We prove that if there is an exact solution to $\textnormal{\textsc{Subset Sum}}$, then the reward can be very large.

\begin{lemma}\label{lem:complete}
If the \textnormal{\textsc{Subset Sum}} instance has a solution, then there exists
$\lambda\in\mathbb{R}_+^2$ with $\lambda_1+\lambda_2=1$ such that $f(S(\lambda))\ge D$.
\end{lemma}

\begin{proof}
Let $\sigma\in\{0,1\}^k$ be a binary encoding of the \textnormal{\textsc{Subset Sum}} solution, so $\sum_{i:\,\sigma_i=1}w_i=T$.
Choose $x\in I_\sigma$ and set $\lambda=(x,1-x)$. Then
$\lambda_1+\lambda_2=1$. By \Cref{lem:exact-encoding}, the active state
vertices in $S(\lambda)$ are exactly the $\sigma$-states.
The anchors have zero cost, and hence belong to $S(\lambda)$. The support of
$r_0$ from its anchor and the active $p_i$ vertices is
\begin{equation*}
\frac L2-\frac{L(W-T)}{100W^2}
        +\sum_{i:\,\sigma_i=0}\frac{Lw_i}{100W^2}
        =\frac L2-\frac{L(W-T)}{100W^2}
        +\frac{L(W-T)}{100W^2}=\frac L2.
\end{equation*} 
Since $c(r_0)=L/4$ and $\lambda_1=x\ge1/2$, the vertex $r_0$ is stable against
$S(\lambda)$, and therefore $r_0\in S(\lambda)$.

Similarly, the support of $r_1$ from its anchor and the active $q_i$ vertices is
\begin{equation*}
\frac L2-\frac{LT}{100W^2}
        +\sum_{i:\,\sigma_i=1}\frac{Lw_i}{100W^2}
        =\frac L2.
\end{equation*}
Since $x\le1/2+\Delta$, we have $\lambda_2=1-x\ge1/2-\Delta$, and therefore
\begin{equation*}
\lambda_2\frac L2
        \ge
        \left(\frac12-\Delta\right)\frac L2
        =
        \frac L4-\frac{\Delta L}{2}
        \ge
        \left(\frac14-\Delta\right)L
        =c(r_1).
\end{equation*}
Thus $r_1\in S(\lambda)$. The interval $I_\sigma$ is contained in
$[1/2+\Delta L/4,1/2+\Delta]$, so \Cref{lem:amplification} gives
$a,b\in S(\lambda)$ and hence $f(S(\lambda))\ge D$.
\end{proof}

\subsubsection{Soundness.}

Conversely, we now prove that if there is no solution $\textnormal{\textsc{Subset Sum}}$ then the maximum reward remains bounded well below the best achievable reward in the feasible case. 

\begin{lemma}[Soundness]\label{lem:sound}
If the \textnormal{\textsc{Subset Sum}} instance has no solution, then every
$\lambda\in\mathbb{R}_+^2$ satisfying $\lambda_1+\lambda_2\leq 1+\varepsilon$ satisfies $f(S(\lambda))\le B$.
\end{lemma}

\begin{proof}
Fix $\lambda\in\mathbb{R}_+^2$ satisfying
$\lambda_1+\lambda_2\leq 1+\varepsilon$.  Let $q=\sum_{i\in[k]:\,q_i\in S(\lambda)}w_i$ and  $p=\sum_{i\in[k]:\,p_i \in S(\lambda)}w_i$. 
The active $q_i$'s correspond to a subset of items. Since the
\textnormal{\textsc{Subset Sum}} instance has no solution, this subset does not
have weight $T$, and hence $q\neq T$. Moreover, By
\Cref{lem:no-double-state}, no item $i \in [k]$ contributes to both
$p$ and $q$, so $p+q\leq W$, and therefore  $p\leq W-q$.

We next show that $r_0$ and $r_1$ cannot both be active. Suppose, for
contradiction, that $r_0,r_1\in S(\lambda)$. Let $\rho=1/(10^5W^4)$ and $t=(q-T)/(100W^2)$.
Since $q,T$ are integers and $q\neq T$, we have $1/(100W^2) \leq \vert t \vert \leq 1/(100W)$. The support of $r_0$ from anchor-verification and state-verification edges is $\frac L2+\frac{L(p-(W-T))}{100W^2}$. The only other possible support at $r_0$ is the source-side edge of the final
path from $r_0$ to $b$, whose weight is at most $L\rho$ by
\Cref{lem:construction-bounds}. Since $p\leq W-q$, the total support of $r_0$
is at most
\begin{equation*}
\frac L2+\frac{L(p-(W-T))}{100W^2}+ 1024 \Delta^2 \cdot 12\Delta D\leq   \frac L2-\frac{L(q-T)}{100W^2}+L\rho
        =
        L\left(\frac12+\rho-t\right).
\end{equation*}
Because $r_0$ is active and $c(r_0)=L/4$, stability implies $\lambda_1 \geq \frac{1/4}{\frac12+\rho-t}$.

Similarly, the support of $r_1$ from its anchor and state-verification edges is
\begin{equation*}
        \frac L2-\frac{LT}{100W^2}
        +
        \sum_{i:\,q_i \in S(\lambda)}\frac{Lw_i}{100W^2}
        =
        \frac L2+\frac{L(q-T)}{100W^2}.
\end{equation*}
Again, the only other possible support is the source-side edge of the final
path from $r_1$ to $a$, whose weight is at most $L\rho$. Hence the total support
of $r_1$ is at most
\begin{equation*}
        \frac L2+\frac{L(q-T)}{100W^2}+L\rho
        =
        L\left(\frac12+\rho+t\right).
\end{equation*}
Because $r_1$ is active and $c(r_1)=(1/4-\Delta)L$, its stability implies $\lambda_2
        \geq
        \frac{1/4-\Delta}{\frac12+\rho+t}$.

The assumptions of \Cref{lem:symmetric-reciprocal} hold for this choice of
$\rho$ and $t$. Therefore
\begin{equation*}
        \lambda_1+\lambda_2
        \geq
        \frac{1/4}{\frac12+\rho-t}
        +
        \frac{1/4-\Delta}{\frac12+\rho+t}
        >
        1+\varepsilon,
\end{equation*}
contradicting the assumption $\lambda_1+\lambda_2\leq 1+\varepsilon$. Thus
$r_0$ and $r_1$ cannot both be active.

Therefore, at least one of $r_0,r_1$ is absent from $S(\lambda)$. By
\Cref{lem:amplification}, this implies that $f(S(\lambda))\leq B$.
\end{proof}

\subsubsection{Rescaling and hardness}

\begin{lemma}\label{prop:quadratic-gap}
After scaling costs by $\beta=1/(1+\varepsilon/2)$, the constructed instance has optimal principal utility at least $2\cdot3^kB$ if \textnormal{\textsc{Subset Sum}} is feasible, otherwise, every equilibrium at every feasible contract has reward at most $B$.
\end{lemma}

\begin{proof}
We first check that the constructed instance is valid and has polynomial
encoding size. The verification-anchor weights are positive because
$1\leq T\leq W-1$ and
\begin{equation*}
        \frac{L(W-T)}{100W^2}\leq \frac{L}{100}<\frac L2,
        \qquad
        \frac{LT}{100W^2}\leq \frac{L}{100}<\frac L2.
\end{equation*}
All other edge weights are visibly nonnegative. The only costs whose positivity
is not immediate are the costs of $r_1$ and of the vertices involving the factor
$1/2-4\Delta$. Since $\Delta=\frac{3^{-3k}}{10^{20}kW^4}\leq 10^{-20}$, 
we have $1/4-\Delta>0$ and $1/2-4\Delta>0$. Hence all costs are nonnegative.
The costs are additive by definition. The reward is monotone and supermodular
because it is a nonnegative sum of functions
$S\mapsto \mathbbm{1}[\{u,v\}\subseteq S]$, which are monotone and supermodular. The graph has polynomially many vertices and edges, and all numbers appearing
in the construction have polynomial bit length in the encoding size of the
\textnormal{\textsc{Subset Sum}} instance.
We normalize rewards to $[0,1]$ after adding the final action below. For the normalization of the rewards to $[0,1]$, simply divide $f$ and the scaled costs by the sum of all the edge weights. Overall, this is a valid instance with polynomial encoding size.  

We now convert the auxiliary payment instance into a feasible-contract instance.
The payment vectors $\lambda$ used in the reduction are not required to be
feasible contracts. Set $\beta=\frac{1}{1+\varepsilon/2}$, and scale every cost by $\beta$, leaving rewards unchanged. If
$\alpha=\beta\lambda$, then each agent's utility in the scaled instance at
contract $\alpha$ equals $\beta$ times the corresponding utility in the
auxiliary instance at payments $\lambda$: $\alpha_i f-\beta c_i
        =
        \beta\lambda_i f-\beta c_i
        =
        \beta(\lambda_i f-c_i)$.
Multiplying all agents' utilities by the positive scalar $\beta$ does not
change best-responses, hence pure Nash equilibria. Therefore the selected equilibrium at contract
$\alpha$ in the scaled instance is the same profile as the selected equilibrium
at payments $\lambda$ in the auxiliary instance.

Suppose first that the \textnormal{\textsc{Subset Sum}} instance is feasible.
By \Cref{lem:complete}, there exists $\lambda\in\mathbb{R}_+^2$ with
$\lambda_1+\lambda_2=1$ and $f(S(\lambda))\geq D$.
Set $\alpha=\beta\lambda$. Then $\alpha$ is feasible, because
$\alpha_1+\alpha_2=\beta<1$. Moreover,
\begin{equation*}
        1-\alpha_1-\alpha_2
        =
        1-\beta
        =
        \frac{\varepsilon/2}{1+\varepsilon/2}
        \geq
        \frac{\varepsilon}{4},
\end{equation*}
where the last inequality uses $\varepsilon<1$. Hence the optimal principal
utility in the scaled instance satisfies
\begin{equation*}
        \mathrm{OPT}_{\text{feasible}}
        \geq
        \frac{\varepsilon}{4}D
        =
        2\cdot 3^kB,
\end{equation*}
by the definition $D=8\cdot3^kB/\varepsilon$.

Suppose now that the \textnormal{\textsc{Subset Sum}} instance is infeasible.
Let $\alpha$ be any feasible contract in the scaled instance, so
$\alpha_1+\alpha_2\leq1$, and define $\lambda=\alpha/\beta$. Then
\begin{equation*}
        \lambda_1+\lambda_2
        =
        \frac{\alpha_1+\alpha_2}{\beta}
        \leq
        \frac{1}{\beta}
        =
        1+\frac{\varepsilon}{2}
        <
        1+\varepsilon.
\end{equation*}
By \Cref{lem:sound}, for all $\lambda_1 +\lambda_2 \leq 1+ \varepsilon$, $f(S(\lambda))$ the selected-equilibrium reward is at most $B$. Since
$\alpha$ is feasible, the principal's residual share is at most $1$. Therefore,
\begin{equation*}
        \mathrm{OPT}_{\text{infeasible}}\leq \max_{\lambda_1+\lambda_2 \leq 1 +\varepsilon} (1-\beta(\lambda_1+\lambda_2))f(S(\lambda)) \leq \max_{\lambda_1+\lambda_2 \leq 1 +\varepsilon}f(S(\lambda)) \leq B.
\end{equation*}

Since the selected equilibrium maximizes reward by \Cref{lem:selected-quadratic}, the same reward bound holds for every equilibrium at every feasible contract.
\end{proof}

We now use \Cref{prop:quadratic-gap} to prove \Cref{thm:supermodular-hardness}. 

\begin{proof}[Proof of \Cref{thm:supermodular-hardness}]
Take the scaled instance from \Cref{prop:quadratic-gap}. Add one action $g$ to agent~1, with cost $2B$, and replace the reward by
\begin{equation*}
        \widehat f(S)
        =
        \sum_{u\in U,\,v\in V}
        f_{uv}\mathbbm{1}[\{u,v,g\}\subseteq S].
\end{equation*}
Thus the reward is $f(S\setminus\{g\})$ when $g$ is taken, and $0$ otherwise. It remains monotone and supermodular, and the construction has polynomial encoding size.

Suppose first that the \textnormal{\textsc{Subset Sum}} instance is infeasible. If an equilibrium at a feasible contract contained $g$, then removing $g$ would leave an equilibrium of the original scaled instance: deviations that retain $g$ have unchanged utility differences. Its reward would therefore be at most $B$. Agent~1's utility would be at most $B-2B<0$, contradicting her option to choose the empty set. Hence every equilibrium has reward $0$, and the optimal principal utility is $0$.

Suppose now that the \textnormal{\textsc{Subset Sum}} instance is feasible. Take a contract $\alpha$ and its greatest equilibrium $S$ with $p f(S)\geq2\cdot3^kB$, where $p=1-\alpha_1-\alpha_2$. Set $\alpha'=(\alpha_1+p/2,\alpha_2)$, and let $S'$ be its greatest equilibrium in the original scaled instance. By Topkis' monotonicity theorem \cite{Topkis1979equilibriumsub}, $S\subseteq S'$. Best-response optimality, monotonicity of $f$, and agent~1's nonnegative utility at $S$ give
\begin{equation*}
        \alpha_1'f(S')-\beta c(S_1')
        \geq
        \alpha_1'f(S_1\cup S_2')-\beta c(S_1)
        \geq
        \frac p2f(S)
        \geq3^kB>2B.
\end{equation*}
Consequently, $S'\cup\{g\}$ is an equilibrium of the modified instance: deviations retaining $g$ have unchanged utility differences, whereas deviations omitting $g$ give agent~1 utility at most $0$. The principal's utility is $(p/2)f(S')>0$.

Dividing the reward and all costs by $\sum_{u\in U,\,v\in V}f_{uv}$ normalizes rewards to $[0,1]$ without changing equilibria or the distinction between positive and zero principal utility. Any finite multiplicative approximation would therefore decide \textnormal{\textsc{Subset Sum}} by testing whether the returned payoff is positive. This payoff can be computed in polynomial time by \Cref{thm:polytime-optimal-equilibrium-supermodular}, contradicting $\mathrm{P}\neq\mathrm{NP}$.
\end{proof}

\subsection{Proof of \Cref{lem:three-edge-path}} \label{app:three-edge-path}

\begin{proof}
We first prove (1) the  forward implication. Since $S(\lambda)$ is stable, every
vertex already in $S(\lambda)$ is stable. Add $y$ and $z$ to $S(\lambda)$. All
old vertices weakly gain support, because all edge weights are nonnegative, and
therefore, remain stable. The vertex $z$ receives its full support $1024\Delta^2M+32\Delta M$.
Since $c(z)=\left(\frac12-4\Delta\right)(1024\Delta^2M+32\Delta M)$ and $\lambda_t\geq 1/2-4\Delta$, $\lambda_t (1024\Delta^2M+32\Delta M) \geq c(z)$, so the vertex $z$ is stable. Similarly, the vertex $y$ receives its full support $32\Delta M+M$.
Since $c(y)=\left(\frac12-4\Delta\right)(32\Delta M+M)$ and $\lambda_s\geq 1/2-4\Delta$, the vertex $y$ is stable. Thus $S(\lambda)\cup{y,z}$ is $\lambda$-stable. Since $S(\lambda)$ is the union of
all $\lambda$-stable profiles, it follows that $y,z\in S(\lambda)$. \medskip

We now prove (2) the no-transmission implication. Suppose, for contradiction, that
$y\in S(\lambda)$. First suppose that $z\notin S(\lambda)$. Then the support of $y$ is at most
$M$. Stability of $y$ implies $\lambda_s M
\geq
\left(\frac12-4\Delta\right)(32\Delta M+M)$. Dividing by $M$ gives
\begin{equation*}
\lambda_s
\geq
\left(\frac12-4\Delta\right)(1+32\Delta)
=
\frac12+12\Delta-128\Delta^2
\geq
\frac12+11\Delta,
\end{equation*}
because $\Delta<1/10^{20}$ is small enough. Since $\lambda_t\geq 1/2-4\Delta$, we obtain
\begin{equation*}
\lambda_s+\lambda_t
\geq
\left(\frac12+11\Delta\right)
+
\left(\frac12-4\Delta\right)
=
1+7\Delta
>
1+\varepsilon,
\end{equation*}
contradicting the assumption.

It remains to consider the case $z\in S(\lambda)$. Since
$s\notin S(\lambda)$, the support of $z$ is at most $32\Delta M$. Stability of
$z$ gives $\lambda_t\cdot32\Delta M
\geq
\left(\frac12-4\Delta\right)(1024\Delta^2M+32\Delta M)$.
Dividing by $32\Delta M$ yields as before
\begin{equation*}
\lambda_t
\geq
\left(\frac12-4\Delta\right)(1+32\Delta)
\geq
\frac12+11\Delta.
\end{equation*}
Also, since $y$ is active, its support is at most $M+32\Delta M$, and hence
\begin{equation*}
\lambda_s
\geq
\frac{c(y)}{M+32\Delta M}
=
\frac12-4\Delta.
\end{equation*}
Thus
\begin{equation*}
\lambda_1+\lambda_2
=
\lambda_s+\lambda_t
\geq
\left(\frac12-4\Delta\right)
+
\left(\frac12+11\Delta\right)
=
1+7\Delta
>
1+\varepsilon,
\end{equation*}
again contradicting the assumption. Therefore $y\notin S(\lambda)$, and the
edge $yt$ of weight $M$ is inactive.
\end{proof}

\subsection{Proof of \Cref{lem:active-state-large-payment}} \label{app:active-state-large-payment}

\begin{proof}
Consider a state vertex, say $q_i \in U$ for $i \in [k]$. Its incoming state-state path values come only from earlier items, and their total value is less than $\sum_{j<i}2\cdot3^{-j}<1$. The state-verification edge has weight at most $Lw_i/(100W^2)\leq L/(100W)$. There are at most $k-i \leq k$ outgoing edges due to the path gadget with each of weight $1024 \Delta^2 (2 \cdot 3^{-i})$, for a total weight of at most $2048 k \Delta^2$.  Thus, every state
vertex has total non-anchor incident weight at most $1+\frac{L}{100W}+2048k\Delta^2<2$.

The stability condition for $q_i$ to be active is $\lambda_1 \geq T_{\theta_i}(s)$, and we have shown that $q_i$ has non-anchor support at most $s \leq 2$, and offset $\theta=1-3^{-i}\in [-1,1]$. By
\Cref{lem:threshold-estimates}, $T_\theta(s)\ge 1/2-4\Delta$. Therefore
$\lambda_1\ge 1/2-4\Delta$. The proof is identical for $p_i \in V$. 
\end{proof}

\subsection{Proof of \Cref{lem:exact-encoding}} \label{app:exact-encoding}

\begin{proof}
The proof is with respect to the full construction, verification vertices thus need to be considered. For $i\in{0,\ldots,k}$, let $\mathcal P_i$ contain the two
anchors, the intended state vertices among the first $i$ items, that is,
$p_j$ if $\sigma_j=0$ and $q_j$ if $\sigma_j=1$ for every $j\leq i$, and the
internal vertices of every three-edge path whose two endpoints are intended
state vertices among the first $i$ items. We prove by induction on $i$ that
$\mathcal P_i$ is $\lambda$-stable and that, for every $j\leq i$, the state
vertex opposite to $\sigma_j$ does not belong to $S(\lambda)$.

The base case $i=0$ holds because $\mathcal P_0$ consists only of the two
zero-cost anchors. Since $x\in I_\sigma$, we have
$x\geq 1/2$ and $x\leq 1/2+\Delta$. Hence $\lambda_1=x\geq \frac12-4\Delta$, and $\lambda_2=1-x\geq \frac12-\Delta\geq \frac12-4\Delta$, so \Cref{lem:three-edge-path} applies in full. Assume the induction statement has been proved through item $i-1$, and let $Q_i=\sum_{\substack{j<i\ \sigma_j=1}}2\cdot3^{-j}$. The total value of earlier indices is $\sum_{j<i}2\cdot3^{-j}=1-3\cdot3^{-i}$, so the earlier indices with $\sigma_j=0$ have total value
$1-3\cdot3^{-i}-Q_i$.

Suppose first that $\sigma_i=0$. We add $p_i$ to the prefix profile, together
with the internal vertices of all three-edge paths from earlier intended
$q_j$'s to $p_i$. These paths provide total support $Q_i$ to $p_i$ by induction and \Cref{lem:three-edge-path}. Additional
support can only help stability, so it is enough to check stability using
support $Q_i$. By \Cref{lem:encoder_interval_bound},
\begin{equation*}
\lambda_2=1-x
\geq
\frac12-\Delta(Q_i+3^{-i})+\frac{\Delta L}{4}.
\end{equation*}
By \Cref{lem:threshold-estimates},
\begin{equation*}
T_{-3^{-i}}(Q_i)
\leq
\frac12-\Delta(Q_i+3^{-i})+20\Delta^2.
\end{equation*}
Using $20\Delta^2\leq \Delta L/20<\Delta L/4$, we get
$\lambda_2\geq T_{-3^{-i}}(Q_i)$. Thus $p_i$ is stable.

Suppose now that $\sigma_i=1$. We add $q_i$ to the prefix profile, together
with the internal vertices of all three-edge paths from earlier intended
$p_j$'s to $q_i$. These paths provide total support
$1-3\cdot3^{-i}-Q_i$ to $q_i$. By \Cref{lem:encoder_interval_bound},
\begin{equation*}
\lambda_1=x
\geq
\frac12+\Delta(Q_i+2\cdot3^{-i})+\frac{\Delta L}{4}.
\end{equation*}
By \Cref{lem:threshold-estimates},
\begin{align*}
T_{1-3^{-i}}(1-3\cdot3^{-i}-Q_i)
\leq
\frac12+\Delta\bigl(1-3^{-i}-(1-3\cdot3^{-i}-Q_i)\bigr)
+20\Delta^2  \
=
\frac12+\Delta(Q_i+2\cdot3^{-i})+20\Delta^2.
\end{align*}
Again $20\Delta^2<\Delta L/4$, so
$\lambda_1\geq T_{1-3^{-i}}(1-3\cdot3^{-i}-Q_i)$. Thus $q_i$ is stable.

In either case, the old vertices in $\mathcal P_{i-1}$ weakly gain support and
therefore remain stable. Each new internal vertex of a three-edge path is also
stable: its two endpoints are present, it receives its full support, its cost is
$(1/2-4\Delta)$ times that full support, and both payment coordinates are at
least $1/2-4\Delta$. Hence $\mathcal P_i$ is $\lambda$-stable. \medskip

It remains to exclude the opposite state. Suppose first that $\sigma_i=0$, and
suppose for contradiction that $q_i\in S(\lambda)$. Since
$\lambda_1+\lambda_2=1\leq 1+\varepsilon$, and since $q_i$ is an active state
vertex, \Cref{lem:active-state-large-payment} implies that the payment
coordinate on the side of $q_i$ is at least $1/2-4\Delta$. Thus
\Cref{lem:three-edge-path} applies to every path from an inactive earlier
$p_j$ to the active target $q_i$. By the induction hypothesis, the inactive
earlier $p_j$'s are exactly those with $\sigma_j=1$. Therefore they transmit no
path value to $q_i$.

The only earlier state paths that can transmit value to $q_i$ are those coming
from intended earlier $p_j$'s, whose total value is at most
$1-3\cdot3^{-i}-Q_i$. All remaining non-anchor support at $q_i$ is at most $\delta(s)\coloneqq \frac{L}{100W}+2048k\Delta^2$,
where the first term comes from the state-verification edge, and the
second term from the small backflow of the three edge paths when $q_i$ is a source. Since
$T_\theta(s)$ is decreasing in $s$, stability of $q_i$ gives
\begin{align*}
x=\lambda_1
\geq
T_{1-3^{-i}}(1-3\cdot3^{-i}-Q_i+\delta(s)) \
\geq
\frac12+\Delta(Q_i+2\cdot3^{-i})-\Delta \delta(s)-20\Delta^2,
\end{align*}
where the last inequality uses \Cref{lem:threshold-estimates}. On the other
hand, \Cref{lem:encoder_interval_bound} gives
\begin{equation*}
x
\leq
\frac12+\Delta(Q_i+3^{-i})-\frac{\Delta L}{4}.
\end{equation*}
The lower bound exceeds the upper bound by at least
\begin{equation*}
\Delta3^{-i}+\frac{\Delta L}{4}-\Delta \delta(s)-20\Delta^2
\geq
\Delta L+\frac{\Delta L}{4}-\frac{\Delta L}{20}
>0,
\end{equation*}
where the inequality uses \Cref{lem:error_upperbound_estimate}. This is a
contradiction. Therefore $q_i\notin S(\lambda)$.

The case $\sigma_i=1$ is symmetric. Suppose for contradiction that
$p_i\in S(\lambda)$. By \Cref{lem:active-state-large-payment} and
\Cref{lem:three-edge-path}, inactive earlier $q_j$'s transmit no path value to
the active target $p_i$. Hence the only earlier state paths that can transmit
value to $p_i$ are those coming from intended earlier $q_j$'s, whose total value
is at most $Q_i$. All other non-anchor support is again at most $e$. Therefore
stability of $p_i$ gives
\begin{align*}
1-x=\lambda_2
\geq
T_{-3^{-i}}(Q_i+e) \
\geq
\frac12-\Delta(Q_i+3^{-i})-\Delta \delta(s)-20\Delta^2.
\end{align*}
On the other hand, \Cref{lem:encoder_interval_bound} gives
\begin{equation*}
1-x
\leq
\frac12-\Delta(Q_i+2\cdot3^{-i})-\frac{\Delta L}{4}.
\end{equation*}
Again the lower bound exceeds the upper bound by at least
\begin{equation*}
\Delta3^{-i}+\frac{\Delta L}{4}-\Delta \delta(s)-20\Delta^2
\geq
\Delta L+\frac{\Delta L}{4}-\frac{\Delta L}{20}
>0,
\end{equation*}
a contradiction. Thus $p_i\notin S(\lambda)$.
This completes the induction. 
\end{proof}

\subsection{Proof of \Cref{lem:no-double-state}} \label{app:no-double-state}

\begin{proof}
Suppose, for contradiction, that both $p_i$ and $q_i$ are active for some item.
Let $i$ be the smallest such item. Then, for every $j<i$, at most one of
$p_j,q_j$ is active. By \Cref{lem:active-state-large-payment}, the payment coordinate on the side of
each active state vertex is at least $1/2-4\Delta$. Therefore
\Cref{lem:three-edge-path} applies to every state path whose source is inactive
and whose target is either $p_i$ or $q_i$. It follows that inactive earlier
states transmit no path value to $p_i$ or $q_i$.

Let $P_i$ be the total path value transmitted from active earlier $p_j$'s to
$q_i$, and let $Q_i$ be the total path value transmitted from active earlier
$q_j$'s to $p_i$. Since at most one earlier state is active for each item
$j<i$, we have
\begin{equation*}
P_i+Q_i
\leq
\sum_{j<i}2\cdot3^{-j}
=
1-3\cdot3^{-i}.
\end{equation*}
The remaining non-anchor support at either $p_i$ or $q_i$ is at most $\delta(s):=\frac{L}{100W}+2048k\Delta^2$, coming from the state-verification edges and from source-side edges
of the three-edge paths. By \Cref{lem:error_upperbound_estimate}, $2\Delta \delta(s)+40\Delta^2\leq \frac{\Delta L}{10}$.

Let $s_p$ and $s_q$ be the actual non-anchor supports of $p_i$ and $q_i$,
respectively. Then
\begin{equation*}
s_p+s_q
\leq
P_i+Q_i+2\delta(s)
\leq
1-3\cdot3^{-i}+2\delta(s). 
\end{equation*}
Since both $p_i$ and $q_i$ are active, their stability conditions imply $\lambda_2\geq T_{-3^{-i}}(s_p)$ and  $\lambda_1\geq T_{1-3^{-i}}(s_q)$.
By the two-threshold part of \Cref{lem:threshold-estimates},
\begin{align*}
\lambda_1+\lambda_2
\geq
1+\Delta(1-2\cdot3^{-i}-s_q-s_p)-40\Delta^2 
&\geq
1+\Delta\bigl(1-2\cdot3^{-i}-(1-3\cdot3^{-i})-2\delta(s)\bigr)-40\Delta^2 \\
&=
1+\Delta3^{-i}-2\Delta \delta(s)-40\Delta^2 
\geq
1+\Delta L-\frac{\Delta L}{10}.
\end{align*}
Since $\varepsilon=\Delta L/8$, the last quantity is larger than
$1+\varepsilon$, contradicting $\lambda_1+\lambda_2\leq1+\varepsilon$.
Therefore at most one of $p_i,q_i$ is active for every item $i$.
\end{proof}

\subsection{Proof of \Cref{lem:amplification}} \label{app:amplification}

\begin{proof}
For the forward implication, suppose $\lambda=(x,1-x)$ with
$x\in[\frac12+\Delta L/4,\frac12+\Delta]$ and $r_0,r_1\in S(\lambda)$. We add
$a,b$ and the internal vertices of the two final three-edge paths to
$S(\lambda)$. Existing vertices only gain support. The support of $a$ is at
least $(1+12\Delta)D$, from $ab$ and from the value $12\Delta D$ sent by the
path from $r_1$ to $a$. Since $x\ge1/2$,
\begin{equation*}
\lambda_1(1+12\Delta)D
        \ge
        \frac12(1+12\Delta)D
        =
        \left(\frac12+6\Delta\right)D
        \ge c(a).
\end{equation*}
The support of $b$ is also at least $(1+12\Delta)D$, and
$1-x\ge1/2-\Delta$, so
\begin{equation*}
\lambda_2(1+12\Delta)D
        \ge
        \left(\frac12-\Delta\right)(1+12\Delta)D
        =
        \left(\frac12+5\Delta-12\Delta^2\right)D
        \ge c(b),
\end{equation*}
because $12\Delta^2\le\Delta$. The two internal vertices of each final path are
stable because their endpoints are present and their costs are $(1/2-4\Delta)$
times their full supports, while both payment coordinates are at least
$1/2-4\Delta$. Thus we have enlarged $S(\lambda)$ to a stable profile
containing $a,b$. By maximality of $S(\lambda)$, $a$ and $b$ must be in $S(\lambda)$, and this activates the edge with reward $D$. \medskip

Now suppose that \(\lambda_1+\lambda_2\le 1+\varepsilon\). Because 
\begin{equation*}
\left(\frac12-3\Delta\right)(1+12\Delta)
        =
        \frac12+3\Delta-36\Delta^2
        \le
        \frac12+4\Delta,
\end{equation*}
this implies the estimate $\frac{1/2+4\Delta}{1+12\Delta}\ge \frac12-3\Delta$.

First, if at least one of \(a,b\) is inactive, then both are inactive. Indeed,
assume by symmetry that \(b\notin S(\lambda)\). If \(a\in S(\lambda)\), then
\(a\) receives no support from \(b\), and its remaining support is at most
\(12\Delta D\), coming from the final path. Stability of \(a\) would imply $\lambda_1\cdot 12\Delta D
        \ge
        \left(\frac12+4\Delta\right)D$, and hence  $\lambda_1
        \ge
        \frac{1/2+4\Delta}{12\Delta}
        >
        1+\varepsilon$ contradicting $\lambda_1+\lambda_2 \leq 1+\varepsilon$. The symmetric argument applies if \(a\notin S(\lambda)\).

Next suppose that \(r_1\notin S(\lambda)\). We claim that \(a\notin
S(\lambda)\). By contradiction, suppose that \(a\in S(\lambda)\), and by the preceding paragraph
we must also have \(b\in S(\lambda)\). Since \(a\) has support at most
\((1+12\Delta)D\), stability of \(a\) gives
\begin{equation*}
\lambda_1
        \ge
        \frac{1/2+4\Delta}{1+12\Delta}
        \ge
        \frac12-3\Delta.
\end{equation*}
Therefore \Cref{lem:three-edge-path} applies to the final path from \(r_1\) to
\(a\). Since \(r_1\notin S(\lambda)\), this path does not transmit its value
\(12\Delta D\) to \(a\). Hence \(a\)'s support is at most \(D\), coming only
from the edge \(ab\), and stability of \(a\) implies $\lambda_1\ge \frac12+4\Delta$. Also, since \(b\in S(\lambda)\) and \(b\) has support at most
\((1+12\Delta)D\), $\lambda_2
        \ge
        \frac{1/2+4\Delta}{1+12\Delta}
        \ge
        \frac12-3\Delta$.
Thus
\begin{equation*}
\lambda_1+\lambda_2
        \ge
        \left(\frac12+4\Delta\right)
        +
        \left(\frac12-3\Delta\right)
        =
        1+\Delta
        >
        1+\varepsilon,
\end{equation*}
a contradiction. Therefore \(a\notin S(\lambda)\). Symmetrically, if
\(r_0\notin S(\lambda)\), then \(b\notin S(\lambda)\). Consequently, if at least one of \(r_0,r_1\) is absent from \(S(\lambda)\),
then at least one of \(a,b\) is absent, and hence both \(a\) and \(b\) are
absent.

It remains to rule out the internal vertices of the two final paths from $r_0$ to $b$ and $r_1$ to $a$. Consider
one final path \(s-z-y-t\), of value \(M=12\Delta D\), where
\(t\in\{a,b\}\). Since \(t\notin S(\lambda)\), if the internal vertex \(y\)
adjacent to \(t\) were active, then its support would be at most
\(32\Delta M\), while
\[
c(y)=\left(\frac12-4\Delta\right)(32\Delta M+M).
\]
Thus its payment coordinate would be larger than \(1+\varepsilon\), a
contradiction. Hence \(y\notin S(\lambda)\). If the other internal vertex
\(z\) were active, then, since \(y\notin S(\lambda)\), its support would be at
most \(1024\Delta^2M\), while
\[
c(z)=\left(\frac12-4\Delta\right)(1024\Delta^2M+32\Delta M).
\]
This again forces its payment coordinate to be larger than \(1+\varepsilon\),
a contradiction. Therefore \(z\notin S(\lambda)\).

Thus no vertex of the final amplification gadget is active, except possibly
\(r_0,r_1\). In particular, no edge introduced after the definition of \(B\) is
active. All active edges lie in the pre-amplification part, whose total edge
weight is \(B\), so $f(S(\lambda))\le B$.
\end{proof}

\subsection{Proof of \Cref{prop:many-critical-contracts-supermodular}} \label{app:many-critical-contracts-supermodular}

\begin{proof}
We first work in the auxiliary payment instance, before the final rescaling of
costs by $\beta$. For every $\sigma\in\{0,1\}^k$, let $x_\sigma = \frac12+\Delta Q(\sigma)+\frac{\Delta L}{4}$, $\lambda^\sigma=(x_\sigma,1-x_\sigma)$, and let $S_\sigma\coloneqq S(\lambda^\sigma)$ be the reward-maximizing equilibrium at payments $\lambda^\sigma$. By the
exact-encoding lemma, the active state vertices in $S_\sigma$ are exactly $\{p_i:\sigma_i=0\}\cup\{q_i:\sigma_i=1\}$.
In particular, the profiles $S_\sigma$ are distinct for distinct $\sigma$.

First, we give an upper bound. 
Since $\lambda^\sigma$ incentivizes $S_\sigma$, the coordinatewise-minimal
payment vector $\lambda_{S_\sigma}$ satisfies
\[
        (\lambda_{S_\sigma})_1
        \le
        \lambda^\sigma_1
        =
        x_\sigma
        =
        \frac12+\Delta Q(\sigma)+\frac{\Delta L}{4}.
\]

We now prove a matching lower bound for every nonzero $\sigma$.
Fix $\sigma\neq 0^k$, and let $r=\max\{i:\sigma_i=1\}$ be the last active $q$ vertex, and in particular $q_r\in S_\sigma$. Define $Q_r=\sum_{\substack{j<r\\ \sigma_j=1}}2\cdot 3^{-j}$.
Since $r$ is the last index with $\sigma_r=1$, we have
\[
        Q(\sigma)=Q_r+2\cdot 3^{-r}.
\]

Consider the non-anchor support of $q_r$ in the profile $S_\sigma$. 
By \Cref{lem:three-edge-path}, only the states vertices $p_j$ with $j <r$ yield some support, which is 
\[
        \sum_{\substack{j<r\\ \sigma_j=0}}2\cdot 3^{-j}
        =
        \sum_{j<r}2\cdot 3^{-j}
        -
        \sum_{\substack{j<r\\ \sigma_j=1}}2\cdot 3^{-j}
        =
        1-3\cdot 3^{-r}-Q_r.
\]
As computed multiple times, the remaining non-anchor direct support is at most $\delta(s)=\frac{L}{100W}+2048k\Delta^2$. So the non-anchor support $s_r$ of $q_r$ in $S_\sigma$ satisfies $s_r
        \le
        1-3\cdot 3^{-r}-Q_r+\delta(s)$. 
Now let $\lambda$ be any payment vector that incentivizes $S_\sigma$. Since
$q_r$ is active in $S_\sigma$, it must be stable under $\lambda$. The vertex
$q_r$ is a comparator on side $U$ with offset $1-3^{-r}$, and therefore its
stability implies $ \lambda_1\ge T_{1-3^{-r}}(s_r)$.
Because $T_\theta(s)$ is decreasing in $s$,
\[
        \lambda_1
        \ge
        T_{1-3^{-r}}\!\left(1-3\cdot 3^{-r}-Q_r+\delta(s)\right).
\]
Using the threshold estimate \Cref{lem:threshold-estimates}, $T_\theta(s)
        \ge
        \frac12+\Delta(\theta-s)-20\Delta^2$,
we obtain
\begin{align*}
        \lambda_1
        &\ge
        \frac12
        +
        \Delta\left(
            1-3^{-r}
            -
            \left(1-3\cdot 3^{-r}-Q_r+\delta(s)\right)
        \right)
        -
        20\Delta^2 
        \\
        &=
        \frac12
        +
        \Delta\left(Q_r+2\cdot 3^{-r}\right)
        -
        \Delta \delta(s)
        -
        20\Delta^2 
        \\
        &=
        \frac12+\Delta Q(\sigma)-\Delta \delta(s)-20\Delta^2.
\end{align*}
By the error bound used in the exact-encoding analysis, $ \Delta \delta(s)+20\Delta^2\le \frac{\Delta L}{20}$, every payment vector that incentivizes $S_\sigma$ thus satisfies $\lambda_1
        \ge
        \frac12+\Delta Q(\sigma)-\frac{\Delta L}{20}$.
In particular, the critical payment vector satisfies
\[
        (\lambda_{S_\sigma})_1
        \ge
        \frac12+\Delta Q(\sigma)-\frac{\Delta L}{20}.
\]
Combining the upper and lower bounds, for every nonzero $\sigma$,
\[
        (\lambda_{S_\sigma})_1
        \in
        \left[
            \frac12+\Delta Q(\sigma)-\frac{\Delta L}{20},
            \frac12+\Delta Q(\sigma)+\frac{\Delta L}{4}
        \right].
\]
For $\sigma=0^k$, the upper bound gives $(\lambda_{S_{0^k}})_1
        \le
        \frac12+\frac{\Delta L}{4}$.

It remains to show that these ranges are disjoint for distinct $\sigma$. For two distinct vectors
$\sigma,\tau\in\{0,1\}^k$, the ternary masses $Q(\sigma)$ and $Q(\tau)$ differ
by at least $2L$. Indeed, let $r$ be the first coordinate at which $\sigma$ and
$\tau$ differ, and assume without loss of generality that $\sigma_r=0$ and
$\tau_r=1$. The contribution at coordinate $r$ differs by $2\cdot 3^{-r}$,
while the total possible compensation from later coordinates is at most $\sum_{j>r}2\cdot 3^{-j}
        =
        3^{-r}-L$.
Thus
\[
        Q(\tau)-Q(\sigma)
        \ge
        2\cdot 3^{-r}-(3^{-r}-L)
        =
        3^{-r}+L
        \ge
        2L.
\]

Now suppose $\sigma,\tau\neq 0^k$ and $Q(\sigma)<Q(\tau)$. Then
$Q(\tau)\ge Q(\sigma)+2L$, and hence
\[
        \frac12+\Delta Q(\tau)-\frac{\Delta L}{20}
        >
        \frac12+\Delta Q(\sigma)+\frac{\Delta L}{4}.
\]
Therefore the possible ranges for $(\lambda_{S_\sigma})_1$ and
$(\lambda_{S_\tau})_1$ are disjoint, so
$\lambda_{S_\sigma}\neq\lambda_{S_\tau}$.

Finally, if $\tau\neq 0^k$, then $Q(\tau)\ge 2L$, so
\[
        (\lambda_{S_\tau})_1
        \ge
        \frac12+2\Delta L-\frac{\Delta L}{20}
        >
        \frac12+\frac{\Delta L}{4}
        \ge
        (\lambda_{S_{0^k}})_1.
\]
Thus the all-zero encoding also gives a distinct critical payment vector. We have therefore obtained at least one distinct critical payment vector for
each $\sigma\in\{0,1\}^k$, hence at least $2^k$ critical payment vectors in the
auxiliary instance. By scaling, this induces $2^k$ distinct critical contracts for the actual contract instance. 
\end{proof}

\begin{remark}
The constructed hard instance contains $\Theta(k^2)$ total actions because of
the auxiliary vertices used in the three-edge path gadgets. Thus the proposition
gives $2^k$ distinct critical contracts, equivalently
$2^{\Omega(\sqrt m)}$ as a function of the total number $m$ of actions in this
particular construction.
\end{remark}

\subsection{Additional Technical Lemmas}

\begin{lemma}[Threshold estimates]\label{lem:threshold-estimates}For every $\theta\in[-1,1]$ and $s\in[0,2]$,
$T_\theta(s)$ is decreasing in $s$ and
\begin{equation*}
\left|
        T_\theta(s)-\left(\frac12+\Delta(\theta-s)\right)
\right|
\le 20\Delta^2.
\end{equation*}
In particular, $T_\theta(s)\ge \frac12-4\Delta$. Moreover, for any $\theta_1,\theta_2\in[-1,1]$ and $s_1,s_2\in[0,2]$,
\begin{equation*}
T_{\theta_1}(s_1)+T_{\theta_2}(s_2)
        \ge
        1+
        \Delta(\theta_1+\theta_2-s_1-s_2)
        -40\Delta^2.
\end{equation*}
\end{lemma}

\begin{proof}
The numerator of $T_\theta(s)$ is positive and fixed, while the denominator
$1+2\Delta s$ is increasing in $s$. Hence $T_\theta(s)$ is decreasing in $s$.

A direct calculation gives
\begin{align*}
T_\theta(s)-\left(\frac12+\Delta(\theta-s)\right) =
\frac{\frac12+\Delta\theta}{1+2\Delta s}
-
\left(\frac12+\Delta\theta-\Delta s\right) =
\frac{\frac12+\Delta\theta
-
(1+2\Delta s)(\frac12+\Delta\theta-\Delta s)}{1+2\Delta s} =
\frac{2\Delta^2s(s-\theta)}{1+2\Delta s}.
\end{align*}
Since $s\in[0,2]$, $\theta\in[-1,1]$, and $1+2\Delta s\ge1$, we have
\begin{equation*}
\left|
        \frac{2\Delta^2s(s-\theta)}{1+2\Delta s}
        \right|
        \le 2\Delta^2\cdot 2\cdot 3
        =12\Delta^2
        \le 20\Delta^2.
\end{equation*}
This proves the approximation. It also gives
\begin{equation*}
T_\theta(s)
        \ge
        \frac12+\Delta(-1-2)-20\Delta^2
        =
        \frac12-3\Delta-20\Delta^2
        \ge
        \frac12-4\Delta,
\end{equation*}
because $20\Delta^2\le\Delta$.
For the two-threshold inequality, apply the lower bound twice proves the claim.
\end{proof}

\begin{lemma}[Symmetric reciprocal estimate]
\label{lem:symmetric-reciprocal}

Suppose $0\le \rho\le \frac{1}{10^5W^4}$ and  $\frac{1}{100W^2}\le |t|\le \frac{1}{100W}$.
Then $\frac12+\rho-t>0$, $\frac12+\rho+t>0$, and
\begin{equation*}
\frac{1/4}{\frac12+\rho-t}
        +
        \frac{1/4-\Delta}{\frac12+\rho+t}
        >1+\varepsilon.
\end{equation*}
\end{lemma}

\begin{proof}
Since $|t|\le1/(100W)\le1/100$ and $\rho\ge0$, both denominators are at least
$1/2-1/100>0$. Developing the following inequality,
\begin{equation*}
\frac14\left(
        \frac{1}{\frac12+\rho-t}
        +
        \frac{1}{\frac12+\rho+t}
        \right) -1
        =
        \frac{\frac12+\rho}{2((\frac12+\rho)^2-t^2)} -1=\frac{\frac12+\rho-2((\frac12+\rho)^2-t^2)}
             {2((\frac12+\rho)^2-t^2)}=\frac{t^2-\rho(\frac12+\rho)}{(\frac12+\rho)^2-t^2},
\end{equation*}
this yields 
\begin{align*}
\frac{1/4}{\frac12+\rho-t}
+
\frac{1/4-\Delta}{\frac12+\rho+t}
-1=
\frac{t^2-\rho(\frac12+\rho)}{(\frac12+\rho)^2-t^2}
-
\frac{\Delta}{\frac12+\rho+t}.
\end{align*}

We now lower bound the two terms in the identity. Since
$\rho\le 1/(10^5W^4)\le10^{-5}$, we have $\frac12+\rho\le0.51$. Therefore $\rho\left(\frac12+\rho\right)
        \le
        \frac{0.51}{10^5W^4}$. Also $t^2\ge 1/(10^4W^4)$. Hence
\begin{equation*}
t^2-\rho\left(\frac12+\rho\right)
        \ge
        \frac{1}{10^4W^4}-\frac{0.51}{10^5W^4}
        \ge
        \frac{9}{10^5W^4}.
\end{equation*}
Moreover, $(\tfrac12+\rho)^2-t^2
        \le
        (0.51)^2
        \leq
        1$.
Thus
\begin{equation*}
\frac{t^2-\rho(\frac12+\rho)}{(\frac12+\rho)^2-t^2}
        \ge
        \frac{9}{10^5W^4}.
\end{equation*}
Finally, since $\frac12+\rho+t\ge1/2-1/100>1/4$,
\begin{equation*}
\frac{\Delta}{\frac12+\rho+t}
        \le 4\Delta
        \le
        \frac{4}{10^{20}W^4}.
\end{equation*}
Consequently,
\begin{equation*}
\frac{1/4}{\frac12+\rho-t}
        +
        \frac{1/4-\Delta}{\frac12+\rho+t}
        -1
        \ge
        \frac{9}{10^5W^4}-\frac{4}{10^{20}W^4}
        >
        \frac{8}{10^5W^4}.
\end{equation*}
Since $\varepsilon\le\Delta\le1/(10^{20}W^4)$, this is larger than
$\varepsilon$, proving the claimed inequality.
\end{proof}

\begin{lemma}\label{lem:construction-bounds}
The pre-amplification weight satisfies $B\le 2k/\Delta$, and the edge incident
to $r_0$ or $r_1$ in either final three-edge path has weight at most
$L/(10^5W^4)$.
\end{lemma}

\begin{proof}
The $2k$ state-anchor edges have total weight $k/\Delta$. The two verification
anchors and all state-verification edges have total weight less than $2$. A
state path of value $M$ has total edge weight at most $2M$, because
$1+32\Delta+1024\Delta^2\le2$. The sum of the values of all state paths is at
most $2k$, so these paths have total weight at most $4k$. Consequently,
\begin{equation*}
        B\le \frac{k}{\Delta}+4k+2\le\frac{2k}{\Delta}.
\end{equation*}
Since $\varepsilon=\Delta 3^{-k}/8$,
\begin{equation*}
        D=\frac{8\cdot3^kB}{\varepsilon}
        =\frac{64\cdot3^{2k}B}{\Delta}
        \le \frac{128k3^{2k}}{\Delta^2}.
\end{equation*}
Hence the source-side edge of either final path has weight at most
\begin{equation*}
        1024\Delta^2\cdot12\Delta D\le
        1024\cdot12\Delta^3\cdot
        \frac{128k3^{2k}}{\Delta^2}<1.6\cdot10^6 k3^{2k}\Delta=
        \frac{1.6\cdot10^6 L}{10^{20}W^4}
        \le
        \frac{L}{10^5W^4}. \qedhere
\end{equation*}
\end{proof}

\begin{lemma}\label{lem:encoder_interval_bound}
Let $\sigma \in \{0,1\}^k$, $x\in I_\sigma$, and $Q_i=\sum_{\substack{j<i\\ \sigma_j=1}}2\cdot3^{-j}$.
Then, for every $i\in[k]$,
\begin{equation*}
        \frac12+\Delta\bigl(Q_i+2\sigma_i3^{-i}\bigr)+\frac{\Delta L}{4}
        \le x\le
        \frac12+\Delta\bigl(Q_i+(2\sigma_i+1)3^{-i}\bigr)
        -\frac{\Delta L}{4}.
\end{equation*}
\end{lemma}

\begin{proof}
Indeed,
\begin{equation*}
        Q(\sigma)=Q_i+2\sigma_i3^{-i}
        +\sum_{\substack{j>i\\ \sigma_j=1}}2\cdot3^{-j},
\end{equation*}
and the final sum lies between $0$ and
$\sum_{j>i}2\cdot3^{-j}=3^{-i}-L$.
\end{proof}

\begin{lemma}\label{lem:error_upperbound_estimate}
We have $ \Delta\left(\frac{L}{100W}+2048k\Delta^2\right)+20\Delta^2
        \le \frac{\Delta L}{20}$.
\end{lemma}

\begin{proof}
As $W\ge1$ and $\Delta\le L/(10^{20}k)$, this yields
\begin{equation*}
        \Delta\left(\frac{L}{100W}+2048k\Delta^2\right)+20\Delta^2
        \le \frac{\Delta L}{100}+2048k\Delta^3+20\Delta^2
        \le \frac{\Delta L}{100}+\frac{\Delta L}{100}+
        \frac{\Delta L}{100}
        \le \frac{\Delta L}{20}.  \qedhere
\end{equation*}
\end{proof}

\section{Proof of \Cref{prop:supermodular-additive-fptas}} \label{app:supermodular-additive-fptas} 

\begin{proof}
Enumerate all feasible contracts whose coordinates are nonnegative integer multiples of $\varepsilon/n$, compute their greatest equilibria using \Cref{thm:polytime-optimal-equilibrium-supermodular}, and return the contract and equilibrium with the largest principal utility. There are at most $(\lfloor n/\varepsilon\rfloor+1)^n=O((n/\varepsilon)^n)$ contracts to evaluate, so the running time is polynomial in the input size and $1/\varepsilon$ for fixed $n$.

Let $\contract$ be an optimal contract. If $1-\sum_{i\in[n]}\alpha_i\leq\varepsilon$, then $\mathrm{OPT}\leq\varepsilon$ because $f\leq1$. The zero contract is among the enumerated contracts and gives nonnegative principal utility, so the claimed guarantee follows. Otherwise, round each coordinate upwards by setting $\alpha_i'=(\varepsilon/n)\lceil n\alpha_i/\varepsilon\rceil$. The rounded contract is feasible, since
\begin{equation*}
        \sum_{i\in[n]}\alpha_i'
        \leq
        \sum_{i\in[n]}\alpha_i+\varepsilon
        <1.
\end{equation*}
By Topkis' monotonicity theorem \cite{Topkis1979equilibriumsub}, $S^+(\contract)\subseteq S^+(\contract')$. Monotonicity of $f$ and the bound $f\leq1$ therefore give
\begin{equation*}
        \big(1-\sum_{i\in[n]}\alpha_i'\big)f(S^+(\contract'))
        \geq
        \big(1-\sum_{i\in[n]}\alpha_i-\varepsilon\big)f(S^+(\contract))\geq
        \mathrm{OPT}-\varepsilon.
\end{equation*}
Since $\contract'$ is among the enumerated contracts, the algorithm returns a contract with at least this utility.
\end{proof}

\section{\Cref{sec:supermodular-search}: Missing Proofs}

\subsection{Proof of \Cref{lem:greatest-monotone}} \label{app:greatest-monotone}

\begin{proof}
Fix an agent $i$ and an action profile $X_{-i}$ of the other agents. At share $t$, agent $i$ maximizes $t f(X_i\sqcup X_{-i})-c(X_i)$
over $X_i\subseteq T_i$. This objective is supermodular in $X_i$, so its maximizers form a lattice and in particular admit a greatest element.

Monotonicity in $X_{-i}$ is the standard Topkis best-response monotonicity for supermodular games, used in the proof of \Cref{thm:polytime-optimal-equilibrium-supermodular}. To establish monotonicity in $t$, fix $X_i\subseteq Y_i$ and $t\leq t'$. The change in the advantage of $Y_i$ over $X_i$ when the share increases from $t$ to $t'$ is
\begin{align*}
\Bigl(t'f(Y_i\sqcup X_{-i})-c(Y_i)-\bigl(t'f(X_i\sqcup X_{-i})-c(X_i)\bigr)\Bigr)-\Bigl(t f(Y_i\sqcup X_{-i})-c(Y_i)-\bigl(t f(X_i\sqcup X_{-i})-c(X_i)\bigr)\Bigr)\\
=(t'-t)\bigl(f(Y_i\sqcup X_{-i})-f(X_i\sqcup X_{-i})\bigr)\geq 0,
\end{align*}
where the inequality follows from monotonicity of $f$. Thus the objective has increasing differences in $(X_i,t)$, and Topkis' monotonicity theorem implies that the greatest best response is nondecreasing in $t$.

It remains to prove the monotonicity of the greatest equilibrium. Run the greatest-best-response iteration from the full action profile $T$, as in the proof of \Cref{thm:polytime-optimal-equilibrium-supermodular}, once under $\contract$ and once under $\beta$. Let $S^r(\contract)$ and $S^r(\beta)$ denote the profiles after round $r$. Both iterations start from $T$, so $S^0(\contract)=S^0(\beta)=T$.
Suppose inductively that $S^r(\contract)\subseteq S^r(\beta)$. Then, for every agent $i$,
\begin{align*}
S_i^{r+1}(\contract)=\operatorname{BR}_i^+\bigl(\alpha_i,S^r_{-i}(\contract)\bigr)
\subseteq\operatorname{BR}_i^+\bigl(\alpha_i,S^r_{-i}(\beta)\bigr)
\subseteq\operatorname{BR}_i^+\bigl(\beta_i,S^r_{-i}(\beta)\bigr)
=S_i^{r+1}(\beta).
\end{align*}
Therefore $S^{r+1}(\contract)\subseteq S^{r+1}(\beta)$. The inclusion holds at every round and hence also at the limits, thus $S^+(\contract)\subseteq S^+(\beta)$.
\end{proof}

\subsection{Proof of \Cref{lem:cover-hidden}} \label{app:cover-hidden}

The full cover construction of \Cref{lem:cover-hidden} is given in \Cref{alg:cover-hidden}.
We start by proving a technical lemma.
The following auxiliary lemma contains the main combinatorial construction and the derandomization used in the proof of \Cref{lem:cover-hidden}. It produces small families of candidate sets and ensures that they are sparse enough for the covering procedure to enumerate the remaining deletion choices in polynomial time. 

\begin{algorithm}[t]
\caption{Covering cheaper sets}\label{alg:cover-hidden}
\KwIn{Nonempty sets $E_1,\ldots,E_r\subseteq[d]$ and nonnegative
rational vertex weights $w$.}
\KwOut{A polynomial-size family $\mathcal C$ of sets, none containing any $E_j$.}
$\mathcal C\gets\emptyset$\;
\ForEach{$E\in\{E_1,\ldots,E_r\}$ and positive $g=jw_a$, $a,j\in[d]$}{
  Set $N_i\gets E\setminus E_i$ for every $i\in[r]$\;
  Apply the auxiliary lemma with $g$ and $r$ to the sets $N_i$ with $w(N_i)\leq 2g$\;
  \ForEach{union $Y$ of a subfamily of a returned family}{
    Set $Y_{\mathrm{out}}\gets Y\setminus E$\;
    \If{$Y_{\mathrm{out}}$ meets every $E_i$ with $w(N_i)>2g$}{
        \ForEach{$i$ with $w(N_i)\le 2g$, $D\subseteq Y\cap N_i$, and $v\in Y\cap E\cap E_i$}{
            \If{$Y_{\mathrm{out}}\cup D\cup\{v\}$ meets every known edge}{
                Add $[d]\setminus(Y_{\mathrm{out}}\cup D\cup\{v\})$ to $\mathcal C$\;
            }
        }
    }
}
}
\Return{$\mathcal C$}\;
\end{algorithm}

\begin{lemma}\label{lem:hitting_result}
Consider $d\geq 1$ vertices each with nonnegative rational weight $w_v$ for $v \in [d]$, $g >0$ a rational value threshold, and for $r \geq 1$, $N_1,\dots,N_r \subset [d]$ sets of weight at most $w(N_i)\leq 2g$. There is a deterministic algorithm that runs in polynomial time, and outputs a list of polynomially many families $\mathcal P$, with each family containing $O(\log(2r))$ subsets of $[d]$. The list satisfies both of the following properties.
\begin{enumerate}
\item \textbf{Separation.} For any $r$ inclusions $A_i\subseteq C_i\subseteq V$ with $w(A_i)<g$ and $w(C_i\setminus A_i)\ge g/2$ for every $i$. There is one listed family $\mathcal P$ such that, for each $i$, some $P\in\mathcal P$ satisfies $P\cap A_i=\varnothing$ and $P\cap C_i\ne\varnothing$.
\item \textbf{Sparsity.} For every listed family $\mathcal P$ and every given $N$, $|N\cap\bigcup_{P\in\mathcal P}P|=O(\log(2dr))$.
\end{enumerate}
\end{lemma}
The algorithm depends on the sets $N_1,\dots,N_r$, but constructs the list before the inclusions $A_i\subseteq C_i$ are specified.

\begin{proof}
We give each vertex a collection of parallel lines whose number depends on its weight. Each grid point then represents a set of vertices, namely those with a line through that point. Steps~1 and~2 show that, for any admissible inclusion $A\subseteq C$, at least a constant fraction of the grid points represent sets that miss $A$ and meet $C$. Hence, for any fixed collection of $r$ admissible inclusions, a sufficiently large sample of independent uniform grid points contains, with high probability, a separating point for each inclusion. In Step~3, we derandomize this sampling argument using a construction of \citet[Corollary~C.4]{Goldreich2011}. It lets us list polynomially many small sets of grid points before the inclusions are specified. For any $r$ admissible inclusions, at least one set on the list contains a separating point for each of them.

It remains to ensure sparsity for \emph{every} set on the list. What makes this possible is that the separation argument works for every choice of the intercepts, so we are still free to choose where each vertex's lines lie. We use this freedom to ensure that each listed point set meets lines from only a few vertices of each given $N_i$. Step~4 studies how many vertices of a given $N_i$ have a line meeting a given point set when the lines are placed at random. Combining the bounds for all pairs means via a probabilistic argument that there exists some placement that satisfies all the sparsity requirements at once. In Step~5, we find such a placement deterministically by choosing the intercepts greedily, one vertex at a time. \medskip

\noindent\textbf{The affine plane over $\mathbb{F}_p$.}\enspace Choose a prime $p$ with $8d\le p\le16d$. By Bertrand's postulate, this interval contains a prime, which we can find in time polynomial in $d$ by testing the integers in the interval for primality.

The notation $\mathbb{F}_p$ denotes the integers modulo $p$. We represent its elements by $0,1,\ldots,p-1$ and perform addition and multiplication modulo $p$. Since $p$ is prime, every nonzero element has a multiplicative inverse, so $\mathbb{F}_p$ is a field. The affine plane $\mathbb{F}_p^2$ is the set of all $p^2$ ordered pairs $(x,y)$ with $x,y\in\mathbb{F}_p$. We use lines of the form $y=ax+b$, where $a\in\mathbb{F}_p$ is the slope and $b\in\mathbb{F}_p$ is the intercept. We recall three facts about these lines.
\begin{enumerate}
\item \textbf{Points on a line.} Each line has exactly $p$ points. Each of the $p$ choices of $x$ gives one point $(x,ax+b)$, and different first coordinates give different ordered pairs.
\item \textbf{Lines of one slope.} There are $p$ intercepts. Exactly one line of slope $a$ passes through $(x,y)$, since its intercept must be $y-ax$. In particular, different lines of the same slope are disjoint.
\item \textbf{Intersections.} Two distinct lines meet in at most one point. For different slopes, an intersection must satisfy $(a-a')x=b'-b$. Primality of $p$ makes $a-a'$ invertible modulo $p$, giving one $x$ and then one $y$. The equal-slope case follows from fact~2.
\end{enumerate}

\medskip
\noindent\textbf{1. Represent weights by lines.}\enspace For each vertex $v \in [d]$, draw $h_v$ lines of slope $v$ with
\[
h_v=\min \bigg \{d,\bigg\lceil\frac{2d w_v}{g}\bigg \rceil \bigg\}.
\]
These slopes are distinct as $d<p$. Give to these $h_v$ parallel lines the consecutive intercepts $s_v,s_v+1,\ldots,s_v+h_v-1$ modulo $p$ for some starting intercept $s_v \in \mathbb{F}_p$. Since $h_v\le d<p$, these lines are distinct, even when the intercepts wraps around. The next two steps work for every choice of the starts.

The following consequences motivate the choice of $h_v$. For any vertex set $S \subset [d]$, the number of lines assigned to its vertices is at most 
\begin{equation*}
\sum_{v\in S} h_v \le \sum_{v\in S}\left\lceil\frac{2d w_v}{g}\right\rceil \le \sum_{v\in S}\left(\frac{2d w_v}{g}+1\right) = \frac{2d\,w(S)}{g}+|S|.
\end{equation*}
So, if $w(A)<g$, its vertices receive at most $3d$ lines, and if $w(N)\le2g$, its vertices receive at most $5d$ lines. On the other hand, if $w(S)\ge g/2$, its vertices receive at least $d$ lines. Indeed, if one vertex receives $d$ lines we are done.  Otherwise the cap is inactive for every vertex in $S$, so
\begin{equation*}
\sum_{v\in S} h_v = \sum_{v\in S}\left\lceil\frac{2d w_v}{g}\right\rceil \ge \frac{2d}{g}\sum_{v\in S}w_v = \frac{2d\,w(S)}{g} \ge \frac{2d}{g}\cdot\frac{g}{2}=d.    
\end{equation*}
Hence, a separation in weight around $g$ induces a separation in the number of lines associated to a set. 

\medskip
\noindent\textbf{2. Count the points that separate one inclusion.}\enspace At each grid point $z$ define the vertex set
\[
P_z=\{v\in V : \text{a line assigned to $v$ passes through $z$}\}.
\]
By fact~2, a vertex contributes at most one line through $z$. Thus $|P_z|$ is exactly the number of assigned lines through $z$. 

Fix some $A\subseteq C$ such that $w(A)<g$ and $w(C\setminus A)\ge g/2$. We count points $z$ for which $P_z$ misses $A$ ($P_z \cap A =\emptyset$) and meets $C$ ($P_z \cap C \neq \emptyset$). By the previous calculation, $w(C \setminus A) \geq g/2$ implies that it contains at least $d$ lines assigned to its vertices, choose $d$ of them. Some chosen lines can belong to one vertex, this is fine. On each chosen line, we discard two sets of points.
\begin{enumerate}
\item Discard points on any line assigned to a vertex of $A$. Such a line belongs to a different vertex, so it differs from the chosen line. We have $w(A) <g$, so it is represented by at most $3d$ lines, therefore by fact~3, at most $3d$ points are discarded. Every surviving point now gives a $P_z$ disjoint from $A$.
\item Discard points on any of the other $d-1$ \emph{chosen} lines. By fact~3 again, this costs at most $d-1$ more points. The surviving points on different chosen lines are now disjoint, so we may add their counts. Any unchosen lines of $C\setminus A$ through those points are harmless.
\end{enumerate}
By fact~1 each chosen line starts with $p$ points before deletions, leaving at least $p-3d-(d-1)\ge p/2$ points overall, since $p\ge8d$. Each survivor lies on a line of $C\setminus A$, so $P_z$ meets $C$. Hence, summing over the $d$ lines, we have exhibited at least $\frac{dp}{2} \geq \frac{p^2}{32}$ points, all of which meet $C$ and misses $A$. 

\medskip
\noindent\textbf{3. Construct a hitting family.}\enspace Let's start by some intuition. Suppose each of $r$ regions $R_i$ occupies at least a fraction $\varepsilon=1/32$ of the grid, as in Step~2. A sample of $t$ independent uniform points misses a particular $R_i$ with probability at most $(1-\varepsilon)^t$. The union bound gives $\Pr[\text{the sample misses some }R_i]\le r(1-\varepsilon)^t\le r e^{-\varepsilon t}$.
Taking $t=\lceil\ln(2r)/\varepsilon\rceil=O(\log(2r))$ makes this probability at most $1/2$, so there exists some $t$ samples that hit all $R_i$ jointly. Enumerating all ordered samples would take $(p^2)^t$ trials, which would not be polynomial. The following derandomization construction keeps the samples short and gives a polynomial-size list to enumerate. We use \citet[Corollary~C.4]{Goldreich2011} in the following counting form:

\smallskip 
\noindent \textit{Combined Hitter:}
For every integer $\ell\ge 1$, density $0<\varepsilon\le 1$, and error $0<\delta\le 1/2$, there is an explicit list of sets $Q(\sigma)\subseteq\{0,1\}^\ell$, each efficiently computable from its seed $\sigma\in\{0,1\}^b$, with
\[
b=2\ell+O(\log(1/\delta)),\qquad |Q(\sigma)|=O\!\left(\frac{\log(1/\delta)}{\varepsilon}\right),
\]
such that every set $R\subseteq\{0,1\}^\ell$ of size at least $\varepsilon 2^\ell$ intersects $Q(\sigma)$ for at least a $(1-\delta)$ fraction of all seeds.
\smallskip

We are going to translate this result into our setting. Fix any choice of the starts and any $r$ admissible inclusions $A_i\subseteq C_i$. For each $i$, let $R_i$ consist of the points $z\in\mathbb{F}_p^2$ for which $P_z$ misses $A_i$ and meets $C_i$. We encode the grid points as bit strings. Label the $p^2$ points from $0$ to $p^2-1$ and write each label in binary using $\ell=\lceil\log_2(p^2)\rceil$ bits, padding with leading zeros when needed. By Step~2, each $R_i$ contains at least $p^2/32$ points. Since $p^2\le 2^\ell<2p^2$, the strings encoding points of $R_i$ make up at least a fraction $\frac{p^2/32}{2^\ell}\ge\frac{1}{64}$ of all $\ell$-bit strings.

Now let $\varepsilon=1/64$ and $\delta=1/(2r)$, and apply the hitter construction. For each seed $\sigma$, decode the labels in $Q(\sigma)$ smaller than $p^2$ and keep the resulting point set $T(\sigma)$. Write $t=\max_\sigma |T(\sigma)|=O(\log(2r))$. The number of seeds is
\[
2^b\le 2^{2\ell}(2r)^{O(1)}\le 4p^4(2r)^{O(1)} .
\]
As $p\le16d$, enumerating all seeds takes polynomial time in $d,r$. This list of point sets can already be constructed before any starts are fixed.

For each $i$, the hitting guarantee implies that $T(\sigma)$ misses $R_i$ for at most a $1/(2r)$ fraction of the seeds, since every string encoding a point of $R_i$ is retained when forming $T(\sigma)$. By the union bound, at most a fraction $r\cdot 1/(2r)=1/2$ of the seeds miss at least one of the $r$ regions. Thus, for at least half the seeds, $T(\sigma)$ contains a separating point for each of the $r$ inclusions.

For each seed $\sigma$, form the family $\mathcal P(T(\sigma))=\{P_z:z\in T(\sigma)\}$, which has at most $t=O(\log(2r))$ members. Whenever $T(\sigma)$ meets every $R_i$, this family witnesses separation for all $r$ inclusions. 

The above separation argument holds for every choice of the starts. It remains to choose the starts so that for every listed $T$, the induced family satisfies the sparsity condition.

\medskip
\noindent\textbf{4. Sparsity for a random intercept start.}\enspace Fix one of the given sets $N$ and one of the point sets $T=T(\sigma)$ constructed in Step~3. Choose the starting intercept $s_v$ of each vertex $v$ independently and uniformly from $\mathbb{F}_p$. Let $Z$ be the number of vertices of $N$ whose lines meet $T$. This is precisely the quantity we want to bound, since a vertex belongs to $\bigcup_{z\in T}P_z$ if and only if one of its lines passes through a point of $T$.

Fix a vertex $v\in N$, and let $q_v$ be the probability that one of its lines meets $T$. To bound $q_v$, first consider a single point $(x,y)\in T$. By fact~2, the unique line of slope $v$ through this point has intercept $y-vx$. Thus one of the lines of $v$ passes through $(x,y)$ precisely when its block of $h_v$ consecutive intercepts contains $y-vx$. If $h_v>0$, the starts that give such a block are $y-vx,\ y-vx-1,\ \ldots,\ y-vx-h_v+1 \pmod p$.  
These are $h_v$ distinct starts since $h_v<p$. Each point of $T$ therefore accounts for $h_v$ possible starts, and a start may be counted for more than one point. Hence at most $|T|h_v$ of the $p$ starts give a line meeting $T$, so
\[
q_v\le \frac{|T|h_v}{p}\le \frac{t h_v}{p}.
\]

We now bound $\mathbb{E}[2^Z]$. For each $v\in N$, let $I_v$ equal $1$ if one of its lines meets $T$ and $0$ otherwise. Then $Z=\sum_{v\in N}I_v$, so $2^Z=\prod_{v\in N}2^{I_v}$. The factor contributed by $v$ equals $2$ with probability $q_v$ and $1$ otherwise. Its expectation is therefore $2q_v+(1-q_v)=1+q_v$. Since each factor depends only on the start of its own vertex, the factors are independent. It follows that
\[
\mathbb{E}[2^Z]=\prod_{v\in N}(1+q_v)\le \exp\!\left(\sum_{v\in N}q_v\right)\le \exp\!\left(\frac{t}{p}\sum_{v\in N}h_v\right)\le e^{5dt/p}.
\]
The first inequality uses $1+x\le e^x$. The last uses Step~1: since $w(N)\le 2g$, the vertices of $N$ receive at most $5d$ lines in total.

\medskip
\noindent\textbf{5. Greedily choosing the intercepts.}\enspace We need a single choice of the starts that satisfies sparsity for every given set $N$ and every listed point set $T$. Form all pairs of a given set and a point set $T(\sigma)$, and index these pairs as $(N_i,T_i)$ for $i\in[M]$. There are at most $r\,2^b$ pairs, so $M$ is polynomial in $d,r$. Let $Z_i$ count the vertices of $N_i$ whose lines meet $T_i$.

Choose one random start per vertex as in Step~4. If a vertex belongs to several sets $N_i$, its start is the same in every pair. The counts $Z_i$ may therefore be dependent, but linearity of expectation still gives
\[
\mathbb{E}\bigg[\sum_{i=1}^M 2^{Z_i}\bigg]=\sum_{i=1}^M\mathbb{E}[2^{Z_i}]\le M e^{5dt/p}.
\]
Consequently, there exists a choice of the starts for which the sum is at most $M e^{5dt/p}$. For this choice, every individual term $2^{Z_i}$ is also at most $M e^{5dt/p}$.

We find the starts efficiently by the method of conditional expectations, a standard method in derandomization. Suppose we have fixed the starts of vertices $1,\ldots,j-1$. For each $a\in\mathbb{F}_p$, try $s_j=a$, using the same start in every pair. For each $i$ and each $v\in N_i$, let $q_{iv}(a)$ be the probability that a line of $v$ meets $T_i$ under these choices. For $v\le j$, the start is fixed, so $q_{iv}(a)$ is either zero or one. For $v>j$, we compute this probability by trying all $p$ possible starts of $v$. Since the remaining starts are independent, the product calculation from Step~4 gives
\begin{equation*}
\mathbb{E}\bigg[\sum_{i=1}^M 2^{Z_i}\,\mid\,\text{earlier choices},\,s_j=a\bigg]=\sum_{i=1}^M\prod_{v\in N_i}(1+q_{iv}(a)).  
\end{equation*}
Choose $a_j\in\mathbb{F}_p$ minimizing the displayed sum, which can be done in polynomial time\footnote{For each trial start, the sum has $M$ terms, each a product of at most $d$ factors. Each $q_{iv}(a)$ can be computed by testing at most $p$ starts against the points of $T_i$. Since each probability is a multiple of $1/p$, multiplying the sum by $p^d$ gives an integer of size at most $M(2p)^d$, which only requires $O(d \log(p)+\log(M))$ bits to write. Thus the $p=O(d)$ trial values can be compared exactly in polynomial time.}, and fix $s_j=a_j$. Then
\begin{align*}
\mathbb{E}\bigg[\sum_{i=1}^M 2^{Z_i}\,\mid \,\text{earlier choices},\,s_j=a_j\bigg]
&= \min_{a\in\mathbb{F}_p}\sum_{i=1}^M\prod_{v\in N_i}(1+q_{iv}(a)) \\
&\le \frac{1}{p}\sum_{a\in\mathbb{F}_p}\sum_{i=1}^M\prod_{v\in N_i}(1+q_{iv}(a)) \\
&= \mathbb{E}\bigg[\sum_{i=1}^M 2^{Z_i}\,\mid\,\text{earlier choices}\bigg].
\end{align*}
The last equality follows from the law of total expectation, since each of the $p$ starts is equally likely before we fix $s_j$. Chaining these inequalities over all $d$ vertices gives, for the resulting assignment,
\begin{equation*}
\sum_{i=1}^M 2^{Z_i}
= \mathbb{E}\bigg[\sum_{i=1}^M 2^{Z_i}\,\mid\,s_1=a_1,\ldots,s_d=a_d\bigg]
\le \cdots \le \mathbb{E}\bigg[\sum_{i=1}^M 2^{Z_i}\bigg]
\le M e^{5dt/p}.    
\end{equation*}
Taking logarithms gives $Z_i\le \log_2(\sum_{i=1}^M 2^{Z_i})\leq \log_2 M+5dt/(p\ln 2)=O(\log(2dr))$ for every pair $i$, since $M$ is polynomial in $d,r$, $p\ge 8d$, and $t=O(\log(2r))$.  Step~3 gives separation for these greedy starts, and the bound on each $Z_i$ gives sparsity.
\end{proof}

\begin{proof}[Proof of \Cref{lem:cover-hidden}]
Fix a set $X^*$ cheaper than every known edge. We want to delete at least one vertex from each known edge while leaving $X^*$ intact. We try each known edge $E$ as a reference and distinguish deletions inside $E$ from deletions outside it. For a suitable choice of $E$, the weight of $X^*\setminus E$ is smaller than the weight of $ E_i \setminus X^*$ which can be deleted in every known edge. This allows us to use \Cref{lem:hitting_result} to find, among the candidate sets $Y$ we enumerate (that are polynomially many), one that avoids $X^*\setminus E$. For this $Y$, we can delete all of $Y\setminus E$ at once. Inside $E$, we must still choose which vertices of $Y$ to delete. We distinguish the known edges according to how much weight they miss from $E$. An edge that misses a large part of $E$ must carry enough weight outside $E\cup X^*$ for us to hit it safely there, given that $X^*$ is cheaper than both $E$ and $E_i$, as it must have sufficient weight somewhere. We can therefore safely arrange for the outside deletions to hit all such edges. Hence, for each remaining edge $E_i$, the weight of the difference $E\setminus E_i$ must be small. We give this set difference to \Cref{lem:hitting_result}, whose sparsity guarantee ensures that each candidate deletion set $Y$ contains only $O(\log(2dr))$ vertices from it. This keeps the remaining choices inside $E$ few enough to enumerate in polynomial time.

We first explain how \Cref{alg:cover-hidden} constructs its list using only the known edges and the vertex weights. We then fix an arbitrary set $X^*$ cheaper than every known edge and show that at least one query on the list contains $X^*$.

\medskip
\noindent\textbf{1. Choose a reference edge and a weight scale.}\enspace 
Of course, we don't know the `right' edge $E_i$ to choose, and neither do we know the correct weight scale, as both depend on $X^*$ which is unknown. Instead, we simply try each known edge $E$ as the reference edge, and for each choice of $E$, we try every positive value $g=jw_a$ with $a,j\in[d]$. This gives at most $rd^2$ choices of $E$ and $g$. This weight list always contains a $2$-approximation value $g$ satisfying $w(S)\le g<2w(S)$ for every set $S$ of positive weight. Indeed, let $a$ be a heaviest vertex of $S$. Since $w_a\le w(S)\le d w_a$, taking $j=\lceil w(S)/w_a\rceil$ gives $j\in[d]$ and $w(S)\le jw_a<w(S)+w_a\le 2w(S)$.

For each $i\in[r]$, let $N_i=E\setminus E_i$ be the set of vertices of the reference edge that are missing from $E_i$. Call $N_i$ light if $w(N_i)\le 2g$ and heavy if $w(N_i)>2g$. The algorithm can compute all these sets and their weights.

\medskip
\noindent\textbf{2. List the candidate vertices for deletion.}\enspace For the current choice of $E$ and $g$, apply the auxiliary lemma with parameter $r$, using $N_i$ as its $i$th input set if $N_i$ is light and $\varnothing$ otherwise. For each family $\mathcal P$ returned by the lemma, and for each subfamily $\mathcal A \subseteq \mathcal P$, let $Y= \cup_{P \in \mathcal{A}} P$. The vertices of $Y$ are the candidates for deletion.

Each family has $O(\log(2r))$ members, so there are at most $2^{O(\log(2r))}$ subfamilies to try. Moreover, every such union $Y$ is contained in the union of the whole family. The sparsity guarantee of the auxiliary lemma therefore gives for every light $N_i$,
\begin{equation*}
|N_i \cap Y|=\big |N_i\cap\bigcup_{P\in\mathcal A}P\big|\le \big |N_i\cap\bigcup_{P\in\mathcal P}P\big|=O(\log(2dr)).   
\end{equation*}
This ensures that we will be able to efficiently consider deletions inside $E$.

\medskip
\noindent\textbf{3. Enumerate the deletions inside the reference 
edge.}\enspace
Fix a candidate set $Y$ and let $Y_{\mathrm{out}}=Y\setminus E$. Delete all vertices in $Y_{\mathrm{out}}$. If $Y_{\mathrm{out}}$ misses an edge with a heavy difference, discard $Y$. Otherwise, every edge still to be hit has a light difference. At least one such edge remains, since $Y_{\mathrm{out}}$ is disjoint from the reference edge $E$. We enumerate the additional deletions inside $E$ as follows.

For each $i$ with $N_i$ light, try every vertex $v\in Y\cap E\cap E_i$ and every subset $D\subseteq Y\cap N_i$. Delete $v$ and the vertices of $D$ in addition to $Y_{\mathrm{out}}$. If $Y_{\mathrm{out}}\cup D\cup\{v\}$ hits every known edge, add $[d]\setminus(Y_{\mathrm{out}}\cup D\cup\{v\})$ to the list of queries. Basically, we try the case in which $v$ is the only deleted vertex of $E_i$, so all other deletions inside $E$ must lie in $Y\cap(E\setminus E_i)=Y\cap N_i$.

Every query on the list is the complement of a deletion set that hits every known edge, so it omits at least one vertex from each known edge. For each $Y$, checking whether $Y_{\mathrm{out}}$ hits every edge with a heavy difference takes $O(rd)$ time. If this check succeeds, we try at most $r$ choices for $i$ and $d$ choices for $v$. For each light $N_i$, sparsity gives $|Y\cap N_i|=O(\log(2dr))$, so there are at most $2^{O(\log(2dr))}$ choices for $D$. Thus we test at most $rd\,2^{O(\log(2dr))}$ deletion sets of the form $Y_{\mathrm{out}}\cup D\cup\{v\}$, each in $O(rd)$ time. There are at most $rd^2$ choices of $E$ and $g$. For each choice, the auxiliary lemma constructs its list of families in polynomial time, and each family gives polynomially many unions $Y$. Finally, since $g=jw_a$ with $j\le d$, its encoding requires at most $O(\log(2d))$ more bits than that of $w_a$. The entire list is therefore constructed in time polynomial in $d$, $r$, and the encoding length of the weights.\medskip

It remains to show that the query list satisfies the stated properties of the lemma. By construction, only deletions that hit every edge were kept, so the queries do not contain any of the $E_i$. Now, consider some set $X^* \subseteq [d]$ with $w(X^*)<\min_j w(E_j)$. We will show that it is contained in at least one of the queries.

Among the known edges, consider $E=\argmin_{E_i} w(E_i \setminus X^*)$ the edge minimizing the weight outside $X^*$, and let $g_*=w(E\setminus X^*)$. Since $w(X^*)<w(E)$, and by the choice of $E$, we have for all $i \in [r]$
\begin{equation*}
w(X^*\setminus E)=w(X^*)-w(X^*\cap E)<w(E)-w(X^*\cap E)=w(E\setminus X^*)=g_* \leq w(E_i \setminus X^*)
\end{equation*}
Thus the weight we must protect outside $E$ is smaller than the weight available for safe deletions in each known edge. Since $g_*=w(E\setminus X^*)>0$, Step~1 applied to $E\setminus X^*$ guarantees that the algorithm tries a scale $g$ with $g_*\le g<2g_*$. Fix this choice of $E$ and $g$.

For a given candidate set $Y$, the algorithm deletes all of $Y_{\mathrm{out}}=Y\setminus E$, but only selected vertices of $Y\cap E$. The outside deletions therefore preserve $X^*$ exactly when $Y$ misses $X^*\setminus E$. Inside $E$, we may allow $Y$ to contain vertices of $X^*$, since we enumerate which vertices of $Y\cap E$ to delete. Thus the set we need the auxiliary lemma to protect is $X^*\setminus E$.

We also need $Y$ to contain a vertex that can be safely deleted from each known edge $E_i$. If $N_i$ is light, this vertex may lie outside $E$, in which case it is deleted immediately, or inside $E$, in which case it is available for the subsequent enumeration. If $N_i$ is heavy, the outside deletions alone must hit $E_i$. Accordingly, for the fixed set $X^*$, let
\begin{equation*}
F_i=
\begin{cases}
E_i\setminus X^*, & \text{if $N_i$ is light},\\
E_i\setminus(E\cup X^*), & \text{if $N_i$ is heavy}.
\end{cases}
\end{equation*}
We use these sets only in the analysis to show that some listed $Y$ misses $X^*\setminus E$ and meets every $F_i$. Since each $F_i$ is disjoint from $X^*\setminus E$, we can apply the separation guarantee to the inclusions $X^*\setminus E \subseteq (X^*\setminus E)\cup F_i$ once we verify that $w(F_i)\ge g/2$ for every $i$.

We have the required bound on $X^*\setminus E$, as $w(X^*\setminus E)<g_*\leq g$.
If $N_i$ is light, the choice of the reference edge gives $w(F_i)=w(E_i\setminus X^*)\ge g_*>g/2$ by the above calculations. If $N_i$ is heavy, the union $E\cup X^*$ has weight $w(X^*)+g_*$. Since $N_i=E\setminus E_i$, the part of $E_i$ inside $E\cup X^*$ is contained in $(E\cup X^*)\setminus N_i$. Hence the weight of $E_i \cap (E \cup X^*)$ is therefore at most $w(E\cup X^*)-w(N_i)=w(X^*)+g_*-w(N_i)$. Because $w(E_i)>w(X^*)$, and $w(N_i) > 2g$ as it is heavy, the weight outside the union satisfies
\begin{equation*}
w(F_i)=w(E_i)-w\bigl(E_i\cap(E\cup X^*)\bigr)>w(N_i)-g_*>2g-g_*\ge g.
\end{equation*}
We stress that here $g$ is known (this is the scale guess), and therefore we can construct the list of families from the auxiliary lemma based on it. Now apply the separation guarantee to the $r$ inclusions $X^*\setminus E\subseteq (X^*\setminus E) \cup F_i$. These inclusions satisfy the hypotheses of \Cref{lem:hitting_result}: $w(X^*\setminus E)<g$, and each $F_i$ is disjoint from $X^*\setminus E$ and has weight at least $g/2$. 

By \Cref{lem:hitting_result}, there is one listed family $\mathcal P$ containing, for each $i$, a member that misses $X^*\setminus E$ and meets $F_i$. Choose one such member for each $i$ and let $Y$ be their union. Step~2 enumerates this union, which misses $X^*\setminus E$ and meets every $F_i$. Thus $Y_{\mathrm{out}}=Y\setminus E$ avoids $X^*$ and hits every edge with a heavy difference, since for each such edge $F_i\subseteq E_i\setminus E$.

For each edge $E_i$ still unhit by $Y_{\mathrm{out}}$, any vertex in $Y\cap F_i$ lies inside $E$ and outside $X^*$. Hence adding all vertices of $(Y\cap E)\setminus X^*$ to $Y_{\mathrm{out}}$ hits every known edge. Let $S\subseteq (Y\cap E)\setminus X^*$ be an inclusion-minimal set for which $Y_{\mathrm{out}}\cup S$ hits every known edge. Since $Y_{\mathrm{out}}$ is disjoint from the reference edge $E$, at least one vertex of $S$ is needed to hit $E$, so $S$ is nonempty.

Choose any $v\in S$ and let $D=S\setminus\{v\}$. By minimality of $S$, some edge $E_i$ is missed by $Y_{\mathrm{out}}\cup D$ but hit by $v$. Its difference $N_i$ is light, since $Y_{\mathrm{out}}$ already hits every edge with a heavy difference. Moreover, every vertex of $D$ lies in $Y\cap E$ and outside $E_i$, so $D\subseteq Y\cap N_i$, while $v\in Y\cap E\cap E_i$. Thus Step~3 tries this pair $(D,v)$. The resulting deletion set $Y_{\mathrm{out}}\cup D\cup\{v\}=Y_{\mathrm{out}}\cup S$ hits every known edge and avoids $X^*$, so its complement is a listed query containing $X^*$.
\end{proof}

\subsection{Proof of \Cref{lem:hidden-frontier}} \label{app:hidden-frontier}

\begin{proof}
Write $V(q)=f(S^+(q))$ and $\|q\|_1=\sum_i q_i$. Fix a threshold $V(\mathbf 0)\le R<V(\mathbf 1)$, and call a critical contract $\contract^j$ eligible if $V(\contract^j)>R$. By \Cref{eq:reward-frontier}, a contract has reward above $R$ exactly when it dominates an eligible critical contract. We first show how to find a cheapest such contract.

We construct a sequence of hypergraphs as we discover new critical shares. For each agent, keep a sorted list of shares, initially containing only $0$ and $1$. For each pair of consecutive shares $s<t$ in agent $i$'s list, create a vertex $(i,t)$ of weight $t-s$. Each eligible critical contract $\contract^j$ defines a hidden edge $E_j$ containing exactly the vertices $(i,t)$ with $t\le\alpha_i^j$. The weights telescope, so $w(E_j)$ is the total payment after rounding each share of $\contract^j$ down to the largest listed share at most $\alpha_i^j$. In particular, $w(E_j)\le\|\contract^j\|_1$. Since every contract with reward above $R$ meets the payment requirements of some eligible critical contract, the minimum edge weight is a lower bound on the payment of every such contract.

\medskip \noindent \textbf{Answering an edge query.} To answer a containment query $X$, construct some contract $q$ one agent at a time. If $X$ contains $(i,t)$ for every positive share $t$ in agent $i$'s list, set $q_i=1$. Otherwise, let $t$ be the smallest strictly positive share in that list (so not $0$) for which $(i,t)\notin X$, and set $q_i=t-1/(2D^2)$. This places $q_i$ just below $t$, while keeping it above every critical share smaller than $t$.\footnote{Recall that every critical share has reduced denominator at most $D=2^{(m+2)L}$. The same bound holds for every listed share, since the lists contain only $0$, $1$, and critical shares. Any two distinct rationals with denominators at most $D$ differ by at least $1/D^2$.}

Fix some eligible critical contract $\contract^j$. For each agent $i$, its edge $E_j$ requires every listed vertex up to share $\alpha_i^j$. If $(i,t)$ is the first vertex missing from $X$, then $X$ contains every earlier vertex but misses $(i,t)$ itself. Thus it contains all the vertices required for agent $i$ exactly when $\alpha_i^j<t$, or equivalently, by our choice of $q_i$, when $\alpha_i^j\le q_i$. If no vertex is missing, the same equivalence holds because $q_i=1$.

It follows that $E_j\subseteq X$ exactly when $\contract^j\le q$. Suppose first that $X$ contains a hidden edge $E_j$. Its critical contract $\contract^j$ is eligible, so $V(\contract^j)>R$. Since $\contract^j\le q$, monotonicity gives $V(q)\ge V(\contract^j)>R$. Conversely, suppose $V(q)>R$. There is some critical contract $\contract^j\le q$ that attains $V(\contract^j)=V(q)$. This contract is eligible because its reward exceeds $R$, and its edge $E_j$ is contained in $X$ because $\contract^j\le q$. Thus the answer to the containment query is positive exactly when $V(q)>R$.

\medskip
\noindent\textbf{Finding a cheapest eligible critical contract.}\enspace
Apply \Cref{thm:hidden-edge} to find a minimum-weight hidden edge $X$. For each agent $i$, let $a_i$ be the largest share represented in $X$ for that agent, or zero if none is represented. Then $\|a\|_1=w(X)$. If $V(a)>R$, return $a$: it attains the lower bound and is therefore a cheapest contract with reward above $R$.

Suppose instead that $V(a)\le R$. Since $X$ is a hidden edge, its associated contract $q$ satisfies $V(q)>R$, and there must basically be some critical contract between $a$ and $q$. We find a new critical share by binary search on the first coordinate whose increase takes the reward above $R$. Starting from $a$, change each agent's share from $a_i$ to $q_i$ in turn, evaluating $V$ after each change. As soon as the reward exceeds $R$, keep all other shares fixed and binary-search $(a_i,q_i]$ for the least share at which the reward exceeds $R$.

By \Cref{eq:reward-frontier}, this least share is the smallest $\alpha_i^j$ among the eligible critical contracts whose requirements are met in every other coordinate. It is therefore a critical share, which we can recover exactly in polynomial time.\footnote{After $O(\log D)$ reward evaluations, binary search isolates the share in an interval of length less than $1/(2D^2)$. It is the unique rational in this interval with reduced denominator at most $D$ and can be recovered by continued fractions.} This share is new to agent $i$'s list, since it exceeds $a_i$ and $q_i$ lies strictly below the next listed share.

Insert this share into agent $i$'s list and repeat the search for the same threshold $R$ using the updated hypergraph. Each iteration either returns a cheapest contract with reward above $R$ or adds a previously unlisted critical share. Since each agent has at most $K$ distinct critical shares, there are at most $nK$ additions. Each hypergraph has at most $K$ edges and $n(K+1)$ vertices, and all vertex weights and query coordinates have polynomial encoding length. Thus the search for a cheapest contract above $R$ takes polynomial time.

\medskip
\noindent\textbf{Recovering the Pareto frontier.}\enspace Record $(0,V(\mathbf 0))$ with witness contract $\mathbf 0$ and set $R=V(\mathbf 0)$. While $R<V(\mathbf 1)$, find a cheapest contract $q$ with $V(q)>R$, record $(\|q\|_1,V(q))$ with witness contract $q$, and set $R=V(q)$. Each search strictly increases the recorded reward. By \Cref{eq:reward-frontier}, each such reward is attained at a critical contract, so there are at most $K$ searches.

To see why the records suffice, fix any contract $q$ and consider the first record whose reward is at least $V(q)$. Such a record exists because the procedure reaches $V(\mathbf 1)$. If it is the initial record, its payment is zero. Otherwise, the threshold immediately before that search was below $V(q)$, so $q$ was itself a candidate. The search returned a contract with payment at most $\|q\|_1$ and reward at least $V(q)$. Thus every attainable pair is weakly dominated by a recorded pair. Since every recorded pair has a witness contract, removing dominated records leaves exactly the Pareto frontier.
\end{proof}

}

\section{Substitutes: (W)MRF, Gross-Substitutes, and Subadditive   Rewards}\label{sec:substitutes}
In this section we study several classes of reward functions in the complement-free hierarchy of \cite{lehmann2001combinatorial}. 
We begin, in \Cref{sec:matroid} with the most restricted class --- weighted matroid rank functions --- and show that while for small teams we can easily identify a list of poly-many candidates containing all critical contracts, the Equilibrium Selection Problem and Optimal Contract Problem admit no PTAS, already when $n=2$. We also find that when restricting to unweighted matroid rank functions or partition WMRF the Equilibrium Selection Problem is tractable, implying tractability of the Optimal Contract Problem in small teams. This stands in contrast to the large teams setting, where no PTAS is possible unless \pequalsnp \cite{equalpay}.

In \Cref{sec:gs-oracle-lb} we increase our scope to the more general gross substitutes class, and show that even in cases where there are few critical contracts and the Equilibrium Selection Problem is easy, the Optimal Contract Problem still admits no PTAS, implying hardness that stems solely from the search over the contract space. This contrasts with the single-agent setting, where identifying the critical points is as easy as solving a demand query~\cite{contractsBeyondGS, deo2024supermodular}.

Finally, in \Cref{sec:subadditive-sqrtk} we study the largest complement-free class --- subadditive functions--- and present a poly-time algorithm achieving a tight approximation of $O(\sqrt{n})$, closing the gap between the previously known lower bound of $\Omega(\sqrt{n})$ \cite{duetting2022multi} and upper bound of $O(n)$ \cite{dutting2025black}.

\subsection{Weighted Matroid Rank Functions}\label{sec:matroid}

This section studies reward functions that are \emph{weighted matroid rank functions}: for a matroid on the action set together with nonnegative action weights, the reward of a set of actions is the maximum weight of an independent subset it contains (see \Cref{sec:model}). 

Our first insight is that under WMRF rewards, critical contracts have sufficient structure to be easily identifiable, and under small teams there are only poly-many of them:
\begin{lemma} [Polynomially many critical contracts] \label{lem:few-critical-contracts}
    Let $f$ be a weighted matroid rank function with weights $\{w_j\}_{j \in T}$. In every critical contract $\contract$ of the instance, each coordinate $\alpha_i$ lies in the set
    \[
    \{0\} \;\cup\; \Bigl\{ \tfrac{c_j}{w_j} : j \in T_i, w_j>0 \Bigr\} \;\cup\; \Bigl\{ \tfrac{c_j}{w_j - w_{j'}} : j \in T_i,\ j' \in T \setminus T_i,\ w_j > w_{j'} \Bigr\} \;\cup\; \Bigl\{ \tfrac{c_j - c_{j'}}{w_j - w_{j'}} : j, j' \in T_i,\ w_j > w_{j'} \Bigr\}.
    \]
    This set has size $O(|T_i|\,m) = O(m^2)$, and a critical contract is determined by its $n$ coordinates, so the instance has at most $O(m^{2n})$ critical contracts, where $m = |T|$.
\end{lemma}
The core of the proof is to link each agent's best response problem to a demand query, where for weighted matroid rank functions it is known that a local maximum for single-item changes --- an addition, removal, or one for one exchange --- is a global maximum. Thus, at a critical contract, each agent's best response is pinned by a single tight deviation, which in turn pins the agent's share to one of the listed cost-to-weight ratios. See the proof in \Cref{app:few-critical-contracts}.

\Cref{lem:few-critical-contracts} implies that any hardness in the small teams setting must originate from the hardness of the Equilibrium Selection Problem. We show that indeed both problems admit no PTAS, even for $n=2$ agents.
\begin{theorem} \label{thm:no-ptas}
    When $f$ is a weighted matroid rank function, both the optimal contract problem and the equilibrium selection problem admit no PTAS, even for two agents, unless \pequalsnp.
\end{theorem}
In \Cref{sec:partition} and \Cref{sec:unweighted}
we show that (even for large teams) the Equilibrium Selection Problem can be solved efficiently for partition WMRF and unweighted matroid rank functions. Together with \Cref{lem:few-critical-contracts} this implies tractability of the Optimal Contract Problem in the corresponding small teams settings.

We now turn to the proof of \Cref{thm:no-ptas}, which uses weighted graphic matroid rank functions.
\begin{definition} [Graphic Matroid] \label{def:graphic_matroid}
    Let $G = (V, E)$ be a (multi-)graph. The \emph{graphic matroid} of $G$ is the matroid $\mathcal{M} = (E, \I)$ whose ground set is the edge set $E$, and whose independent sets are the acyclic edge subsets (i.e., forests),
    \[
    \I = \{S \subseteq E : S \text{ contains no cycle}\}.
    \]
    Given nonnegative edge weights $\{w_e\}_{e \in E}$, the associated \emph{weighted graphic matroid rank function} $f : 2^E \to \reals_{\ge 0}$ assigns to each edge subset $S \subseteq E$ the maximum weight of a forest contained in $S$,
    \[
    f(S) = \max_{S' \subseteq S,\, S' \in \I} \sum_{e \in S'} w_e.
    \]
\end{definition}

The proof is via a reduction from \maxethree, which admits no PTAS \cite{Guruswami2004InapproximabilityRF}.

\subsubsection{The \maxethree Problem}
In this section we present the \maxethree problem.
\begin{definition} [\maxethree]
    An instance of the \maxethree problem is a pair $(U=[k], \F=\{F_1,\dots, F_\ell\})$, where $U$ is a universe, and $\F\subseteq 2^U$ is a family of subsets of $U$ of size $3$ (i.e., $F_i \subseteq U$ and $|F_i| = 3$, for any $i\in [\ell]$). 
    For any partition $U=U_1 \uplus U_2$, a set $S\subseteq U$ is said to be \emph{split} if and only if $S\cap U_1, S\cap U_2 \ne \emptyset$.
    The goal is to find a partition that maximizes the number of subsets in $\F$ that are \emph{split}.
\end{definition}
For simplicity we assume $k\le 3\ell$, since otherwise the instance can be reduced by removing elements not in any set.

\citet{Guruswami2004InapproximabilityRF} show that this problem admits no PTAS:
\begin{theorem} [\cite{Guruswami2004InapproximabilityRF}] \label{thm:guruswami}
    For every constant $\eps > 0$, it is NP-hard to distinguish between instances of \maxethree where all the sets can be split by some partition, and those where any partition splits at most a $(\frac{19}{20}+\eps)$-fraction of the sets.
\end{theorem}
\subsubsection{Our Reduction}

Given an instance $(U=[k], \F=\{F_1,\dots, F_\ell\})$ of \maxethree, we construct a contract instance where $2$ agents play edges of a multi-graph $G=(V,E)$, described below. The graph edges are colored red or blue, and are associated with a \emph{weight} and a \emph{cost}. Agent $1$'s available actions are the red edges, and agent $2$'s actions are the blue edges.
The reward function $f$ is the weighted graphic matroid rank function of $G$ (see \Cref{def:graphic_matroid}); that is, for any action profile $S\subseteq E$, the value $f(S)$ is the maximum weight of a forest contained in the edges $S$. For simplicity, the construction we describe has values of $f$ which are greater than $1$. Dividing $f$ and costs by the maximum value of $f$ (i.e., $f(E)$) yields an instance where $f$ takes values in $[0,1]$ as needed and agent and principal utilities are simply divided by the same factor, thus preserving all the results.

The graph $G=(V,E)$ and the edge costs and weights are constructed as follows (see \Cref{fig:graphic_matroid_instance} for an illustration). $V=\{s\} \cup \{v_1,\dots, v_k\} \cup \{b_1,\dots, b_\ell\} \cup \{r_1,\dots, r_\ell\} \cup \{B_1, B_2, R_1, R_2\}$, where we call $s$ the \emph{root vertex}, $\{v_1,\dots, v_k\}$ are the \emph{element vertices}, $\{b_1,\dots, b_\ell\}$ are the \emph{blue set vertices}, $\{r_1,\dots, r_\ell\}$ are the \emph{red set vertices}, and $\{B_1, B_2, R_1, R_2\}$ are the \emph{heavy vertices}.
The edge set consists of four types of edges, each with a red and a blue variant: 

\begin{enumerate}
    \item Red and blue \emph{assignment edges} $\{a_i^R\}_{i\in [k]}, \{a_i^B\}_{i\in [k]}$, where $a_i^R$ and $a_i^B$ both join $s$ to $v_i$. Each of these edges has weight $32$ and cost $6$, i.e., $w(a_i^R)=w(a_i^B)=32$ and $c(a_i^R)=c(a_i^B)=6$ for all $i\in [k]$. The interpretation is that the division of assignment edges along the agents defines the partition, i.e., $i\in U_1$ if agent $1$ has taken $a_i^R$, and $i\in U_2$ if agent $2$ has taken $a_i^B$ (observe that an equilibrium will never contain parallel edges). 
    \item Red and blue \emph{member edges} $\{m^R_{i,j}\}_{j\in [\ell], i\in F_j}, \{m^B_{i,j}\}_{j\in [\ell], i\in F_j}$, where $m^R_{i,j}=\{v_i, r_j\}$ and $m^B_{i,j}=\{v_i, b_j\}$. These edges have weight $24$ and cost $5$, i.e., $w(m_{i,j}^B)=w(m_{i,j}^R)=24$ and $c(m_{i,j}^B)=c(m_{i,j}^R)=5$. The interpretation is that an agent takes a member edge touching a set vertex when his side of the partition contains the corresponding element. A set $F_j$ is then split if both its associated set vertices (the vertices $r_j, b_j$) are each connected to a member edge in the equilibrium.
    \item Red and blue \emph{default edges} $\{d^R_j\}_{j\in [\ell]}, \{d^B_j\}_{j\in [\ell]}$, where $d^R_j = \{s, b_j\}$ and $d^B_j =\{s, r_j\}$. These edges have weight $16$ and cost $1$, i.e., $w(d^R_j) = w(d^B_j)=16$ and $c(d^R_j) = c(d^B_j)=1$. The interpretation is that each agent takes the default edge if and only if the other agent hasn't taken a member edge to the corresponding set vertex, i.e., these \emph{lighter} edges are selected by the agents when the set isn't split. 
    \item Red and blue \emph{heavy edges}, $h^R = \{R_1, R_2\}, h^B = \{B_1,B_2\}$. Both edges have weight $1000 \ell $ and cost $250 \ell$, i.e., $w(h^R)=w(h^B)=1000 \ell$ and $c(h^R)=c(h^B)=250 \ell$. In particular they are by far the \emph{heaviest} edges in $G$. These edges are entirely isolated from other edges, and thus agent $1$ (and symmetrically, agent $2$) weakly prefers taking the heavy edge when $\alpha_1 w(h^R) - c(h^R) \ge 0$, that is, when $\alpha_1 \ge \frac{1}{4}$. These edges are there to ``force'' the value of an optimal contract.
\end{enumerate}

\begin{figure}[t]
\centering
\begin{tikzpicture}[
    scale=1.05,
    every node/.style={font=\small},
    selectorR/.style={red, line width=0.55pt},
    selectorB/.style={blue, line width=0.55pt},
    vtx/.style={circle, fill=black, inner sep=1.7pt},
    tedge/.style={blue, line width=0.95pt},
    fedge/.style={red, line width=0.95pt},
    etedge/.style={red, dashed, line width=1.05pt},
    efedge/.style={blue, dashed, line width=1.05pt},
    indep/.style={line width=2.4pt}
]

\node[vtx, label=below:$s$] (r) at (0,-4.8) {};

\node[vtx, label=left:$v_1$] (v1) at (-2,-2) {};
\node[vtx, label=left:$v_2$] (v2) at (-1,-1.7) {};
\node[vtx, label=below:$v_3$] (v3) at (0,-1.5) {};
\node[vtx, label=right:$v_4$] (v4) at (1,-1.7) {};
\node[vtx, label=right:$v_5$] (v5) at (2,-2) {};

\node[vtx, label=above:$b_1$] (a1) at (-3.10,0.78) {};
\node[vtx, label=above:$b_2$] (a2) at (-2.10,1.15) {};
\node[vtx, label=above:$b_3$] (a3) at (-1.00,0.95) {};

\node[vtx, label=above:$r_1$] (b1) at ( 1.00,0.95) {};
\node[vtx, label=above:$r_2$] (b2) at ( 2.10,1.15) {};
\node[vtx, label=above:$r_3$] (b3) at ( 3.10,0.78) {};

\draw[selectorR] (r) to[bend left=7] (v1);
\draw[selectorB] (r) to[bend right=7] (v1);
\draw[selectorR] (r) to[bend left=7] (v2);
\draw[selectorB] (r) to[bend right=7] (v2);
\draw[selectorR] (r) to[bend left=7] (v3);
\draw[selectorB] (r) to[bend right=7] (v3);
\draw[selectorR] (r) to[bend left=7] (v4);
\draw[selectorB] (r) to[bend right=7] (v4);
\draw[selectorR] (r) to[bend left=7] (v5);
\draw[selectorB] (r) to[bend right=7] (v5);

\draw[tedge] (a1) -- (v1);
\draw[tedge] (a1) -- (v2);
\draw[tedge] (a1) -- (v3);

\draw[tedge] (a2) -- (v2);
\draw[tedge] (a2) -- (v3);
\draw[tedge] (a2) -- (v4);

\draw[tedge] (a3) -- (v3);
\draw[tedge] (a3) -- (v4);
\draw[tedge] (a3) -- (v5);

\draw[fedge] (b1) -- (v1);
\draw[fedge] (b1) -- (v2);
\draw[fedge] (b1) -- (v3);

\draw[fedge] (b2) -- (v2);
\draw[fedge] (b2) -- (v3);
\draw[fedge] (b2) -- (v4);

\draw[fedge] (b3) -- (v3);
\draw[fedge] (b3) -- (v4);
\draw[fedge] (b3) -- (v5);

\draw[etedge] (r) .. controls (-4.65,-3.55) and (-5.05,-1.00) .. (a1);
\draw[etedge] (r) .. controls (-4.20,-3.48) and (-4.45,-0.20) .. (a2);
\draw[etedge] (r) .. controls (-3.70,-3.38) and (-3.25, 0.18) .. (a3);

\draw[efedge] (r) .. controls ( 4.65,-3.55) and ( 5.05,-1.00) .. (b3);
\draw[efedge] (r) .. controls ( 4.20,-3.48) and ( 4.45,-0.20) .. (b2);
\draw[efedge] (r) .. controls ( 3.70,-3.38) and ( 3.25, 0.18) .. (b1);

\node[vtx, label=below:$R_1$] (R1) at (-1.55,-6.0) {};
\node[vtx, label=below:$R_2$] (R2) at (-0.35,-6.0) {};
\node[vtx, label=below:$B_1$] (B1) at ( 0.35,-6.0) {};
\node[vtx, label=below:$B_2$] (B2) at ( 1.55,-6.0) {};
\draw[indep, red]  (R1) -- (R2) node[midway, above] {$h^R$};
\draw[indep, blue] (B1) -- (B2) node[midway, above] {$h^B$};

\begin{scope}[shift={(5.6,-0.5)}]
    \draw[selectorR] (0,0.05) -- (0.8,0.05);
    \draw[selectorB] (0,-0.05) -- (0.8,-0.05) node[right,black,yshift=2pt] {$a_i^{R},a_i^{B}$:\ $w=32,\ c=6$};
    \draw[fedge] (0,-0.55) -- (0.8,-0.55);
    \draw[tedge] (0,-0.65) -- (0.8,-0.65) node[right,black,yshift=2pt] {$m^{R}_{i,j},m^{B}_{i,j}$:\ $w=24,\ c=5$};
    \draw[etedge] (0,-1.15) -- (0.8,-1.15);
    \draw[efedge] (0,-1.25) -- (0.8,-1.25) node[right,black,yshift=2pt] {$d^{R}_{j},d^{B}_{j}$:\ $w=16,\ c=1$};
    \draw[indep, red]  (0,-1.75) -- (0.8,-1.75);
    \draw[indep, blue] (0,-1.85) -- (0.8,-1.85) node[right,black,yshift=2pt] {$h^{R},h^{B}$:\ $w=1000\ell,\ c=250\ell$};
\end{scope}

\end{tikzpicture}
\caption{The constructed graph $G$. Assignment edges $a_i^R$ and $a_i^B$ between $s$ and $v_i$ are drawn as parallel
red/blue pairs (agent $1$ red, agent $2$ blue). Solid blue member edges $m^B_{i,j}=\{v_i,b_j\}$ and
solid red member edges $m^R_{i,j}=\{v_i,r_j\}$ join element vertices to the set vertices. Dashed red
default edges $d^R_j=\{s,b_j\}$ and dashed blue default edges $d^B_j=\{s,r_j\}$ join the root $s$ to the
set vertices. The isolated heavy edges $h^R=\{R_1,R_2\}$ (red) and $h^B=\{B_1,B_2\}$ (blue)
force the value of the optimal contract.}
\label{fig:graphic_matroid_instance}
\end{figure}

Agent $1$ owns the red edges, which we denote $E_1$, and agent $2$ the blue edges $E_2$, so that $E=E_1\uplus E_2$.

We analyse the equilibria of the constructed instance in two steps, aligned with the two directions of the reduction. \Cref{lem:eq-forests} is the shared structural engine, characterizing equilibria as certain maximum-weight spanning trees. Building on it, \Cref{lem:eq-to-partition} shows that every equilibrium projects to a partition and bounds its reward by an affine function of the number of sets that partition splits (the soundness direction), while \Cref{lem:partition-reward} shows that a distinguished profile is an equilibrium attaining that bound (the completeness direction).

Throughout, fix $\varepsilon=\frac1{20}$ and a contract $\contract=(\alpha_1,\alpha_2)\in[\frac14,\frac14+\varepsilon]^2$. For an edge $e\in E_i$, its \emph{utility weight} is $\omega_i(e)=w(e)-\tfrac1{\alpha_i}c(e)$. Since $\tfrac1{\alpha_i}\in[\tfrac{10}3,4]$, the assignment, member, default, and heavy utility weights, namely $32-\tfrac6{\alpha_i}$, $24-\tfrac5{\alpha_i}$, $16-\tfrac1{\alpha_i}$, and $250\ell(4-\tfrac1{\alpha_i})$, satisfy
\begin{equation}\label{eq:reduced-weights}
0<\omega_i(\text{member})<\omega_i(\text{assignment})<\omega_i(\text{default})<16,
\qquad \omega_i(\text{heavy})\ge0\ \ (=0\ \text{iff}\ \alpha_i=\tfrac14).
\end{equation}
The interpretation of this is that the agent always prefers to take a default edge over an assignment edge and an assignment edge over a member edge, and that taking the heavy edge is optimal if and only if $\alpha_i\ge\tfrac14$. Furthermore, the agent never wants to close a cycle with the other agent's edges, since the other agent's edges have weight at least $16$ while the agent's own edges in the main component have utility weight below $16$ by \eqref{eq:reduced-weights}.

We call $\{s\}\cup\{v_i\}_{i}\cup\{r_j,b_j\}_{j}$ the \emph{main} vertices (root, element, and set vertices).

\begin{lemma} [Equilibria as maximum-weight spanning trees] \label{lem:eq-forests}
    A set $S\subseteq E$ is an equilibrium of $\contract$ if and only if $S$ is a forest that spans the main vertices with a tree $T$, contains $h^R$ if $\alpha_1>\frac14$ and $h^B$ if $\alpha_2>\frac14$ (while at $\alpha_i=\frac14$ the corresponding heavy edge may be freely included or omitted), and satisfies, for each agent $i$, the exchange criterion: 
    \begin{quote}
        no non-tree edge of agent $i$ is heavier, in utility weight $\omega_i$, than the lightest agent-$i$ edge on its tree-path in $T$; a tree-path with no agent-$i$ edge imposes no constraint.
    \end{quote}
\end{lemma}
Fixing the opponent's edges, agent $i$'s best response is a maximum-weight forest under the weighting that scores his own edges at their utility weight $\omega_i$ and the opponent's at full weight. Such a forest is a spanning tree of the connected main component together with each positive-weight heavy edge, and the matroid exchange property turns its optimality into the stated tree-path criterion. See the proof in \Cref{app:eq-forests}.

\begin{lemma} [Every equilibrium induces a partition] \label{lem:eq-to-partition}
    Let $\varepsilon=\frac1{20}$ and let $\contract\in[\frac14,\frac14+\varepsilon]^2$. Every equilibrium $S$ of $\contract$ corresponds to a partition $[k]=U_1\uplus U_2$ via $U_1=\{i:a_i^R\in S\}$ and $U_2=\{i:a_i^B\in S\}$, in the following way:
    Each element vertex $v_i$ is connected to exactly one assignment edge (either $a^R_i$ or $a^B_i$). Each red set vertex $r_j$ is connected either to a red member edge $m^R_{i,j}$ with $i\in U_1\cap F_j$ or to the blue default edge $d^B_j$ (the latter forced when $U_1\cap F_j=\emptyset$), and symmetrically each blue set vertex $b_j$ is connected either to a blue member edge $m^B_{i,j}$ with $i\in U_2\cap F_j$ or to the red default edge $d^R_j$ (forced when $U_2\cap F_j=\emptyset$). Consequently, writing $x$ for the number of sets in $\F$ split by $(U_1,U_2)$,
    \[
    f(S)\ \le\ 2040\,\ell+32\,k+8\,x.
    \]
\end{lemma}
The exchange criterion of \Cref{lem:eq-forests} forces each element vertex onto a single assignment edge---which defines the partition---and each set vertex onto a member edge when its side meets the set and onto a default edge otherwise. Summing the resulting weights, with a split set contributing at most $48$ and an unsplit set at most $40$, gives the bound. See the proof in \Cref{app:eq-to-partition}.

Although \Cref{lem:eq-to-partition} shows that every equilibrium defines a partition, a partition need not determine a unique equilibrium. Even more, different equilibria which define the same partition may have different rewards. We therefore analyze an equilibrium of maximum reward among those corresponding to a given partition $[k]=U_1\uplus U_2$, which we denote by $S(U_1,U_2)$. For each set $F_j\in\F$, let its \emph{red connector} be the member edge $m^R_{i,j}$ with $i=\min(U_1\cap F_j)$ if $U_1\cap F_j\neq\emptyset$, and the default edge $d^B_j$ otherwise; define the \emph{blue connector} symmetrically, using $U_2$ and $d^R_j$. Set
\[
S(U_1, U_2)=\{h^R,h^B\}\ \cup\ \{a_i^R \mid i\in U_1\}\ \cup\ \{a_i^B\mid i\in U_2\}\ \cup\ \bigcup_{j\in[\ell]}\{\text{red and blue connector of }j\}.
\]
That is, both agents take their heavy edge and the assignments prescribed by the partition, and each red set vertex $r_j$ (resp.\ blue set vertex $b_j$) is attached by a member edge whenever its side meets $F_j$, and by a default edge otherwise. The following lemma shows that $S(U_1,U_2)$ is an equilibrium and that its reward attains the upper bound of \Cref{lem:eq-to-partition}.

\begin{lemma}\label{lem:partition-reward}
    Let $\varepsilon=\frac1{20}$. For every partition $[k]=U_1\uplus U_2$ and every contract $\contract\in[\frac14,\frac14+\varepsilon]^2$, the profile $S(U_1,U_2)$ is an equilibrium of $\contract$, and its reward is
    \[
    f(S(U_1,U_2)) = 2040\,\ell+32\,k+8\,x,
    \]
    where $x$ is the number of subsets in $\F$ split by the partition $(U_1,U_2)$. 
\end{lemma}
The profile $S(U_1,U_2)$ is a spanning tree of the main vertices together with both heavy edges, and a direct check of the exchange criterion of \Cref{lem:eq-forests} shows it is an equilibrium. Being a forest, its reward is its total edge weight, which meets the upper bound of \Cref{lem:eq-to-partition} with equality. See the proof in \Cref{app:partition-reward}.

\begin{lemma} \label{lem:opt-contract}
    Let $\varepsilon=\tfrac1{20}$. Any optimal contract and equilibrium $(\constar, S^\star)$ satisfy $\constar = (\tfrac14,\tfrac14)$, and every contract-equilibrium pair $(\contract,S)$ that is $\tfrac{20}{19}$-approximately optimal satisfies $\contract\in[\tfrac14,\tfrac14+\varepsilon]^2$.
\end{lemma}
Inside of the box $[1/4, 1/4+\varepsilon]^2$, as characterized by \Cref{lem:eq-to-partition}, the only critical contract is $(1/4, 1/4)$. If a payment is lower than $1/4$, one of the heavy edges will not be included and the reward will be too low, and if a payment is higher than $1/4+\varepsilon$, the principal's share will be too low. See the proof in \Cref{app:opt-contract}.

We are ready to prove \Cref{thm:no-ptas}.
\begin{proof}[Proof of \Cref{thm:no-ptas}]
    Given a \maxethree instance $(U=[k],\F=\{F_1,\dots,F_\ell\})$, construct the contract instance described above in time polynomial in $k+\ell$; it is a two-agent instance whose reward $f$ is a weighted graphic matroid rank function, and recall that $k\le 3\ell$.  

    Let $x^\star$ be the largest number of sets of $\F$ split by a partition of $U$. By \Cref{lem:eq-to-partition,lem:partition-reward}, the maximum reward of an equilibrium at the contract $(\tfrac14,\tfrac14)$ is $2040\ell+32k+8x^\star$, so the best principal utility there is half of it; and by \Cref{lem:opt-contract} the contract $(\tfrac14,\tfrac14)$ is optimal. Hence the optimum of the optimal contract problem, and that of the equilibrium selection problem at the fixed contract $(\tfrac14,\tfrac14)$ (\Cref{sec:model}), both equal
    \[
    \Phi(x^\star), \qquad\text{where}\qquad \Phi(x):=1020\ell+16k+4x.
    \]

    We give an L-reduction from \maxethree\ to the optimal contract problem in the unnormalized units of the construction. A PTAS for normalized instances gives the same approximation ratio in these units, since dividing rewards and costs by $f(E)$ preserves equilibria and scales every principal utility by the same factor. Its solution map sends a contract-equilibrium pair $(\contract,S)$ to a partition $g(\contract,S)$ as follows. Define $U_1=\{i:a_i^R\in S\}$ and $U_2=\{i:a_i^B\in S\}$. If neither assignment edge of some element $i$ is present, place $i$ in $U_1$ arbitrarily. Since an equilibrium cannot contain both parallel assignment edges---otherwise either owner could delete his edge without changing the reward and save positive cost---this completes $U_1,U_2$ to a partition. In the in-box case, \Cref{lem:eq-to-partition} shows that no completion is needed. Write $x(\pi)$ for the number of sets a partition $\pi$ splits.
    We now prove the two conditions of an L-reduction.

    First, $\Phi(x^\star)$ is at most a constant multiple of $x^\star$. A uniformly random partition places all three elements of a set on one side with probability $2\cdot(\tfrac12)^3=\tfrac14$, hence splits it with probability $\tfrac34$, so some partition splits at least $\tfrac34\ell$ sets and $x^\star\ge\tfrac34\ell$. With $k\le3\ell$ this gives $\ell\le\tfrac43 x^\star$ and $k\le 4x^\star$, so $\Phi(x^\star)=1020\ell+16k+4x^\star\le 1428\,x^\star$.

    Second, every pair $(\contract,S)$ satisfies $x^\star-x(g(\contract,S))\le\tfrac14\big(\Phi(x^\star)-u_P(\contract,S)\big)$. It suffices that $u_P(\contract,S)\le\Phi\big(x(g(\contract,S))\big)$, since then $\Phi(x^\star)-u_P(\contract,S)\ge\Phi(x^\star)-\Phi\big(x(g(\contract,S))\big)=4\big(x^\star-x(g(\contract,S))\big)$. If $\contract\in[\tfrac14,\tfrac14+\varepsilon]^2$, then $\alpha_1,\alpha_2\ge\tfrac14$ and \Cref{lem:eq-to-partition} give $u_P(\contract,S)=(1-\alpha_1-\alpha_2)f(S)\le\tfrac12\big(2040\ell+32k+8\,x(g(\contract,S))\big)=\Phi\big(x(g(\contract,S))\big)$. Otherwise $\contract\notin[\tfrac14,\tfrac14+\varepsilon]^2$, so by \Cref{lem:opt-contract} the pair is not $\tfrac{20}{19}$-approximately optimal, giving $u_P(\contract,S)<\tfrac{19}{20}\Phi(x^\star)<1020\ell\le\Phi\big(x(g(\contract,S))\big)$, where the middle inequality uses $\Phi(x^\star)\le 1072\ell$.

    These are the two conditions of an L-reduction, with constants $1428$ and $\tfrac14$; the equilibrium selection problem reduces the same way, with the contract fixed at $(\tfrac14,\tfrac14)$ so that only the first case above arises. Since \maxethree\ admits no PTAS unless \pequalsnp\ by \Cref{thm:guruswami}, neither the optimal contract problem nor the equilibrium selection problem does.
\end{proof}

\subsection{An Exponential Oracle Lower Bound for Gross Substitutes}\label{sec:gs-oracle-lb}

We now show that efficient equilibrium selection and polynomially many critical contracts do not suffice for tractability with gross-substitutes rewards. We allow access to an oracle returning the entire correspondence of principal-optimal Nash equilibria.

\begin{definition} [Optimal Equilibria Query] \label{def:gs-bne}
For a contract $\contract=(\alpha_1,\alpha_2)$, define $\displaystyle \bne(\contract)\coloneqq\argmax_{S\in\nash(\contract)}u_P(\contract,S)$. An \emph{optimal equilibria} query returns this entire set, including all ties. 
\end{definition}
Note that $\bne(\contract)$  consists of the reward-maximizing equilibria when $\alpha_1+\alpha_2<1$, all equilibria when $\alpha_1+\alpha_2=1$, and the reward-minimizing equilibria when $\alpha_1+\alpha_2>1$. 

The following theorem shows that a bound on the number of critical contracts cannot replace knowledge of their locations for general gross-substitutes rewards. 

\begin{theorem}\label{thm:gs-oracle-lb}

For $m$ large enough, suppose that there is an algorithm with access to an optimal equilibria oracle, such that on every two-agent instance with $m=|T|$ actions, a gross-substitutes reward function, and $O(m)$ critical contracts in $[0,1]^2$, it outputs a $64/63$-approximate contract with probability at least $2/3$. Then on at least one such instance, this algorithm requires $2^{\Omega(m)}$ value or $\bne$ oracle queries.
\end{theorem}
Intuitively, in more than one dimension, finding the critical contracts is hard, even if there are only polynomially many of them. We remark that this is an information theoretic lower bound, rather than a computational complexity one, and is thus independent of computational assumptions (such as \pequalsnp). 

The hard family consists of assignment valuations with $O(m)$-bit rational costs and reward values. Each agent's action set and all action costs are identical across the family. Only the reward function varies between instances, which depends on a hidden unknown set, and is accessed through value and optimal equilibria oracles. 

\paragraph{Proof overview.}
The aim is to hide the location of an approximately optimal contract while keeping the number of critical contracts small. We construct an exponentially large family of instances, indexed by hidden sets $H$. In each instance, a distinguished equilibrium gives the principal a constant-factor advantage over every other equilibrium and arises at only one contract. These contracts are distinct across the family, so achieving the approximation factor in the theorem requires identifying the hidden instance.

To show that the hidden instance is hard to identify, we compare the family with a fixed reference instance. For any fixed value or optimal equilibria query, the answer agrees with the reference answer on all but polynomially many instances. Thus exponentially many queries are needed to distinguish among the exponentially many possible hidden sets. Finally, we show that profiles selecting the same number of actions yield only a constant number of distinct critical contracts. Summing over the possible profile sizes gives a linear bound. \mse{The full proof appears in \Cref{app:gs-construction}.}

\subsection{A Tight Approximation for Subadditive Rewards}
\label{sec:subadditive-sqrtk}

The main result of this section is a poly-time $O(\sqrt n)$-approximation for any number of agents (not necessarily constant), using value and demand queries. This is asymptotically tight by the $\Omega(\sqrt{n})$ lower bound of \cite{duetting2022multi}.

We now introduce Algorithm~\ref{alg:subadditive-block-search}. 
The algorithm searches over two families of contracts. The first family consists of single-agent contracts: for each agent $i$, we ignore all other agents and run the known demand-query approximation for the single-agent combinatorial contract problem on $T_i$. This family captures instances in which one agent alone accounts for a significant fraction of the optimal reward.

The second family consists of uniform block contracts. Set $s=\lceil\sqrt n\rceil$ and $\rho=1/(4s)$. The algorithm partitions $A$ into arbitrary blocks of size $s$ with possibly a smaller block; for each block $B$, it pays every agent in $B$ the same share $\rho$, and pays every other agent zero. Since $|B|\le s$, the total payment under such a contract is at most $1/4$, so the principal keeps at least a $3/4$ share. A demand query computes a global maximizer of the potential $f(S)-\sum_{i\in B}c(S_i)/\rho$. By \cite{multimulti}, the resulting potential maximizer is a pure Nash equilibrium under the block contract. The algorithm evaluates the principal's utility for every candidate it constructs and returns the best one. The number of uniform block candidates is $\lceil n/s \rceil=O(\sqrt n)$. Each uses one demand query. Hence the algorithm runs in polynomial time for arbitrary $n$.

\begin{algorithm}
\caption{Best Single Agent or Uniform Block Contract Algorithm}
\label{alg:subadditive-block-search}
\KwIn{Instance $\langle A,\{T_i\}_{i\in A},f,c\rangle$ with value and demand-query access to $f$.}
\KwOut{A contract and a pure Nash equilibrium.}

$\mathcal C\gets\emptyset$\;

\ForEach{agent $i\in A$}{
    Run the $2$-approximation for the single-agent problem on $T_i$\;
    Add the returned single-agent contract, with all other shares set to $0$, and its induced equilibrium to $\mathcal C$\;
}

Set $s\gets \lceil\sqrt n\rceil$ and $\rho\gets 1/(4s)$\;

Partition $A$ into arbitrary blocks $B_1,\ldots,B_{\lceil n/s\rceil}$ of size $s$ with possibly a last smaller block \;
\ForEach{block $B_j$}{
    Define the uniform block contract $\alpha^{(j)}$ by $\alpha_i^{(j)}=\rho$ for $i\in B_j$ and $\alpha_i^{(j)}=0$ otherwise\;
    Define prices $q^{(j)}$ by $q^{(j)}_\ell=c_\ell/\rho$ for $\ell\in T_i$, $i\in B_j$, and $q^{(j)}_\ell=2$ for $\ell\in T\setminus\bigsqcup_{i\in B_j}T_i$\;
    Query demand to compute $S^{(j)}\in\arg\max_{S\subseteq T}\{f(S)-\sum_{\ell\in S}q^{(j)}_\ell\}$\;
    Add $(\alpha^{(j)},S^{(j)})$ to $\mathcal C$\;
}

\Return a candidate $(\alpha,S)\in\mathcal C$ maximizing $(1-\sum_i\alpha_i)f(S)$\;
\end{algorithm}

\begin{theorem}
\label{thm:subadditive-sqrt-n}
Let $f$ be normalized, monotone, and subadditive. Algorithm~\ref{alg:subadditive-block-search} runs in polynomial time with demand-query access to $f$ and returns a contract $\alpha$ and an equilibrium $S\in\mathrm{NE}(\alpha)$ satisfying
\begin{equation*}
 u_P(\alpha,S)\ge \frac{1}{8000\sqrt n}\cdot \mathrm{OPT}.
\end{equation*}
\end{theorem}

\paragraph{Proof overview:} At a high level, the proof balances two regimes, when the reward is concentrated on a single agent, and when it is spread out to multiple ones. Let $S^\star$ be the equilibrium induced by an optimal contract $\alpha^\star$, and write $S^\star_i$ for the actions taken by agent $i$. First, suppose that there are no highly paid agents with $\alpha^\star_i >1/2$. If some agent has large standalone value $f(S^\star_i)$, then the best single-agent contract already gives the desired approximation.

Otherwise, every agent has small standalone value. In this case, we partition the agents into cardinality blocks of size $s=\lceil\sqrt n\rceil$ with possibly one final smaller block. For each block $B$, we try the uniform contract that pays every agent in $B$ the share $\rho=1/(4s)$. The total payment is then at most $1/4$, so the principal keeps a constant fraction of the reward. We leverage the potential function introduced by \cite{multimulti}, namely, $\Phi(S) = f(S)-\sum_{i\in B} \sum_{j\in S_i} \frac{c_j}{\rho}$ for the block contract $\alpha$, where agents outside $B$ take no actions. A set maximizing this potential is an equilibrium of the contract $\alpha$. We lower bound the reward of an equilibrium from contracts offering a uniform payment to a single partition block by this potential function (since costs are non-negative, the potential is at most the reward). Due to the subadditive nature of $f$, if no agent is highly paid then either some single agent contributes an $\Omega(1/\sqrt{n})$ fraction of the reward, and we can incentivize him alone, or some block of the partition with the uniform payment share $\rho$ induces an equilibrium whose potential is an $\Omega(1/\sqrt{n})$ fraction of the reward, as needed.

The remaining issue is the possibility of one highly paid agent. If the high-share agent is responsible for most of the reward, a single-agent contract for him gives a constant approximation. Otherwise, the remaining agents carry a constant fraction of the optimal value, and all of them have payment share below $1/2$, so the same block contracts work: in the analysis, we evaluate each block's potential at the profile in which the highly paid agent plays nothing, and apply the above reasoning to the remaining agents. \medskip

We use the standard single-agent demand-query approximation for monotone subadditive rewards as a black box\revision{~\cite{multimulti}}.

\begin{theorem*}
\label{thm:single-agent-subadditive}
For the single-agent combinatorial contract problem with monotone subadditive reward function and value and demand-query access, there is a FPTAS, and therefore a $2$-approximation, to the optimal single-agent contract value.
\end{theorem*}

\begin{observation}
\label{obs:subadditive-cost-certificate}
When $f$ is subadditive, if $S\in\mathrm{NE}(\alpha)$, then $c(S_i)\le \alpha_i f(S_i)$ for every agent $i\in A$. 
\end{observation}

\begin{proof}
Since $S$ is an equilibrium, agent $i$ does not want to deviate to $\emptyset$. Therefore $\alpha_i f(S)-c(S_i)\ge \alpha_i f(S_{-i})$, and hence $c(S_i)\le \alpha_i(f(S)-f(S_{-i}))$. By subadditivity, $f(S)=f(S_i\sqcup S_{-i})\le f(S_i)+f(S_{-i})$, so $f(S)-f(S_{-i})\le f(S_i)$. Combining the two inequalities gives $c(S_i)\le \alpha_i f(S_i)$.
\end{proof}

The following observation is the single-agent component used throughout the section. The first part handles low-share agents, showing that the best single agent contract approximates the marginal reward of any Nash equilibrium for low-paid agents. The second part, which was proved by \citet{multimulti} and we state here for completeness, handles a possibly costly agent with a relatively high marginal welfare, in this case, the best single agent contract for this agent approximates well the total reward of any equilibrium.

\begin{observation}
\label{obs:single-agent-recovery-subadditive}
Let $f$ be subadditive and fix a contract $\contract$ and equilibrium $S\in\mathrm{NE}(\contract)$.
\begin{enumerate}
    \item If $\alpha_i\le 1/2$, then the optimal single-agent contract for agent $i$ has value at least $f(S_i)/12$.
    \item \emph{(Case $1$ in Lemma~5.2 of \cite{multimulti})}~If there is an agent $h$ such that $f(S_{-h})\le \frac{1-\alpha_h}{4}f(S_h)$, then the optimal single-agent contract for agent $h$ has value at least $\frac15\left(1-\sum_{i\in A}\alpha_i\right)f(S)$.
\end{enumerate}
\end{observation}

\begin{proof}
For the first claim, consider the single-agent contract that pays only agent $i$ with share $3/4$. Let $D_i\in\arg\max_{D\subseteq T_i}\{\frac34 f(D)-c(D)\}$. Since $S_i$ is feasible for this demand problem,
\begin{equation*}
\frac34 f(D_i)-c(D_i)\ge \frac34 f(S_i)-c(S_i).
\end{equation*}
By \Cref{obs:subadditive-cost-certificate} and $\alpha_i\le 1/2$, we have $c(S_i)\le f(S_i)/2$. Thus $\frac34 f(D_i)-c(D_i)\ge f(S_i)/4$. Since costs are nonnegative, $f(D_i)\ge f(S_i)/3$. The principal keeps a $1/4$ fraction of the reward under the contract $3/4$, and therefore this contract obtains utility at least $f(S_i)/12$. The optimal single-agent contract is at least as good.

\end{proof}

The following main lemma will be applied to all agents or, if there is a unique high-share agent, to all remaining agents. The lemma formalizes the uniform-block intuition: if no one in $R$ has large standalone value, then the algorithm's partition of $A$ into $O(\sqrt n)$ blocks contains a block with large positive potential under a uniform block contract.

\begin{lemma}
\label{lem:low-share-subset-recovery}
Let $f$ be subadditive, and fix a contract $\alpha$ with $\sum_i\alpha_i\leq1$, an equilibrium $S\in\nash(\alpha)$, and a set $R\subseteq A$. Suppose that $\alpha_i\leq1/2$ for every $i\in R$. Then the candidate returned by Algorithm~\ref{alg:subadditive-block-search} has principal utility at least
\begin{equation*}
 \frac{1}{1536\sqrt n}f(S_R).
\end{equation*}
\end{lemma}

\begin{proof}
If $f(S_R)=0$, the claim is immediate. Set $s=\lceil\sqrt n\rceil$ and $\rho=1/(4s)$. 
First suppose that some agent $i\in R$ satisfies $f(S_i)\ge f(S_R)/(32s)$. By \Cref{obs:single-agent-recovery-subadditive}, the optimal single-agent contract for $i$ has value at least $f(S_i)/12$. The algorithm runs a $2$-approximation for this single-agent problem, and therefore obtains value at least
\begin{equation*}
\frac{f(S_i)}{24}\ge \frac{f(S_R)}{768s}\ge \frac{f(S_R)}{1536\sqrt n},
\end{equation*}
where the last inequality uses $s=\lceil\sqrt n\rceil\le 2\sqrt n$.

We now assume that $f(S_i)<f(S_R)/(32s)$ for every $i\in R$. Let $B_1,\ldots,B_{\lceil n/s\rceil}$ be the algorithm's partition of $A$ into blocks of size $s$ with possibly a smaller last block. Since $s=\lceil\sqrt n\rceil$, there are $\lceil n/s\rceil\le s$ blocks.

For each block $B_j$, let $\alpha^{(j)}$ be the uniform block contract with $\alpha_i^{(j)}=\rho$ for $i\in B_j$ and $\alpha_i^{(j)}=0$ otherwise. Since $|B_j|\le s$, we have
\begin{equation*}
\sum_i\alpha_i^{(j)}=|B_j|\rho\le s\cdot \frac1{4s}=\frac14,
\end{equation*}
so the principal keeps at least a $3/4$ fraction of the reward. Let
\begin{equation*}
\Phi_j(Y)=f(Y)-\sum_{i\in B_j}\frac{c(Y_i)}{\rho}
\end{equation*}
be the potential associated with this contract. Evaluating it at the original profile restricted to the agents in $B_j\cap R$, i.e., $S_{B_j\cap R}$, and using \cref{obs:subadditive-cost-certificate} gives
\begin{equation*}
\Phi_j(S_{B_j\cap R})
\ge
f(S_{B_j\cap R})-4s\sum_{i\in B_j\cap R}\alpha_i f(S_i).
\end{equation*}
Summing over the blocks,
\begin{align*}
\sum_j \Phi_j(S_{B_j\cap R})
&\ge
\sum_j f(S_{B_j\cap R})-4s\sum_{i\in R}\alpha_i f(S_i).
\end{align*}
By subadditivity, $\sum_j f(S_{B_j\cap R})\ge f(S_R)$. Moreover, since $f(S_i)<f(S_R)/(32s)$ for every $i\in R$,
\begin{equation*}
\sum_{i\in R}\alpha_i f(S_i)
\le
\frac{f(S_R)}{32s}\sum_{i\in R}\alpha_i
\le
\frac{f(S_R)}{32s}.
\end{equation*}
Therefore,
\begin{equation*}
\sum_j \Phi_j(S_{B_j\cap R})
\ge
f(S_R)-4s\cdot \frac{f(S_R)}{32s}
=
\frac{7f(S_R)}{8}.
\end{equation*}
Since there are at most $s$ blocks and this sum is nonnegative, some block $B_k$ satisfies
\begin{equation*}
\Phi_k(S_{B_k\cap R}) \ge \frac{1}{s} \sum_j \Phi_j(S_{B_j\cap R})\ge \frac{7f(S_R)}{8s}.
\end{equation*}
Let $S^{(k)}$ be the demand-query maximizer for $\alpha^{(k)}$. Then
\begin{equation*}
\Phi_k(S^{(k)})\ge \Phi_k(S_{B_k\cap R})\ge \frac{7f(S_R)}{8s}.
\end{equation*}
By \cite{multimulti}, $S^{(k)}\in\mathrm{NE}(\alpha^{(k)})$. Since costs are nonnegative, $f(S^{(k)})\ge \Phi_k(S^{(k)})$. The principal keeps at least $3/4$ of this reward, so
\begin{equation*}
 u_P(\alpha^{(k)},S^{(k)})
 \ge
 \frac34 f(S^{(k)})
 \ge
 \frac34\cdot \frac{7f(S_R)}{8s}
 =
 \frac{21f(S_R)}{32s}
 \ge
 \frac{f(S_R)}{1536\sqrt n}.
\end{equation*}

The lemma follows.
\end{proof}

We can now prove the main result of the section. 

\begin{proof}[Proof of Theorem~\ref{thm:subadditive-sqrt-n}]
If $\mathrm{OPT}=0$, the guarantee is immediate because every candidate has nonnegative principal utility. Assume henceforth that $\mathrm{OPT}>0$. Let $(\alpha^\star,S^\star)$ be an optimal solution, so $\mathrm{OPT}=(1-\sum_i\alpha^\star_i)f(S^\star)$. Then $\sum_i\alpha^\star_i<1$, in particular every $\alpha^\star_i<1$. By \cref{obs:subadditive-cost-certificate}, $c(S^\star_i)\le\alpha^\star_i f(S^\star_i)$ for all $i\in A$.

First suppose that $\alpha^\star_i\le 1/2$ for every agent. Applying Lemma~\ref{lem:low-share-subset-recovery} with $R=A$ gives a candidate of value at least
\begin{equation*}
\frac{f(S^\star)}{1536\sqrt n}\ge \frac{\mathrm{OPT}}{1536\sqrt n}.
\end{equation*}

It remains to handle the case where some agent $h$ has $\alpha^\star_h>1/2$. Such an agent is unique. If
$f(S^\star_{-h})\le \frac{1-\alpha^\star_h}{4}f(S^\star_h)$
then \Cref{obs:single-agent-recovery-subadditive} shows that the optimal single-agent contract for $h$ has value at least $\mathrm{OPT}/5$. The algorithm runs a $2$-approximation for every single-agent problem, so it obtains value at least $\mathrm{OPT}/10$.

Finally suppose
$f(S^\star_{-h})> \frac{1-\alpha^\star_h}{4}f(S^\star_h)$,
By subadditivity and $1-\sum_i\alpha^\star_i\le 1-\alpha^\star_h$,
\begin{align*}
\mathrm{OPT}
\le (1-\alpha^\star_h)\bigl(f(S^\star_h)+f(S^\star_{-h})\bigr)
<4 f(S^\star_{-h})+(1-\alpha^\star_h)f(S^\star_{-h})
\le 5f(S^\star_{-h}).
\end{align*}
For every $i\ne h$, $\alpha^\star_i\le 1-\alpha^\star_h<1/2$. Together with the cost certificate, this verifies the assumptions of Lemma~\ref{lem:low-share-subset-recovery} for $R=A\setminus\{h\}$. The lemma gives a candidate of value at least
\begin{equation*}
\frac{f(S^\star_{-h})}{1536\sqrt n}
>
\frac{\mathrm{OPT}}{7680\sqrt n}. \qedhere
\end{equation*}
\end{proof}

This approximation result actually closes two additional gaps from the literature. First, we say that a contract is equal-pay if all agents receiving a positive payment share receive the same share. One of the main questions in \cite{equalpay} is to determine the \emph{price of equality}, the worst case ratio between the optimal principal utility and the optimal principal utility restricted to equal pay contracts. The following result improves upon the known $O(n)$ gap from \cite{equalpay}. 

\begin{corollary}
\label{cor:subadditive-equal-pay}
For normalized, monotone, subadditive rewards, Algorithm~\ref{alg:subadditive-block-search}
computes an equal-pay contract and a pure Nash equilibrium with
principal utility at least $\mathrm{OPT}/(8000\sqrt n)$.
Consequently, the worst-case price of equality is $\Theta(\sqrt n)$.
\end{corollary}

\begin{proof}
Every candidate considered by the algorithm is either a single-agent
contract or a uniform block contract, and is therefore equal-pay.
The approximation guarantee follows from
\Cref{thm:subadditive-sqrt-n}, with respect to the unrestricted optimum.
The matching lower bound on the price of equality is established
in~\cite{equalpay}.\footnote{Their construction extends to normalized rewards by setting $f(\emptyset)=0$. This preserves subadditivity and can affect equilibrium conditions only at profiles with at most one active agent, all of which have reward $O(1/\sqrt n)$. A common rescaling of rewards and costs then puts $f$ in $[0,1]$ and preserves the gap.}
\end{proof}

The same algorithm also approximates the principal's optimal utility
when coarse correlated equilibria are allowed. A coarse correlated
equilibrium (CCE) is a distribution over action profiles under which
no agent can improve his expected utility by switching to a fixed
action bundle. Let
\[
\mathrm{OPT}_{\mathrm{CCE}}
=
\max_{\substack{\alpha\geq 0,\ \sum_i\alpha_i\leq 1\\
                 \mu\in\operatorname{CCE}(\alpha)}}
\big(1-\sum_i\alpha_i\big)\mathbb E_{S\sim\mu}[f(S)].
\]

\begin{corollary}
\label{cor:subadditive-cce-pne}
For normalized, monotone, subadditive rewards, the worst-case gap between the optimal principal utilities
under CCE and PNE is $\Theta(\sqrt n)$.
\end{corollary}

\begin{proof}
Fix a feasible contract $\alpha$ and a CCE $\mu$ yielding positive
principal utility $U$. All expectations below are with respect to
$\mu$. Deviating to the empty bundle gives
\[
\mathbb E[c(S_i)]
\leq \alpha_i\mathbb E[f(S)-f(S_{-i})]
\leq \alpha_i\mathbb E[f(S_i)].
\]
The single-agent and block arguments in
\Cref{lem:low-share-subset-recovery} therefore extend in expectation,
since each maximizer performs at least as well as the corresponding
random bundle on average. This handles the case where all shares
are at most $1/2$.

Otherwise, let $h$ be the unique agent with $\alpha_h>1/2$, and let
$D$ be his standalone best response at share $(1+\alpha_h)/2$.
Its optimality and the CCE constraint for deviating to $D$ give
\[
\frac{1-\alpha_h}{2}f(D)
\geq
\frac{1+\alpha_h}{2}\mathbb E[f(S_h)]
-\alpha_h\mathbb E[f(S)].
\]
If
$\mathbb E[f(S_{-h})]\leq
(1-\alpha_h)\mathbb E[f(S_h)]/4$,
subadditivity implies $f(D)\geq \mathbb E[f(S_h)]/2$.
The standalone contract then yields principal utility at least
$U/5$, so the algorithm obtains at least $U/10$.
Otherwise, subadditivity gives
$U<5\mathbb E[f(S_{-h})]$, and the expected low-share bound
applies to the remaining agents. The same case analysis thus guarantees utility at least $U/(8000\sqrt n)$. Maximizing over contracts and their CCE proves the claim.

Finally, \cite{dutting2025black} establishes an
$\Omega(\sqrt n)$ gap already between mixed and pure Nash
equilibria. Since every mixed Nash equilibrium is a CCE,
this gives the matching lower bound.
\end{proof}

\section*{Acknowledgments}

This project has been partially funded by the European Research Council (ERC) under the European Union's Horizon Europe Program (FACT, grant agreement No.~101170373), by an Amazon Research Award, and by the Israel Science Foundation Breakthrough Program (grant No.~2600/24).  T. Ezra is supported by the Israel Science Foundation (grant No. 3256/26).
Maya Schlesinger is supported by the Adams Fellowships Program of The Israel Academy of Sciences and Humanities.
\newpage

\newcommand{\PrintSubstitutesAppendices}{
\section{WMRF Rewards: Missing Proofs}\label{app:matroid-proofs}

\subsection{Proof of \Cref{lem:few-critical-contracts}} \label{app:few-critical-contracts}
\begin{proof}
    Fix a critical contract $\contract$, critical for an equilibrium $S$, and an agent $i$; we show that $\alpha_i$ belongs to the stated set. If $\alpha_i = 0$ this is immediate, so assume $\alpha_i > 0$.

    Since $\contract$ is critical, $\alpha_i$ is the least share under which $S_i$ is a best response to $S_{-i}$, i.e.\ under which $S_i$ maximizes $a\,f(X \uplus S_{-i}) - c(X)$ over $X \subseteq T_i$ at $a = \alpha_i$. The scaled reward $a\,f$ is gross substitutes, and fixing the opponents' actions leaves $X \mapsto a\,f(X \uplus S_{-i})$ gross substitutes; the maximum of a gross-substitutes valuation minus an additive cost is attained at a selection that no single move---adding an action, dropping an action, or exchanging one action for another---can improve~\cite{leme2017gross}. Thus $S_i$ is a best response at share $a$ precisely when no such move increases agent $i$'s utility, a system of inequalities each affine in $a$. The shares at which $S_i$ best-responds therefore form an interval whose lower endpoint is $\alpha_i$; there some move is exactly indifferent and turns strictly profitable as the share falls below $\alpha_i$. Since lowering the share only makes agent $i$'s actions less rewarding, this tight move drops one action $j \in S_i$, possibly moving a single other action into its place (one already present in $S$, or one it adds), lowering both reward and cost. Writing $S'$ for the resulting profile, it is indifferent at $\alpha_i$:
    \[
    \alpha_i \bigl[\, f(S) - f(S') \,\bigr] = c(S_i) - c(S'_i).
    \]

    It remains to evaluate the reward drop $f(S) - f(S')$ and the cost saved $c(S_i) - c(S'_i)$. Dropping one action changes the reward of a weighted matroid rank function by a two-weight difference, $f(S) - f(S') = w_j - w_r$, as $j$'s place in a maximum-weight independent subset of $S$ is taken by at most one other action $r$ ($w_r = 0$ if none). If $j$'s place stays empty, the cost saved is $c_j$ and $\alpha_i = \tfrac{c_j}{w_j}$. If $r \in S_{-i}$ is an opponent's action---which agent $i$ does not pay for---the cost saved is still $c_j$, so $\alpha_i = \tfrac{c_j}{w_j - w_r}$ with $r \in T \setminus T_i$; this is the case in which dropping $j$ lets an opponent's action back into the reward. If $r \in T_i$ is one of agent $i$'s own actions, the cost saved is $c_j - c_r$ and $\alpha_i = \tfrac{c_j - c_r}{w_j - w_r}$, where $r$ is either an action the move newly adds, or---when $S_i \uplus S_{-i}$ is not independent---one already in $S_i$ that is redundant, contributing nothing to the reward and hence, $S_i$ being a best response, free, so that $c_r = 0$. In every case $\alpha_i$ belongs to the stated set. 
    
    Suppose the binding move drops $j$ and adds $k$. If dropping $j$ alone is binding, the drop case already applies. Otherwise, dropping $j$ alone strictly lowers utility, and adding $k$ alone cannot raise it. These two profiles have the same total cost as $S$ and $S'$, which give the same utility, so
\begin{equation*}
f(S\setminus\{j\})+f(S\cup\{k\})<f(S)+f(S').
\end{equation*}
Let $I$ and $I'$ attain $f(S)$ and $f(S')$, respectively. Since dropping $j$ lowers utility, $j\in I$. By symmetric exchange, we can move $j$ from $I$ to $I'$, possibly moving one action back. That action must be $k$. Otherwise, the resulting independent sets would lie in $S\setminus\{j\}$ and $S\cup\{k\}$ and still have total weight $f(S)+f(S')$, contradicting the displayed inequality. Exchanging $j$ and $k$ therefore gives independent sets in $S'$ and $S$ with the same total weight as before. Both must attain their respective maxima, so $f(S)-f(S')=w_j-w_k$.
\end{proof}

The second fact concerns the equilibria themselves: it suffices to search over \emph{independent} profiles, since a principal-optimal equilibrium can always be taken independent.

\begin{observation} [Independent optimal equilibrium] \label{lem:uniform-independent}
    When $f$ is a weighted matroid rank function (not necessarily uniform), for any contract $\contract$ and every equilibrium $S\in \nash(\contract)$, any maximum-weight independent subset $S'\subseteq S$---an independent $S'\subseteq S$ with $f(S') = f(S)$---is again an equilibrium and satisfies $u_P(\contract, S') = u_P(\contract, S)$. In particular, there is a principal-optimal equilibrium that is independent.
\end{observation}

The actions in $S \setminus S'$ do not add to the reward (already $f(S') = f(S)$), so in an equilibrium they must be costless---otherwise their owners would simply drop them. Removing them therefore leaves every agent's utility unchanged and opens no new profitable deviation, so $S'$ is again an equilibrium with $u_P(\contract, S') = u_P(\contract, S)$.

\subsection{Proof of \Cref{lem:eq-forests}} \label{app:eq-forests}
\begin{proof}
    Recall $u_i(\contract,S)=\alpha_i f(S)-c(S\cap E_i)$, where $f(S)$ is the maximum weight of a forest in $S$. Every equilibrium is a forest: if $S$ contained a cycle, its owner could delete a minimum-weight edge $e$ of that cycle, leaving the reward $f(S)$ unchanged---a maximum-weight forest never needs the lightest edge of a cycle---while saving the cost $c(e)>0$, a profitable deviation. We identify the equilibria one agent at a time.

    Fix an agent $i$ and its opponent's edges $S_{-i}$, and let $\Omega_i$ be the edge-weight function on $S_{-i}\cup E_i$ that assigns each opponent edge $e\in S_{-i}$ its full weight, $\Omega_i(e)=w(e)$, and each of agent $i$'s edges $e\in E_i$ its utility weight, $\Omega_i(e)=\omega_i(e)$; for a set of edges $Y$ we write $\Omega_i(Y)=\sum_{e\in Y}\Omega_i(e)$. For any $S_i\subseteq E_i$, if $Y$ is a maximum-weight forest of $S_i\cup S_{-i}$ then
    \[
    \tfrac1{\alpha_i}\,u_i(\contract,S_i\cup S_{-i})
    = f(S_i\cup S_{-i})-\!\sum_{e\in S_i}\!\tfrac1{\alpha_i}\, c(e)
    = w(Y)-\!\sum_{e\in S_i}\!\tfrac1{\alpha_i}\, c(e)
    \le \Omega_i(Y),
    \]
    the inequality merely discarding the cost of edges of $S_i$ that $Y$ leaves unused. Maximizing over $S_i$, the right-hand side is at most the greatest $\Omega_i$-weight of a forest of $S_{-i}\cup E_i$, a bound attained when $S_i$ is the restriction to $E_i$ of a maximum-$\Omega_i$ forest; so agent $i$ best-responds if and only if $S$ is a maximum-$\Omega_i$ forest of $S_{-i}\cup E_i$.

    These forests have a simple shape. The ground set $S_{-i}\cup E_i$ consists of the main component---which agent $i$'s own edges already span, since the red edges join $s$ to every element and set vertex, as do the blue---together with the two isolated heavy edges, and every non-heavy utility weight is positive by \eqref{eq:reduced-weights}. A maximum-$\Omega_i$ forest is thus a maximum-$\Omega_i$ spanning tree $T$ of the main component plus each isolated heavy edge of positive weight, so, as each agent controls its own heavy edge, the equilibrium contains $h^R$ when $\alpha_1>\frac14$ and $h^B$ when $\alpha_2>\frac14$, while at $\alpha_i=\frac14$ the owner's heavy edge has utility weight $0$ and may be freely included or omitted.

    Such a $T$ is $\Omega_i$-maximum precisely when no non-tree edge is $\Omega_i$-heavier than the lightest edge on the tree-path joining its endpoints. Since every opponent edge has full weight at least $16$ while every agent-$i$ edge in the main component has utility weight below $16$ by \eqref{eq:reduced-weights}, a maximum-$\Omega_i$ forest contains all of $S_{-i}$: indeed, if some opponent edge in the main component were omitted, adding it to $T$ would create a cycle, and because $S_{-i}$ is acyclic this cycle would contain an agent-$i$ edge; replacing such an agent-$i$ edge by the omitted opponent edge would strictly increase the $\Omega_i$-weight. Isolated opponent heavy edges, if present in $S_{-i}$, are included as well since they have positive weight. Hence the only non-tree edges are agent $i$'s own $E_i\setminus T$, and no opponent edge on the tree-path of such an edge can be its lightest. The criterion therefore reduces to the form stated in the lemma: no non-tree edge of agent $i$ is $\omega_i$-heavier than the lightest agent-$i$ edge on its tree-path, while a tree-path with no agent-$i$ edge imposes no constraint.
\end{proof}

\subsection{Proof of \Cref{lem:eq-to-partition}} \label{app:eq-to-partition}
\begin{proof}
    By \Cref{lem:eq-forests}, $S$ is a forest whose main part is a tree $T$ spanning the main vertices. We first locate the assignment edges. The parallel edges $a_i^R$ and $a_i^B$ cannot both lie in $T$, since together they form a cycle; nor can both be absent, for then the tree-edge of $T$ at the element vertex $v_i$ would be a member edge---say a red one, of utility weight $\omega_1(\text{member})$, on the path from $s$ to $v_i$---and the non-tree red assignment edge $a_i^R$, of weight $\omega_1(\text{assignment})>\omega_1(\text{member})$ by \eqref{eq:reduced-weights}, would be $\omega_1$-heavier than the lightest agent-$1$ edge on its tree-path, which \Cref{lem:eq-forests} forbids at an equilibrium (the blue-member case is symmetric, via agent $2$). So each element vertex $v_i$ carries exactly one assignment edge, joining it to the root $s$, which is what defines $U_1$ and $U_2$.

    The same criterion pins down the set vertices. The red set vertex $r_j$ is incident only to the red member edges $m^R_{i,j}$ ($i\in F_j$) and the blue default edge $d^B_j$, and $\deg_T(r_j)=1$, since two tree edges at $r_j$ would close a cycle together with the assignment edges attaching the relevant element vertices to $s$. If its tree-edge were a red member edge $m^R_{i,j}$ with $i\in U_2$, the path from $s$ to $r_j$ would be $s\,\text{--}\,v_i\,\text{--}\,r_j$, whose only agent-$2$ edge is the blue assignment edge $a_i^B$ of weight $\omega_2(\text{assignment})$ (the red member edge is agent $1$'s and does not count); the non-tree blue default edge $d^B_j$, of weight $\omega_2(\text{default})>\omega_2(\text{assignment})$ by \eqref{eq:reduced-weights}, would then be $\omega_2$-heavier than $a_i^B$ on its tree-path, again forbidden by \Cref{lem:eq-forests}. Hence $r_j$'s member edge has $i\in U_1\cap F_j$, and when $U_1\cap F_j=\emptyset$ no such edge exists and $r_j$ is attached instead by the blue default edge $d^B_j$; the blue set vertex $b_j$ is symmetric.

    It remains to bound the reward. Since $S$ is a forest, $f(S)$ is the total weight of its edges. The two heavy edges contribute at most $2\cdot1000\ell=2000\ell$ and the $k$ assignment edges exactly $32k$, while each set vertex is attached by a single edge, of weight $24$ if it is a member edge and $16$ if a default edge. As $r_j$ can use a member edge only when $U_1\cap F_j\neq\emptyset$ and $b_j$ only when $U_2\cap F_j\neq\emptyset$, a split set contributes at most $24+24=48$ and a non-split set at most $24+16=40$ (the side not meeting $F_j$ is forced onto a default edge). Summing over the $\ell$ sets gives at most $48x+40(\ell-x)=40\ell+8x$, so altogether $f(S)\le 2000\ell+32k+40\ell+8x=2040\ell+32k+8x$.
\end{proof}

\subsection{Proof of \Cref{lem:partition-reward}} \label{app:partition-reward}
\begin{proof}
    The main part of $S(U_1,U_2)$ has $k+2\ell$ edges --- one assignment edge per element vertex and one connector per set vertex --- on the $k+2\ell+1$ main vertices, and it is connected, hence a spanning tree $T$; the two heavy edges are isolated and both present, as \Cref{lem:eq-forests} requires when $\alpha_i>\frac14$ and permits when $\alpha_i=\frac14$ (where $\omega_i(\text{heavy})=0$). By that lemma, $S(U_1,U_2)$ is then an equilibrium exactly when it meets the exchange criterion: for each agent, no non-tree edge is $\omega_i$-heavier than the lightest same-agent edge on its tree-path. We check this for agent $1$; agent $2$ is symmetric. Agent $1$'s non-tree red edges are of three kinds. A non-tree red assignment edge $a_i^R$ ($i\in U_2$) has as its tree-path the single blue assignment edge $a_i^B$; and a non-tree red default edge $d^R_j$---off the tree exactly when $U_2\cap F_j\neq\emptyset$---reaches $b_j$ through a blue assignment and a blue member edge. Both tree-paths consist entirely of agent-$2$ edges, so neither kind constrains agent $1$. The remaining kind is a non-tree red member edge $m^R_{i,j}$, of weight $\omega_1(\text{member})$; every agent-$1$ edge on its tree-path is an assignment or member edge, hence of weight at least $\omega_1(\text{member})$ by \eqref{eq:reduced-weights}, so the criterion holds here too. Thus $S(U_1,U_2)$ is an equilibrium.

    Since $S(U_1,U_2)$ is a forest, its reward is the total weight of its edges, which by construction realizes the bound of \Cref{lem:eq-to-partition} with equality: both heavy edges contribute $2000\ell$, the $k$ assignments $32k$, and each set vertex is attached by a member edge (weight $24$) exactly when its side meets $F_j$ and by a default edge (weight $16$) otherwise, so a split set contributes $48$ and a non-split set $40$, giving $48x+40(\ell-x)=40\ell+8x$ over the $\ell$ sets. Summing, $f(S(U_1,U_2))=2000\ell+32k+40\ell+8x=2040\ell+32k+8x$, so $S(U_1,U_2)$ is reward-maximal among the equilibria inducing $(U_1,U_2)$.
\end{proof}

\subsection{Proof of \Cref{lem:opt-contract}} \label{app:opt-contract}
\begin{proof}
    Let $U_1^\star, U_2^\star$ be a partition of $U$ that splits the most sets in $\F$, and let $x^\star$ be the number of sets split by this partition, i.e., an optimal solution to the original $\maxethree$ problem instance and its value. Recall that we assume every element of $U$ lies in some set of $\F$, so that $k\le 3\ell$. By \Cref{lem:partition-reward}, $S(U_1^\star, U_2^\star)\in \nash((1/4, 1/4))$, and furthermore 
    \[
    u_P((1/4, 1/4), S(U_1^\star, U_2^\star)) = (1-1/4-1/4)f(S(U_1^\star, U_2^\star)) = \frac{1}{2}(2040\ell + 32k + 8x^\star).
    \]

    By \Cref{lem:eq-to-partition}, any equilibrium $S$ at an in-box contract $\contract\in[\tfrac14,\tfrac14+\varepsilon]^2$ has reward $f(S)\le 2040\ell+32k+8x\le f(S(U_1^\star,U_2^\star))$, where $x\le x^\star$ is the number of sets split by the induced partition. Since $\alpha_1,\alpha_2\ge\tfrac14$ there, the principal keeps a share $1-\alpha_1-\alpha_2\le\tfrac12$, so $u_P(\contract,S)\le\tfrac12 f(S(U_1^\star,U_2^\star))=u_P((1/4,1/4),S(U_1^\star,U_2^\star))$, with equality only at $\contract=(1/4,1/4)$. Thus $(1/4,1/4)$ is the unique optimum inside the box; it remains to rule out contracts outside it.

    Let $(\contract, S)$ be a $\tfrac{20}{19}$-approximately optimal contract-equilibrium pair. This implies
    \[
    u_P(\contract,S) \ge \tfrac{19}{20} \cdot u_P((1/4, 1/4), S(U_1^\star, U_2^\star))= \tfrac{19}{20} \cdot \tfrac12(2040\ell+32k+8x^\star)\ge \tfrac{19}{20}\cdot 1020\ell = 969\ell.
    \]

    Assume towards contradiction that $\contract \notin [\tfrac14,\tfrac14+\varepsilon]^2$.
    Consider a maximum-weight forest contained in \(S\) after deleting the heavy edges. Extend this weighted forest, within the full main graph, to a spanning tree of the main vertices. This can only increase total weight. Root the resulting tree at \(s\). Each of the $k$ element vertices then has a parent edge of weight at most $32$ (an assignment or member edge) and each of the $2\ell$ set vertices a parent edge of weight at most $24$ (a member or default edge), so the main component contributes reward at most $32k+48\ell\le 144\ell$ (as $k\le 3\ell$). Moreover each heavy edge is isolated, so an agent with share below $\tfrac14$ drops it in every equilibrium, its net utility $\alpha_i\cdot 1000\ell-250\ell$ being negative. We now consider three cases for $\contract$.
    
    If $\alpha_1, \alpha_2 < 1/4$, then $S$ contains no heavy edges, so $u_P(\contract, S) \le f(S) \le 144\ell < 969\ell$, a contradiction.

    If exactly one of $\alpha_1, \alpha_2$ is less than $1/4$, then $S$ contains at most one heavy edge, so $f(S)\le 144\ell+1000\ell$ and, as $1-\alpha_1-\alpha_2\le 1-1/4=3/4$, $u_P(\contract, S) \le \tfrac34(1144\ell)=858\ell < 969\ell$, a contradiction.

    If both $\alpha_1, \alpha_2 \ge 1/4$ and at least one exceeds $1/4+\varepsilon$, then $1-\alpha_1-\alpha_2 < 1-\tfrac14-(\tfrac14+\varepsilon)=\tfrac9{20}$; both heavy edges may belong to $S$, so $f(S)\le 144\ell+2000\ell=2144\ell$ and $u_P(\contract, S) < \tfrac9{20}(2144\ell) = 964.8\ell < 969\ell$, a contradiction.

    This concludes the proof.
\end{proof}

\subsection{Poly-Time Optimal Equilibria under Weighted Partition Matroid Rank Functions}\label{sec:partition}

This subsection proves the tractability of the Equilibrium Selection Problem under weighted partition matroid rank rewards, the special case of weighted matroid rank functions in which the ground set is partitioned into blocks with individual capacities, and a set is independent when it respects every block's capacity. In sharp contrast to the graphic-matroid hardness of \Cref{thm:no-ptas}, here the equilibrium selection problem is solved \emph{exactly} in polynomial time, for a fixed feasible contract and for \emph{any} number of agents; and as a corollary the optimal contract problem is solved in polynomial time for a constant number of agents.

\begin{definition} [Partition Matroid] \label{def:partition_matroid}
    Let the ground set $T$ be partitioned into blocks $T = P_1 \uplus \dots \uplus P_p$, and let $d_1,\dots,d_p$ be nonnegative integer capacities. The \emph{partition matroid} is the matroid $\mathcal{M} = (T, \I)$ whose independent sets respect every block's capacity,
    \[
    \I = \{S \subseteq T : |S \cap P_r| \le d_r \text{ for all } r \in [p]\}.
    \]
    Given nonnegative weights $\{w_j\}_{j \in T}$, the associated \emph{weighted partition-matroid rank function} $f : 2^T \to \reals_{\ge 0}$ decomposes over the blocks,
    \[
    f(S) = \sum_{r \in [p]} f_r(S \cap P_r),
    \qquad
    f_r(X) = \max_{X' \subseteq X,\, |X'| \le d_r} \sum_{j \in X'} w_j ,
    \]
    where $f_r(X)$ is the sum of the $\min(|X|, d_r)$ largest weights in $X$, i.e.\ the weighted rank of the uniform matroid of rank $d_r$ on the block $P_r$.
\end{definition}

Note that the blocks $P_r$ are unrelated to the agents' action sets $T_i$: a block may contain actions of several agents, and an agent's actions may be spread across many blocks.

We prove the tractability of the Equilibrium Selection Problem in two steps. First, we show that the game separates across the blocks of the partition, so that computing an optimal equilibrium reduces to solving one \emph{uniform-matroid} instance per block (\Cref{sec:partition-reduction}). Second, we give a polynomial-time algorithm for a single uniform matroid (\Cref{sec:partition-uniform}).

\subsubsection{Reduction to Uniform Matroids}\label{sec:partition-reduction}

Fix a contract $\contract$. For a block $P_r$, let the \emph{induced game on $P_r$} be the game in which each agent $i$ chooses a subset of $T_i \cap P_r$ and the reward is $f_r$.

\begin{observation} [Separability] \label{obs:partition-separable}
    The game decomposes across blocks: $S \in \nash(\contract)$ if and only if $S \cap P_r$ is an equilibrium of the induced game on every block $P_r$. Consequently a principal-optimal equilibrium is obtained by taking a principal-optimal equilibrium of each block independently and forming their union.
\end{observation}

This holds because the reward and the costs are both additive across blocks. Indeed, $f(S) = \sum_r f_r(S \cap P_r)$ gives
\[
u_i(\contract, S) = \sum_{r} \bigl( \icontract \, f_r(S \cap P_r) - c(S_i \cap P_r) \bigr),
\]
in which agent $i$'s choice inside block $P_r$ affects only the $r$-th summand; hence no agent can profitably deviate if and only if none can within any single block. Likewise $u_P(\contract, S) = \bigl(1 - \sum_i \icontract\bigr) \sum_r f_r(S \cap P_r)$ is a fixed scalar times a sum of per-block rewards, so it is maximized by choosing, in each block independently, an equilibrium that is principal-optimal there.

By \Cref{obs:partition-separable} it suffices to solve a single block, i.e.\ a uniform matroid, we next describe how to do this.

\subsubsection{Optimal Equilibria for Weighted Uniform Matroid Functions}\label{sec:partition-uniform}

We now show how to solve the equilibrium selection problem when $f$ is a weighted uniform matroid rank function of rank $d$; by \Cref{obs:partition-separable}, applying this within each block solves the partition case. The reward $f(S)$ is then the sum of the $\min(|S|, d)$ largest weights in $S$. Fix a feasible contract $\contract$. If $d=0$, the empty profile is optimal, so assume $d\ge1$. For agent $i$ such that\footnote{For an agent with $\alpha_i=0$, define $\omega_i(j)=0$ when $c_j=0$, and $\omega_i(j)=-\infty$ otherwise. Such an agent may take any subset of his zero-cost actions as a best response. All lemmas below also hold for this special case, but we usually omit it for simplicity of presentation.} $\alpha_i > 0$ and action $j \in T_i$ we write
\[
\omega_i(j) = w_j - \frac{1}{\icontract}\, c_j
\]
for the \emph{utility weight} of $j$. Scaling agent $i$'s utility by $\tfrac1{\icontract}$ gives $\tfrac1{\icontract}\,u_i(\contract, S) = f(S) - \sum_{j \in S_i} \tfrac1{\icontract}\,c_j$, so on an independent profile it counts each of agent $i$'s selected actions at its utility weight. Here a profile $S$ is \emph{independent} if $|S| \le d$, so that $f(S) = \sum_{j \in S} w_j$ and every selected action is counted. 
\begin{proposition} [Uniform-matroid algorithm] \label{prop:uniform-algo}
    The reward-maximizing equilibrium under a weighted uniform matroid rank function of rank $d$ can be computed in time $O(m^2 d)$.
\end{proposition}

We prove \Cref{prop:uniform-algo} through two structural facts. The first is general: by the independent-optimal-equilibrium observation (\Cref{lem:uniform-independent}) it suffices to search over independent profiles. The second is specific to the uniform matroid: we characterize the independent equilibria in terms of the utility weights and the lightest selected weight.

\begin{lemma} [Equilibrium characterization] \label{lem:uniform-char}
    An independent profile $S$ with $|S| = s \le d$ and, when $s=d$, lightest selected weight $\theta = \min_{g \in S} w_g$ is an equilibrium if and only if, for every agent $i$:
    \begin{enumerate}
        \item \emph{(no drop)} $\omega_i(j) \ge 0$ for every selected $j \in S_i$;
        \item \emph{(prefix)} $\omega_i(j') \le \omega_i(j)$ for every selected $j \in S_i$ and unselected $j' \in T_i \setminus S$; equivalently, ordering agent $i$'s actions by decreasing utility weight, $S_i$ is a prefix of this order (a tie in utility weight may be ordered so that the selected actions come first);
        \item \emph{(no add)} for every unselected $j \in T_i \setminus S$,
        \[
        \omega_i(j) \le 0 \quad\text{if } s < d,
        \qquad\text{and}\qquad
        \omega_i(j) \le \theta \quad\text{if } s = d.
        \]
    \end{enumerate}
\end{lemma}
A weighted matroid rank reward minus additive costs is maximized greedily, so an agent best-responds exactly when no single drop, swap, or add improves his utility. Evaluating these three moves on an independent profile yields conditions~1--3. See the proof in \Cref{app:uniform-char}.

The algorithm rests on a simple picture. Since each agent plays a prefix of his actions in utility-weight order, an equilibrium is fixed once we decide how many actions each agent takes, and a dynamic program can search these counts for the profile of largest reward. The one thing standing in the way of a purely per-agent computation is the full-capacity case, where the agents are coupled through the lightest selected weight $\theta$; enumerating the possible values of $\theta$ removes this coupling. This is the algorithm of \Cref{prop:uniform-algo}, which we now prove.

\begin{proof}[Proof of \Cref{prop:uniform-algo}]
    By \Cref{lem:uniform-independent} an optimal equilibrium may be taken independent, and by \Cref{lem:uniform-char} the independent equilibria are exactly the profiles in which each agent selects a prefix of his actions in utility-weight order (condition~2) meeting conditions~1 and~3. Sort each agent's actions in $T_i$ by $(\omega_i(\cdot)$ descending, $w$ descending$)$, so that a prefix length $k_i$ per agent determines his selected set and its weight. The independent equilibria split into two cases, according to whether the profile leaves capacity to spare or fills it.

    When $s < d$, every agent takes all his actions of non-negative utility weight (condition~3); if the resulting size is at most $d$, all unselected actions have negative utility weight, so the profile is an equilibrium, giving a single candidate. When $s = d$, condition~3 instead reads $\omega_i(j) \le \theta$, whose only cross-agent term is the lightest selected weight $\theta$, so we enumerate $\theta$ over the at most $m$ distinct weights. Fixing $\theta$ makes conditions 1--3 per-agent: agent $i$'s admissible prefix lengths $k$ are those for which the top $k$ actions all have weight at least $\theta$ and non-negative utility weight, and which include every action with $\omega_i(j) > \theta$. It remains to pick one admissible prefix per agent with lengths summing to exactly $d$ and at least one selected action of weight exactly $\theta$, maximizing the total weight; this is the dynamic program
    \[
    \mathrm{DP}[i, t] = \max_{k \text{ admissible for } i} \Bigl( \mathrm{DP}[i-1,\, t-k] + w\bigl(\text{agent } i\text{'s length-}k \text{ prefix}\bigr) \Bigr),
    \]
    over agents $i$ and counts $t = 0, \dots, d$ (with an extra bit tracking whether a weight-$\theta$ action has been chosen), returning the maximum-weight admissible profile at $t = d$.

    By \Cref{lem:uniform-char} every profile produced is an equilibrium, and every independent equilibrium with spare capacity is weakly dominated in reward by the maximal nonnegative-utility profile (or by a full-capacity extension), while the full-capacity cases are covered by the threshold enumeration and the reward-maximizing tie-breaking; by \Cref{lem:uniform-independent} restricting to independent profiles is without loss of generality, so the best candidate is optimal. There are $O(m)$ thresholds of $\theta$ to check. For a fixed $\theta$ the table has $O(nd)$ cells, but its complexity is better counted per agent: computing the row $\mathrm{DP}[i,\cdot]$ over all counts $t \le d$ takes $O(|T_i|\,d)$ time, one update per pair of a target count $t \le d$ and one of agent $i$'s at most $|T_i|+1$ prefix lengths. Summing over agents, $\sum_i |T_i|\,d = md$ since $\sum_i |T_i| = m$, so each threshold is handled in $O(md)$, for a total of $O(m^2 d)$.
\end{proof}
\subsection{Proof of \Cref{lem:uniform-char}} \label{app:uniform-char}
\begin{proof}
    If $\alpha_i=0$, best responses are exactly the subsets of zero-cost actions, as encoded by the convention for $\omega_i$. Now fix an agent $i$ with $\alpha_i>0$ and the remaining agents' actions $S_{-i}$. Up to the positive factor $\icontract$, agent $i$ best-responds by choosing $X \subseteq T_i$ to maximize $f(X \cup S_{-i}) - \sum_{j \in X} \tfrac1{\icontract} c_j$, a weighted matroid rank function---that of the uniform matroid with $S_{-i}$ contracted---minus an additive cost. Such a function is maximized greedily, so local optimality already implies global optimality: $X$ is a best response if and only if no single insertion, deletion, or one-for-one exchange improves it\cite{leme2017gross}. It therefore suffices to rule out a profitable drop, swap, or add for every agent. We check each move on the independent profile $S$, where $f(S) = \sum_{j \in S} w_j$, measuring agent $i$'s gain in the scaled utility $\tfrac1{\icontract} u_i$; there each selected action of his contributes its utility weight and the reward contributes weights in full.

    Dropping a selected $j \in S_i$ costs $w_j$ in reward but refunds the scaled cost $\tfrac1{\icontract} c_j$, a net change of $-\omega_i(j)$, so no drop helps precisely when $\omega_i(j) \ge 0$ for every $j \in S_i$---condition~1. Swapping a selected $j$ for an unselected $j' \in T_i$ leaves the profile size at $s$, so both actions remain counted and the scaled utility shifts by $\omega_i(j') - \omega_i(j)$; no swap helps precisely when every selected action has utility weight at least that of every unselected one---condition~2.

    Only additions feel the capacity. When $s < d$ there is room to spare, so adding an unselected $j$ gains $\omega_i(j)$ and is unprofitable exactly when $\omega_i(j) \le 0$. When $s = d$ the profile is full, and adding $j$ evicts the lightest counted action, of weight $\theta$: if $w_j > \theta$ the reward rises by $w_j - \theta$ and the scaled utility by $\omega_i(j) - \theta$, whereas if $w_j \le \theta$ nothing is displaced and the move merely wastes $\tfrac1{\icontract} c_j$. Either way the addition is unprofitable exactly when $\omega_i(j) \le \theta$ for every unselected $j$, an inequality that holds automatically once $w_j \le \theta$ since then $\omega_i(j) \le w_j \le \theta$. These two cases give the two clauses of condition~3. As a drop, a swap, or an add is the only deviation an agent has, conditions 1--3 hold for all agents if and only if $S$ is an equilibrium.
\end{proof}

\subsection{Poly-Time Optimal Equilibria under Unweighted Matroid Rank Functions}\label{sec:unweighted}

The hardness of \Cref{thm:no-ptas} rests on the weights: they are what let a single graphic-matroid instance encode a set-splitting problem. Without them the difficulty disappears. We show that for an \emph{unweighted} matroid rank function---the rank function of any matroid, which gives every action unit weight---the equilibrium selection problem is solved exactly in polynomial time, for a fixed feasible contract and \emph{any} number of agents; and, as with partition matroids, the optimal contract problem then follows in polynomial time when the number of agents is constant. This locates the source of hardness in the interaction of the weights with the matroid structure, rather than in the matroid structure alone.

\begin{definition} [Unweighted Matroid Rank Function] \label{def:unweighted}
    Given a matroid $\mathcal{M} = (T, \I)$, the associated \emph{unweighted matroid rank function} $f : 2^T \to \reals_{\ge 0}$ maps each set to the size of the largest independent subset it contains,
    \[
    f(S) = \max_{S' \subseteq S,\, S' \in \I} |S'| = r(S),
    \]
    the \emph{rank} of $S$. We write $\rho = r(T)$ for the rank of the matroid.
\end{definition}

This is the same as the WMRF definition (see \Cref{sec:model}) with all weights equal to one. If $\rho=0$, the empty profile is optimal. For $\rho>0$, dividing $f$ and the costs by $\rho$ places the reward in $[0,1]$ and scales every utility by $\tfrac1\rho$, so we may work with unit weights directly.

Fix a contract $\contract$ with $\sum_i\alpha_i\leq1$. The principal's utility $u_P(\contract, S) = (1 - \sum_i \alpha_i)\,f(S)$ depends on the profile only through the reward $f(S) = r(S)$, so a principal-optimal equilibrium is simply one of largest rank: which agent takes which action is immaterial, only how many independent actions are taken. Call an action \emph{affordable} to its owner when his share covers its cost, $c_j \le \alpha_i$ for $j \in T_i$, and let $A$ be the set of affordable actions.

The optimal reward is exactly $r(A)$. On one hand, no equilibrium can exceed it: because the matroid is unweighted, any single action contributes at most one unit of reward, so an action costing its owner more than his share, $c_j > \alpha_i$, earns him at most $\alpha_i < c_j$ and he would rather drop it; hence every equilibrium $S$ uses only affordable actions, $S \subseteq A$, and $r(S) \le r(A)$. On the other hand, $r(A)$ is attained by a minimum-cost basis of $A$---a maximum independent subset of the affordable actions of least total cost, produced by the matroid greedy. Its rank is $r(A)$; every action in it is affordable, so no owner wishes to drop one; it is a maximum independent set of $A$, so no owner can afford an action that raises the rank; and its minimum cost leaves no owner a cheaper affordable action to substitute. By the local-to-global optimality of a matroid rank reward minus additive costs \cite{leme2017gross}, each agent best-responds, so this profile is an equilibrium, of reward $r(A)$.

A principal-optimal equilibrium is therefore computed by a single greedy pass over the affordable actions, thus proving tractability of the Equilibrium Selection Problem.

\section{The Gross-Substitutes Hard Family}\label{app:gs-construction}

\subsection{Construction of the hard family}

Fix $k\ge32$, and set $\beta\coloneqq1/3$. Let $E\coloneqq\{a_1,\ldots,a_{2k}\}$. We construct one instance for each set $H$ consisting of $a_1$ and $k-1$ other actions from $E$. There are $N\coloneqq\binom{2k-1}{k-1}$ such sets.

We order these sets lexicographically and denote them by $H_0,\ldots,H_{N-1}$ and define $\theta_j\coloneqq(j+1)/(N+1)$ and $t_j\coloneqq1/(k^2+1+\theta_j)$. Then $0<\theta_j<1$, the shares $t_j$ are pairwise distinct, and both $\theta_j$ and $t_j$ have $O(k)$-bit size. We will construct the reward function and costs such that the instance indexed by $H_j$ will have $(\beta,t_j)$ as its unique optimal contract. When the index is clear from the context, we write $H=H_j$, $\theta=\theta_j$, and $t=t_j$.

The agents' action sets are $T_1\coloneqq E\cup\{a_0\}$ and $T_2\coloneqq\{b_1,b_2\}$, so the total number of actions is $m=2k+3$. Given a profile $S\subseteq T$, we also use $a_0,b_1,b_2$ for their membership indicators, for example, $a_0=1$ means $a_0\in S$. 

\paragraph{Assignment graph.}
For each set $H$, we first construct a corresponding assignment valuation  function. It is defined by a weighted bipartite graph $G_H$, with the actions in $T$ on one side and a set of positions on the other. There are $k$ identical bulk positions $u_1,\ldots,u_k$, two additional positions $v_1,v_2$, and a shared position $\rho$.  The graph and its edge weights are shown in \Cref{fig:gs-core-graph}. In the table, the column $u_s$ gives the weights for each of the $k$ bulk positions, and ``$-$'' denotes the absence of an edge. Let $g_H(S)$ denote the maximum matching weight between actions in $S$ and the positions, vertices may remain unmatched.  This is an assignment valuation and therefore satisfies the gross substitutes property.

\begin{figure}[htbp]
\centering
\begin{minipage}[c]{0.35\linewidth}
\centering
\renewcommand{\arraystretch}{1.15}
\[
\begin{array}{c|cccc}
 &u_s&v_1&v_2&\rho\\
\hline
a_i\in H&4&2-\theta&\theta&-\\
a_i\notin H&4&4&2&-\\
b_1&-&4&2&-\\
a_0&4&-&-&2k^2\\
b_2&-&-&-&2k^2
\end{array}
\]
\end{minipage}\hfill
\begin{minipage}[c]{0.62\linewidth}
\centering
\begin{tikzpicture}[
    x=1cm,y=1cm,
    action/.style={
        circle,draw=black!75,fill=white,
        minimum size=7mm,inner sep=1pt,font=\small
    },
    position/.style={
        rectangle,draw=black!75,fill=black!5,
        minimum size=7mm,inner sep=1pt,font=\small
    },
    edge/.style={draw=black!60,line width=.6pt},
    note/.style={font=\scriptsize,text=black!70}
]

\node[action] (b1) at (0,.9) {$b_1$};
\node[action,minimum size=11mm] (H) at (1.75,.9) {$H$};
\node[action,minimum size=11mm] (Out) at (3.5,.9) {$E\setminus H$};
\node[action] (a0) at (5.35,.9) {$a_0$};
\node[action] (b2) at (7.2,.9) {$b_2$};

\node[position] (v1) at (0,-.9) {$v_1$};
\node[position] (v2) at (2.1,-.9) {$v_2$};
\node[position] (u) at (4.6,-.9) {$u_s$};
\node[position] (rho) at (7.2,-.9) {$\rho$};

\draw[edge] (b1) -- (v1);
\draw[edge] (b1) -- (v2);
\draw[edge] (H) -- (v1);
\draw[edge] (H) -- (v2);
\draw[edge] (H) -- (u);
\draw[edge] (Out) -- (v1);
\draw[edge] (Out) -- (v2);
\draw[edge] (Out) -- (u);
\draw[edge] (a0) -- (u);
\draw[edge] (a0) -- (rho);
\draw[edge] (b2) -- (rho);

\node[note,anchor=east] at (-.65,.9) {actions};
\node[note,anchor=east] at (-.65,-.9) {positions};
\node[note,anchor=south] at (1.75,1.55) {$k$ actions};
\node[note,anchor=south] at (3.5,1.55) {$k$ actions};
\node[note,anchor=north] at (4.6,-1.35) {$k$ positions};
\end{tikzpicture}
\end{minipage}
\caption{The assignment weights and graph $G_H$. The nodes $H$ and $E\setminus H$ represent the actions in and outside the hidden set; $u_s$ represents any of the $k$ identical bulk positions.}
\label{fig:gs-core-graph}
\end{figure}
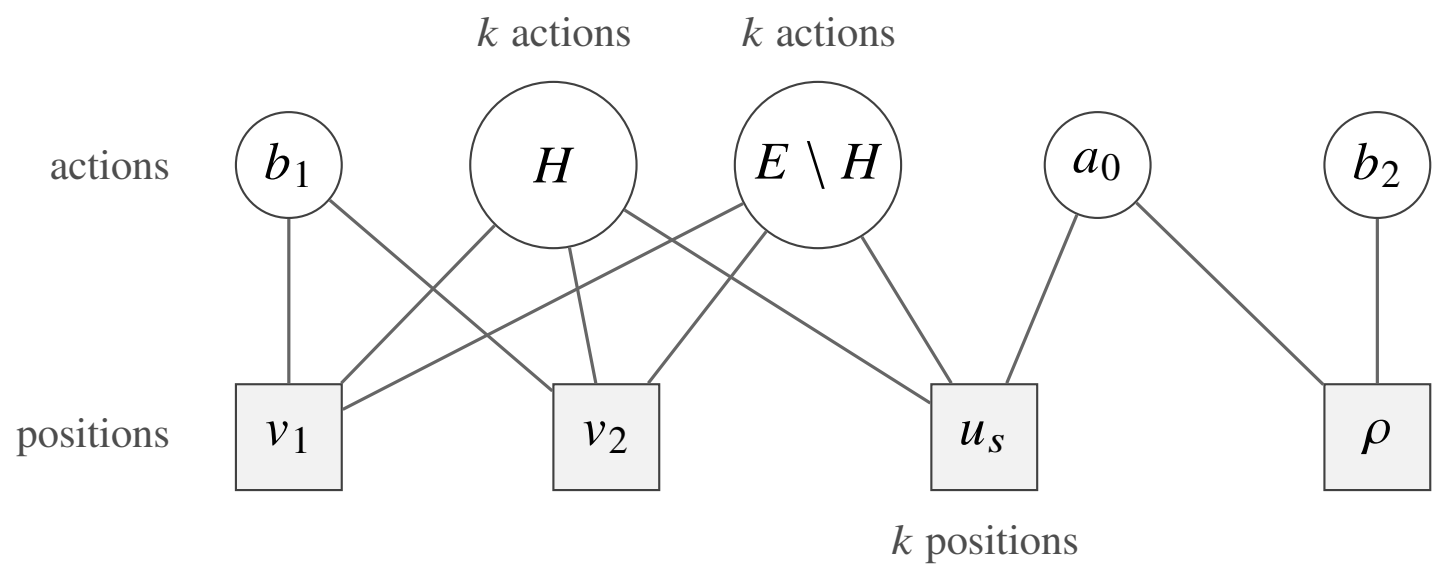

The next lemma provides a closed form expression for $g_H$. Its dependence on $H$ is captured by a single loss term:
\begin{equation}\label{eq:gs-loss}
\ell_H(S)\coloneqq
\begin{cases}
2+\theta,
    & S=H\cup\{a_0,b_2\},\\
2-\theta,
    & H\cup\{a_0,b_2\}\subseteq S,\quad\text{and}\quad
      \bigl|S\setminus(H\cup\{a_0,b_2\})\bigr|=1,\\
0,
    & \text{otherwise}.
\end{cases}
\end{equation}
Thus a loss occurs only when the selected actions are exactly $H\cup\{a_0,b_2\}$, or when this exact set and one additional action are selected together. 

\begin{lemma}\label{lem:gs-core-value}
For every profile $S \subset T $, $\displaystyle g_H(S)=2k^2\max\{a_0,b_2\}+\min\bigl\{4\bigl(|S|-\max\{a_0,b_2\}\bigr),\,4k+6\bigr\}
-\ell_H(S)$
\end{lemma}

\begin{proof}
We do a case disjunction to compute the maximum weight matching. 

If $b_2$ is selected, we can assign it to $\rho$ as it has the same edge weight as $a_0$, also leaving $a_0$ available for a bulk position. If $a_0$ is selected but $b_2$ is not, assigning $a_0$ to $\rho$ is optimal because $2k^2>4$. Thus $\rho$ contributes $2k^2\max\{a_0,b_2\}$, leaving $|S|-\max\{a_0,b_2\}$ actions for the other positions.

Each remaining edge has weight at most $4$. The bulk positions and $v_1$ each contribute at most $4$, while $v_2$ contributes at most $2$. The remaining matching weight is therefore at most $4\bigl(|S|-\max\{a_0,b_2\}\bigr)$ as there are only $|S|-\max\{a_0,b_2\}$ actions remaining, and at most $4k+6$ which is the best you can get if you match all the positions. Overall, the remaining weight is upper bounded by 
\[
\min\bigl\{4\bigl(|S|-\max\{a_0,b_2\}\bigr),\,4k+6\bigr\}.
\]
We now show that this bound is attained except in the two cases defining $\ell_H$.

If at most $k$ actions remain, assign them to bulk positions, except that $b_1$, when selected, is assigned to $v_1$. Every assigned edge has weight $4$, attaining the bound. This always works. 

If exactly $k+1$ actions remain, the bound is attained whenever at least one of them belongs to $(E\setminus H)\cup\{b_1\}$. Assign one such action to $v_1$, choosing $b_1$ whenever it is selected, and assign the others to bulk positions. If not, all the actions are in $H\cup\{a_0\}$, which is of size exactly $k+1$, therefore the remaining actions are exactly $H\cup\{a_0\}$.
 
If at least $k+2$ actions remain and at least two belong to $(E\setminus H)\cup\{b_1\}$, assign two such actions to $v_1,v_2$. Choose $b_1$ for $v_1$ whenever it is selected. At least $k$ other actions remain available to fill the $k$ bulk positions and attain the bound $4k+6$. Otherwise, at most one of the remaining actions belongs to $(E\setminus H)\cup\{b_1\}$. Every other remaining action belongs to $H\cup\{a_0\}$, which has $k+1$ elements. There are therefore at most $k+2$ remaining actions in total. Since we assumed that at least $k+2$ remain, equality must hold. Thus the remaining actions are exactly $H\cup\{a_0\}$ together with one action from $(E\setminus H)\cup\{b_1\}$.

In both exceptional cases, $a_0$ remains after the assignment to $\rho$. Thus $b_2$ must have been selected and assigned to $\rho$. We now compute the matching value for these two profiles.
\begin{itemize}
\item $S=H\cup\{a_0,b_2\}$. The remaining actions are exactly $H\cup\{a_0\}$. Since $2-\theta>\theta$, the best matching is to assign one action of $H$ to $v_1$, then $a_0$ and the rest of $H$ to the bulk positions, yielding $4k+2-\theta$. The loss compared to the original bound $4k+4$ is therefore $2+\theta$.

\item $H\cup\{a_0,b_2\}\subseteq S$ and $\bigl|S\setminus(H\cup\{a_0,b_2\})\bigr|=1$. The additional action belongs to $(E\setminus H)\cup\{b_1\}$. After assigning $b_2$ to $\rho$, there are $k+2$ actions left. Here, the optimal matching is obtained by assigning the extra action to $v_1$, an action of $H$ to $v_2$, and the remaining $k$ actions to bulk positions, yielding $4k+4+\theta$. The loss from the original bound of $4k+6$ is therefore $2-\theta$.
\end{itemize}
These are exactly the losses defining $\ell_H$. Adding the contribution of $\rho$ proves the lemma.
\end{proof}

\paragraph{Reward and costs.}
We define the reward of the instance indexed by $H$ by adding modular bonuses to the matching value:
\begin{equation}\label{eq:gs-reward}
f_H(S)\coloneqq g_H(S)+2k^2b_1+(k^2+1)b_2+4k^2a_0+w(S),
\end{equation}
where  $w(S)\coloneqq|S\cap(E\setminus\{a_1\})|$ is a modular function. Because $f_H$ is simply $g_H$ added to a modular function, it is itself an assignment valuation function \citep{leme2017gross}, and therefore still satisfies the gross substitutes property. 
The action costs are $c_{a_0}=2k^2$, $c_{b_1}=2$, $c_{b_2}=1$, $c_{a_1}=2/3=2 \beta $, and $c_{a_i}=1=2\beta +\beta$ for $i=2,\ldots,2k$. The bundle costs are therefore
\begin{equation}\label{eq:gs-bundle-costs}
c(S\cap T_1)=\beta\bigl(2|S\cap E|+w(S)+6k^2a_0\bigr),\quad 
c(S\cap T_2)=2b_1+b_2.
\end{equation}

\paragraph{Reference instance.}
For comparison, consider the instance obtained by giving every action in $E$ weight $4$ at $v_1$ and weight $2$ at $v_2$ in the corresponding assignment graph, instead of $2-\theta$ and $\theta$. All other edges, bonuses, and costs remain unchanged. Let $f_\star$ denote the associated reward function. This instance does not depend on $H$.

The matching calculation in \Cref{lem:gs-core-value} now attains its upper bound for every profile. Thus the two reward functions differ only by the loss term:
\begin{equation}\label{eq:gs-reference-difference}
f_H(S)=f_\star(S)-\ell_H(S).
\end{equation}
This immediately limits how many instances a value query can distinguish from the reference instance.

\begin{lemma}\label{lem:gs-value-locality}
For every fixed profile $S$, the value $f_H(S)$ differs from $f_\star(S)$ for at most $k+1$ choices of $H$.
\end{lemma}

\begin{proof}
A nonzero loss requires $H\subseteq S\cap E$ and $|S\cap E|\in\{k,k+1\}$. If $|S\cap E|=k$, the only possible choice is $H=S\cap E$. If $|S\cap E|=k+1$, there are at most $k+1$ possible choices of a $k$-element subset $H$. In every other case, the loss is zero.
\end{proof}

We now need for each $H$ to characterize the equilibrium for the associated reward and cost functions. 

\subsection{Equilibrium analysis}

We first bound the change in reward when $a_0$ is added. This will determine when agent~1 can benefit from selecting it. 

\begin{lemma}\label{lem:gs-a0-marginal}
For every $S\subseteq T\setminus\{a_0\}$ and $f\in\{f_H,f_\star\}$, 
\[
f(S\cup\{a_0\})-f(S)=6k^2
    \quad \text{if }b_2\notin S, \qquad 
4k^2\le f(S\cup\{a_0\})-f(S)\le4k^2+4 \quad 
    \text{if }b_2\in S.
\]
\end{lemma}

\begin{proof}
If $b_2\notin S$, we can assign $a_0$ to $\rho$ without changing the matching of the other actions. This is optimal because the edge to $\rho$ has weight $2k^2>4$. The matching value therefore increases by exactly $2k^2$. Adding the bonus $4k^2$ gives the first claim.

If $b_2\in S$, we may keep $b_2$ assigned to $\rho$. Adding $a_0$ cannot decrease the matching value, and removing it from an optimal matching loses at most $4$. Together with its bonus $4k^2$, this gives the second claim.
\end{proof}

We now describe how agent~1's choices depend on its share and compare the equilibria of the two instances.

\begin{lemma}\label{lem:gs-agent1-beta}
Fix the actions chosen by agent~2. For both reward functions $f_H$ and $f_\star$:
\begin{enumerate}[label=(\roman*)]
\item If $\alpha_1<\beta$, every best response omits $a_0$.
\item If $\alpha_1>\beta$, every best response selecting between $1$ and $2k-1$ actions from $E$ omits $a_1$.
\item If $\alpha_1=\beta$, the best responses are exactly the choices satisfying $|S\cap E|+b_1\in\{k+1,k+2\}$ with $a_0$ omitted when $b_2$ is selected, and either selecting or omitting $a_0$  otherwise.
\item For $\alpha_1\ne\beta$, the two instances have the same Nash equilibria, with the same rewards. For $\alpha_1=\beta$, this remains true for profiles omitting $a_0$.
\end{enumerate}
\end{lemma}

\begin{proof}
If $\alpha_1<\beta$, \Cref{lem:gs-a0-marginal} shows that selecting $a_0$ increases agent~1's utility by at most $6k^2\alpha_1-c_{a_0} < 6k^2\cdot \frac{1}{3} -2k^2=  0$, thus every best response omits $a_0$, proving~(i). \medskip

Now suppose $\alpha_1>\beta$. Fix whether $a_0$ is selected and the number of selected actions from $E$. Agent~2's actions are also fixed. By \Cref{lem:gs-core-value}, all terms in $f_H(S)$ other than $w(S)-\ell_H(S)$ are then fixed. Since $c(S\cap T_1)=\beta\bigl(2|S\cap E|+w(S)+6k^2a_0\bigr)$, 
agent~1's utility is
\[
\alpha_1f_H(S)-c(S\cap T_1)
=\text{fixed terms}+(\alpha_1-\beta)w(S)-\alpha_1\ell_H(S).
\]
Thus only the last two terms depend on which actions from $E$ are selected. 
In the reference instance $f_\star$, the loss term is absent. At any fixed size between $1$ and $2k-1$, omitting $a_1$ gives the largest possible value of $w(S)$ and makes $\ell_H(S)=0$, since $a_1\in H$ implies that $a_1 \notin S \implies H\not \subset S \implies \ell_H(S)=0$. At a fixed number of actions, including $a_1$ reduces $w(S)$ by one and cannot make the loss negative. This proves~(ii). \medskip 

Finally, suppose $\alpha_1=\beta$. If $b_2$ is absent, \Cref{lem:gs-a0-marginal} shows that adding $a_0$ changes agent~1's utility by $6k^2\beta-2k^2=0$. If $b_2$ is selected, the change is at most $\beta(4k^2+4)-2k^2=\frac{4-2k^2}{3}<0$. Hence agent~1 is indifferent about selecting $a_0$ when $b_2$ is absent and must omit it when $b_2$ is selected.
After the preceding choice of $a_0$, we have $a_0b_2=0$, so $\max\{a_0,b_2\}=a_0+b_2$, $|S|-\max\{a_0,b_2\}=|S\cap E|+a_0+b_1+b_2-(a_0+b_2)=|S\cap E|+b_1$ and $\ell_H(S)=0$. For either reward function $f$, this gives
\begin{align*}
\frac{\beta f(S)-c(S\cap T_1)}{\beta}
&=f_\star(S)-2|S\cap E|-w(S)-6k^2a_0\\
&=2k^2(a_0+b_2)
  +\min\bigl\{4(|S\cap E|+b_1),\,4k+6\bigr\}+2k^2b_1+(k^2+1)b_2+4k^2a_0+w(S)\\
&\qquad-2|S\cap E|-w(S)-6k^2a_0\\
&=\min\bigl\{4(|S\cap E|+b_1),\,4k+6\bigr\}
  -2|S\cap E|+2k^2b_1+(3k^2+1)b_2.
\end{align*}
The last two terms depend only on agent~2's actions, which are fixed, so they can be dropped from agent~1's objective. Adding an action from $E$ increases this expression by $2$ when $|S\cap E|+b_1\le k$, leaves it unchanged when $|S\cap E|+b_1=k+1$, and decreases it by $2$ thereafter. It is therefore maximized exactly when $|S\cap E|+b_1\in\{k+1,k+2\}$, regardless of which actions from $E$ are selected. This proves~(iii). 
\medskip

We now prove the final assertion about Nash equilibria under $f_H$ and $f_\star$. If $\alpha_1<\beta$, by part~(i), $a_0$ is omitted so $\ell_H(S)=0$. If $\alpha_1 > \beta$, by part~(ii), either $a_1$ is omitted or all of $E$ is selected, in which case $\ell_H(S)=0$. In either case, $\ell_H=0$ on every profile that remains, regardless of agent~2's actions. The rewards and costs therefore agree on all profiles needed to check equilibrium. Hence the two instances have the same Nash equilibria, with the same rewards.

At $\alpha_1=\beta$, \Cref{lem:gs-a0-marginal} shows that removing $a_0$ never decreases agent~1's utility. Thus, for a profile omitting $a_0$, it suffices to consider deviations that also omit $a_0$. On all these profiles, $\ell_H=0$, so the same conclusion applies. This proves~(iv). 
\end{proof}

Consider the profile $S_H\coloneqq H\cup\{a_0,b_1\}$, which we will use to establish the approximation gap. We first identify the contract supporting it.

\begin{lemma}\label{lem:gs-high-share-isolation}
The profile $S_H$ is a Nash equilibrium under $f_H$ if and only if $\alpha=(\beta,t)$. Moreover, it is the only equilibrium under $f_H$ in which agent~1 selects $a_0$ and agent~2 selects only $b_1$. No equilibrium with these choices exists under $f_\star$.
\end{lemma}

\begin{proof}
We first determine when $S_H$ is an equilibrium. At this profile, agent~1 selects $a_0$ and exactly the $k$ actions of $H$, including $a_1$. Parts~(i) and~(ii) of \Cref{lem:gs-agent1-beta} therefore force $\alpha_1=\beta$. Conversely, at this share, $H\cup\{a_0\}$ is a best response against $\{b_1\}$ by part~(iii), since $|H|+1=k+1$ and $b_2$ is absent.

For agent~2, write $S_1=H\cup\{a_0\}$. By \Cref{lem:gs-core-value} and the definition of the bonuses,
\[
f_H(S_1\cup\{b_1\})-f_H(S_1\cup\{b_2\})=k^2+1+\theta, \qquad 
f_H(S_1\cup\{b_1,b_2\})-f_H(S_1\cup\{b_1\})=k^2+1+\theta.
\]
The corresponding cost differences are both $1$. Thus choosing $\{b_1\}$ requires
$\alpha_2(k^2+1+\theta)\ge1$ and
$\alpha_2(k^2+1+\theta)\le1$,
which force $\alpha_2=t$. At this share, all three nonempty choices give agent~2 the same utility. Not playing any actions is suboptimal as 
\[
t\bigl(f_H(S_1\cup\{b_1\})-f_H(S_1)\bigr)-2=t(2k^2+4)-2=2t(1-\theta)>0.
\]
Hence $\{b_1\}$ is a best response exactly when $\alpha_2=t$. Together with the condition for agent~1, this proves that $S_H$ is an equilibrium if and only if $\alpha=(\beta,t)$. \medskip 

We now show that no other equilibrium under $f_H$ has agent~1 selecting $a_0$ and agent~2 selecting only $b_1$. Fix an arbitrary choice $S_1\subseteq T_1$ containing $a_0$, and suppose that $\{b_1\}$ is a best response for agent~2. Necessarily $\alpha_2>0$, since otherwise agent~2 prefers the empty set. The costs of $\{b_2\}$, $\{b_1\}$, and $\{b_1,b_2\}$ are $1,2,3$, respectively. Comparing $\{b_1\}$ with the other two choices and adding the best-response inequalities gives
\begin{equation} \label{eq:best_rep_ineq}
f_H(S_1\cup\{b_1,b_2\})
+f_H(S_1\cup\{b_2\})
-2f_H(S_1\cup\{b_1\})\le0.
\end{equation}

Since $a_0\in S_1$, the contribution of $\rho$ and the bonus for $a_0$ total $6k^2$ in all three profiles. The bonuses for $b_1$ and $b_2$ are $2k^2$ and $k^2+1$, respectively. Also, adding these actions does not change $w(S_1)$, and $\ell_H(S_1\cup\{b_1\})=0$ because $b_2$ is absent. Thus \Cref{lem:gs-core-value} gives
\begin{align*}
&f_H(S_1\cup\{b_1\})
=8k^2+\min\{4|S_1|,\,4k+6\}+w(S_1),\\
&f_H(S_1\cup\{b_2\})
=7k^2+1+\min\{4|S_1|,\,4k+6\}+w(S_1)
-\ell_H(S_1\cup\{b_2\}), \\
&f_H(S_1\cup\{b_1,b_2\})
=9k^2+1+\min\{4(|S_1|+1),\,4k+6\}+w(S_1)
-\ell_H(S_1\cup\{b_1,b_2\}).    
\end{align*}
Substituting these expressions, the terms involving $w(S_1)$ cancel, and the constant terms give $(9k^2+1)+(7k^2+1)-2(8k^2)=2$. The expression on the left of \Cref{eq:best_rep_ineq} therefore equals
\[
2+\min\{4(|S_1|+1),\,4k+6\}
  -\min\{4|S_1|,\,4k+6\}-\ell_H(S_1\cup\{b_2\})
       -\ell_H(S_1\cup\{b_1,b_2\}).
\]
If $S_1=H\cup\{a_0\}$, the difference between the minima is $2$, and the losses sum to $(2+\theta)+(2-\theta)=4$. The expression is therefore zero.
If $S_1$ consists of $H\cup\{a_0\}$ and one additional action from $E\setminus H$, the minima are equal and the losses are $2-\theta$ and $0$. The expression is therefore $\theta>0$.
In every other case, both losses vanish and the difference between the minima is nonnegative, so the expression is at least $2$. The necessary inequality therefore forces $S_1=H\cup\{a_0\}$, giving the profile $S_H$. \medskip 

Finally, under $f_\star$, the same calculation has no loss terms. The expression is then at least $2$ for every $S_1$, contradicting the necessary best-response inequality. Thus no equilibrium under $f_\star$ has agent~1 selecting $a_0$ and agent~2 selecting only $b_1$.
\end{proof}

\subsection{Approximation gap}

We now show that $S_H$ gives the principal a constant-factor advantage over every other equilibrium.

\begin{lemma}\label{lem:gs-approximation-gap}
Under $f_H$, the profile $S_H$ is the unique principal-optimal equilibrium at $(\beta,t)$. Every other contract has principal value strictly less than $63/64$ of the value at $(\beta,t)$. Consequently, $(\beta,t)$ is the unique $64/63$-approximate contract.
\end{lemma}

\begin{proof}
By \Cref{lem:gs-core-value} and the definition of $f_H$,
\[
f_H(S_H)
=2k^2+(4k+4)+2k^2+4k^2+(k-1)
=8k^2+5k+3.
\]
Since $t<1/k^2$ and $f_H(S_H)<10k^2$, we have $t f_H(S_H)<10$. Thus the principal's utility at this equilibrium satisfies
\[
\begin{aligned}
u_P((\beta,t),S_H)
=\left(\frac23-t\right)(8k^2+5k+3)>\frac{16}{3}k^2+\frac{10}{3}k-8
&\ge\frac{16}{3}k^2.
\end{aligned}
\]
This provides a lower bound for the utility of $S_H$.\medskip

We next bound the utility of every other equilibrium. Since $\ell_H(S)\ge0$, $w(S)\le2k-1$, and the capped matching term is at most $4k+6$, \Cref{lem:gs-core-value} gives
\[
f_H(S)\le k^2\bigl(2\max\{a_0,b_2\}+2b_1+b_2+4a_0\bigr)+6k+6\le k^2\left(2\max\{a_0,b_2\}+2b_1+b_2+4a_0+\frac14\right).
\]
The last inequality uses $6k+6\le8k\le k^2/4$, since $k\ge32$.

Let $S\ne S_H$ be an equilibrium at a contract $\alpha$. If $a_0=1$ and $b_2=0$, \Cref{lem:gs-high-share-isolation} forces $b_1=0$. Also, selecting $a_0$ requires $\alpha_1\ge\beta$ by \Cref{lem:gs-agent1-beta}. Suppose now that both $a_0$ and $b_2$ are selected. Removing the action $a_0$ saves its cost $2k^2$ but reduces the agent's utility by $\alpha_1\big(f_H(S)-f_H(S\setminus\{a_0\})\big)$. Hence, since $S$ is an equilibrium, $\alpha_1\bigl(f_H(S)-f_H(S\setminus\{a_0\})\bigr) \geq 2k^2$. Now, since $b_2$ and $a_0$ are present,  \Cref{lem:gs-a0-marginal} gives $\alpha_1\bigl(f_H(S)-f_H(S\setminus\{a_0\})\bigr)
\le\alpha_1(4k^2+4)$. Combining both inequalities,
\begin{equation*}
\alpha_1\ge\frac{2k^2}{4k^2+4}\ge\frac49.
\end{equation*}
Combining these observations with the reward bound gives
\[
\renewcommand{\arraystretch}{1.3}
\begin{array}{c|c|c|c}
\text{case}
&\text{reward bound}
&\text{necessary share}
&\text{principal-utility bound}\\
\hline
a_0=0
&(5+\tfrac14)k^2
&\alpha_1\ge0
&(5+\tfrac14)k^2\\
a_0=1,\ b_2=0
&(6+\tfrac14)k^2
&\alpha_1\ge\tfrac13
&(4+\tfrac14)k^2\\
a_0=b_2=1
&(9+\tfrac14)k^2
&\alpha_1\ge\tfrac49
&(5+\tfrac14)k^2
\end{array}
\]
For the last column, we use $u_P(\alpha,S)\le(1-\alpha_1)f_H(S)$. The leading terms are multiplied by $1$, $2/3$, and $5/9$, respectively, while the contribution of the additional $k^2/4$ is at most $k^2/4$. Thus every equilibrium $S\ne S_H$ satisfies $u_P(\alpha,S)\le21k^2/4$. \medskip 
Comparing this with the utility at $S_H$ gives
\[
\frac{u_P(\alpha,S)}{u_P((\beta,t),S_H)}
<\frac{21/4}{16/3}
=\frac{63}{64}.
\]
Since $S_H$ occurs only at $(\beta,t)$, this is the unique
$64/63$-approximate contract.
\end{proof}

\subsection{Best-Nash Queries}

We next show that, away from $(\beta,t)$, the two instances have the same principal-optimal equilibria. Let $\bne_\star$ be the best Nash correspondence associated to the instance with $f_\star$. 

\begin{lemma}\label{lem:gs-bne-locality}
For every $\contract\in[0,1]^2$, $\displaystyle \bne_H(\contract)=
\begin{cases}
\{S_H\},&\contract=(\beta,t_H),\\
\bne_\star(\contract),&\contract\ne(\beta,t_H).
\end{cases}$.
\end{lemma}

\begin{proof}
Let $\contract=(\alpha_1,\alpha_2)$ and $t=t_H$. At $(\beta,t)$, the claim follows from \Cref{lem:gs-approximation-gap}. If $\alpha_1\ne\beta$, all equilibria and their rewards agree by \Cref{lem:gs-agent1-beta}.

It remains to consider $\alpha_1=\beta$ and $\alpha_2\ne t$. By \Cref{lem:gs-agent1-beta,lem:gs-high-share-isolation}, every equilibrium containing $a_0$ has agent~2 select no actions and agent~1 select either $k+1$ or $k+2$ actions from $E$. Adding $b_2$ increases the reward by at least its bonus $k^2+1$, so such an equilibrium requires $\alpha_2(k^2+1)\le1$.
If $\alpha_2>1/(k^2+1)$, all equilibria therefore omit $a_0$. Their equilibrium conditions and rewards agree by \Cref{lem:gs-agent1-beta}, proving the claim in this case.

It remains to consider $\alpha_2\le1/(k^2+1)$, still with $\alpha_1=\beta$ and $\alpha_2\ne t$. We will show that, in both instances, the principal-optimal equilibria are exactly the profiles in which agent~2 selects no actions and agent~1 selects $a_0$ together with $k+2$ actions from $E\setminus\{a_1\}$.
We first verify that every such profile is an equilibrium. Fix one of them and denote it by $S$. Since agent~2 selects no actions, $b_1=b_2=0$. Thus $|S\cap E|+b_1=k+2$, and part~(iii) of \Cref{lem:gs-agent1-beta} shows that agent~1's choice, including $a_0$, is a best response. For agent~2, observe that $S$ omits $a_1$, so the loss remains zero after any of its deviations. By \Cref{lem:gs-core-value}, the matching value is already $2k^2+4k+6$ and does not increase when agent~2 adds actions. Selecting $b_1$ or $b_2$ therefore contributes only its modular bonus, giving utility gains
\[
2k^2\alpha_2-2<0,
\qquad
(k^2+1)\alpha_2-1\le0,
\]
respectively. These gains add when both actions are selected, so selecting no actions is a best response. Thus every profile of the stated form is an equilibrium in both instances, with reward
\[
f_H(S)=f_\star(S)
=6k^2+(4k+6)+(k+2)
=6k^2+5k+8.
\]

We now show that every other equilibrium gives the principal strictly less utility. Write $f$ for either reward function. As established above, any equilibrium $S'$ containing $a_0$ has agent~2 select no actions and satisfies $|S'\cap E|\in\{k+1,k+2\}$. Its reward is therefore
\[
f(S')=6k^2+\min\{4|S'\cap E|,\,4k+6\}+w(S')\le6k^2+5k+8.
\]
Equality holds exactly when $|S'\cap E|=k+2$ and $a_1\notin S'$, which gives precisely the profiles described above.
If $S'$ omits $a_0$, \Cref{lem:gs-core-value} and the definition of the reward functions instead give
\[
f(S')
\le 5k^2+1+(4k+6)+(2k-1)
=5k^2+6k+6
<6k^2+5k+8.
\]
Finally, the principal's retained share satisfies $1-\alpha_1-\alpha_2
=\frac23-\alpha_2
\ge\frac23-\frac{1}{k^2+1}>0$, so that maximizing the reward is equivalent to maximizing the principal's utility. The profiles described above are therefore exactly the principal-optimal equilibria in both instances, completing the proof.
\end{proof}

\subsection{Critical Contracts}

We next show that all the instances in this family have polynomially many critical contracts, even on the whole $[0,1]^2$.

\begin{lemma}\label{lem:gs-critical-count}
Each instance $f_H$, as well as the reference instance $f_\star$, has $O(k)=O(m)$ critical contracts.
\end{lemma}

\begin{proof}
Group agent~1's choices according to the number of selected actions from $E$, whether $a_0$ and $a_1$ are selected, and whether the selected actions from $E$ form $H$, form $H$ together with one additional action, or neither. There are $O(k)$ such groups.

Within each group, the cost is identical, and \Cref{lem:gs-core-value} shows that the reward is identical for each of agent~2's four choices. Indeed, the cardinalities and membership indicators determine the reference reward and costs, while the last condition determines the loss. Thus choices within the same group give agent~1 the same utility against every choice of agent~2, at every contract. They also give agent~2 the same utility comparisons between its four choices. Consequently, for each fixed choice of agent~2, the profiles within a group are equilibria at exactly the same contracts. They therefore have the same critical contract, whenever one exists.

There are $O(k)$ groups and only four choices for agent~2, so there are $O(k)$ distinct critical contracts. The same argument applies to $f_\star$, where the loss is always zero.
\end{proof}

In fact, using similar reasoning, a principal-optimal equilibrium can be computed in polynomial time for this family. In the reference instance, the grouping used in the proof of \Cref{lem:gs-critical-count} simplifies: it suffices to record how many actions agent~1 selects from $E$ and whether he selects $a_0$ and $a_1$. For each of these $O(k)$ types and each of agent~2's four choices, we can check whether either agent has a profitable deviation by trying every other type for agent~1 and every other choice for agent~2. Among the pairs that pass this test, return a representative maximizing the principal's utility.

For a hidden instance, \Cref{lem:gs-bne-locality} shows that the reference answer remains valid unless $\alpha_1=\beta$ and $\alpha_2=t_j$ for some $j$. The formula for $t_j$ determines at most one such index. Given $j$, we can compute the $(k-1)$-subset of $E\setminus\{a_1\}$ of lexicographic rank $j$ in polynomial time and add $a_1$ to obtain $H_j$. By \Cref{lem:gs-core-value}, the value of $H_j\cup\{a_0,b_2\}$ differs from its reference value on instance $H_j$ and agrees with it on every other instance. One value query therefore determines whether to return $S_{H_j}$ or the reference answer. Thus one principal-optimal equilibrium can be computed in polynomial time on this family.

\subsection{Oracle Lower Bound}

All instances, including $f_\star$, have the same reward on the full action set, $f_H(T)=f_\star(T)=9k^2+6k+6$. Hence we normalize the instance by dividing every reward and cost by this quantity. This does not change the equilibria, optimal-equilibria query answers, or approximation ratios. Value-query answers are divided by the same known factor. 

\begin{proof}[Proof of \Cref{thm:gs-oracle-lb}]
We prove a lower bound for deterministic algorithms when the instance is drawn uniformly from $\{f_{H_0},\ldots,f_{H_{N-1}}\}$. By Yao's principle, this yields a worst-case lower bound for randomized algorithms. Consider a deterministic algorithm making at most $Q$ value and best-Nash queries. Run the algorithm against the reference instance $f_\star$. Record all queries made by the algorithm during its run on $f_\star$. Let $\mathcal D$ be the set of indices $j$ for which $f_{H_j}$ and $f_\star$ give different answers to at least one recorded query. Each value query has a different answer for at most $k+1$ hidden instances, and each best-Nash query has a different answer for at most one. Taking the union over the at most $Q$ queries gives $|\mathcal D|\le(k+1)Q$.

For every $j\notin\mathcal D$, the algorithm follows exactly the reference execution. Indeed, as long as the answers agree, its next query is the same, and that query again receives the reference answer by the definition of $\mathcal D$. Thus all these instances produce the same transcript and the same output contract.
By \Cref{lem:gs-approximation-gap}, success on $f_{H_j}$ requires the output to be exactly $(\beta,t_j)$. Since the shares $t_j$ are pairwise distinct, the common output succeeds on at most one instance with $j\notin\mathcal D$. Even counting every instance indexed by $\mathcal D$ as a success, the algorithm therefore succeeds on at most $|\mathcal D|+1\le(k+1)Q+1$ of the $N$ hidden instances.

Thus, when the hidden instance is drawn uniformly from the $N$ possibilities, every deterministic algorithm making at most $Q$ queries succeeds with probability at most 
\begin{equation*}
\Pr[\text{success}]
=\frac{\text{number of instances on which the algorithm succeeds}}{N}
\le\frac{|\mathcal D|+1}{N}
\le\frac{(k+1)Q+1}{N}.
\end{equation*}
Therefore, an algorithm that succeeds with probability at least $2/3$ on every instance must satisfy $\frac23\le\frac{(k+1)Q+1}{N}$, and hence 
\[
Q\ge\frac{(2/3)N-1}{k+1}.
\]
Overall, as $N=\binom{2k-1}{k-1}\ge\frac{2^{2k-1}}{2k}$, the required number of queries is $2^{\Omega(k)}=2^{\Omega(m)}$. The construction has $m=2k+3$ actions. For even $m$, add one positive-cost action that never changes the reward. It is absent from every equilibrium, and a value query containing it is answered by deleting it from the query.

This rules out a PTAS using polynomially many value and best-Nash queries, even for normalized gross-substitutes rewards with $O(m)$ critical contracts.
\end{proof}

}

\appendix

\section{AI Use Disclosure}

The extension of the critical-points framework to multi-agent critical
contracts, the identification of the three sources of computational
hardness, and the conceptual framing of this paper were developed
entirely by the authors.

Generative AI tools made substantive contributions to the development
of several results. Most of the hardness results were obtained with
assistance from ChatGPT using GPT-5.2 and GPT-5.4, through interactions
requiring extensive guidance from the authors. Initial versions of
the algorithmic results were obtained by the authors and subsequently
strengthened with AI assistance. In particular, the authors first
obtained the subadditive approximation result for a constant number
of agents; the extension removing this restriction, presented in
Section~4.3, was obtained with GPT-5.4. Similarly, for supermodular
rewards with polynomially many critical contracts, the authors
initially developed a deterministic search algorithm for a constant
number of agents. AI assistance was subsequently used to remove
the restriction on the number of agents, yielding first a randomized
algorithm obtained via GPT 6., and then a de randomized version currently presented in Section~3.3. obtained via GPT 6.2. 

The overall organization and explanatory approach of the paper are
the authors' own. AI tools were also used to assist with the
preparation of some figures and to polish the writing. The authors
take full responsibility for the correctness, originality, and
integrity of all results, proofs, figures, text, and references
in the manuscript.

\PrintComplementsAppendix
\PrintSubstitutesAppendices

\bibliographystyle{abbrvnat}
\bibliography{Bibliography}
\end{document}